\documentclass[10pt,a4paper]{article}
\usepackage{authblk}
 \usepackage[utf8]{inputenc}
 \usepackage{amsmath}
 \usepackage{amsfonts}
 \usepackage{amssymb}
 \usepackage{makeidx}
 \usepackage{graphicx}
 \usepackage{lmodern}
 \usepackage{kpfonts}
\numberwithin{equation}{section}
\usepackage{fullpage}
\usepackage{cite}
\usepackage[T1]{fontenc}
\usepackage{mathtools}
\newcounter{mysubequations}
\renewcommand{\themysubequations}{(\roman{mysubequations})}
\newcommand{\mysubnumber}{\refstepcounter{mysubequations}\themysubequations}

\title{  $h(1) \oplus su(2)$ vector algebra eigenstates with eigenvalues in the matrix domain}

 \author{Nibaldo Edmundo Alvarez Moraga\footnote{Email address: nibaldo.alvarez.m@mail.pucv.cl} }
 
  \affil{ Autonomous Center of Theoretical Physics and Applied Mathematics, \\
2805 Place de Darlington, Montreal (Quebec), H3S1L4, Canada}

\begin{document}

   \maketitle

\begin{abstract}
A new set of  $ h(1)  \oplus  su(2)$ vector algebra eigenstates on the matrix domain is obtained by defining them as  eigenstates  of a generalized annihilation operator formed from a  linear combination of the generators of this algebra which  eigenvalues are distributed as the elements of a square complex normal matrix.   A combined method is used to compute  these eigenstates, namely, the method of exponential operators and that of a system of first-order linear differential equations. We compute these states for all possible combination of generators and classify them in different categories  according to a generalized commutation relation as well as according to the value of a characteristic parameter related to the  $su(2)$ algebra  eigenvalues.   Proceeding  in this way, we found a subset   of  generalized vector coherent states in the matrix domain which can be easily separated from the general set of Schr\"odinger-Robertson minimum uncertainty intelligent states.  In particular, for a special choice of the matrix eigenvalue parameters we found the so-called vector coherent states  with  matrices associated to the Heisenberg-Weyl group as well as a generalized version of them,  and also a direct  connection with the coherent state quantization of quaternions.  
 \end{abstract}
 
\section{Introduction}
Coherent states\cite{Schro}  expressed as eigenvectors of the lowering harmonic oscillator operator verifying important physical properties were
first introduced by J. R. Klauder \cite{JRK}. From there, several definitions of coherent states related  to this  and other  quantum physical system were introduced in the literature as well as generalization of this concept of theoretical interest  in Physics and  Mathematics  became to know, all of them assuring the accomplishment of some desired properties such as minimum uncertainty states,  resolution of the identity, temporal stability  or  action-identity.   Thus,  some consider them as states generated from the action of a unitary irreducible representation of the group  on a fixed state of the Hilbert space representation,  others define these states  as eigenstates of  ladder operators generating a representation of the associated Lie algebra or as states associated to physical Hamiltonians having a non-degenerate spectra \cite{JRG}--\cite{Bar-Gir}.   The concept of algebra eigenstates associated to an arbitrary  Lie group was introduced by C. Brif who defined them as eigenstates of elements of the corresponding complex Lie algebra \cite{Cbr-96,Cbr-97}. He used this method, among others,  to compute the algebra eigenstates of the $ SU(2) $ and $ SU(1,1) $ Lie groups and shown  that  these states  included different subsets of  Perelomov’s generalized coherent states as well as  of the so called intelligent  states that minimize the Schr\"odinger-Robertson uncertainty relation.   An adaptation of this last concept to the $h(1) \oplus su(2)$ Lie algebra allowed us to generate  new sets of generalized coherent and squeezed states associated with  physical systems whose Hamiltonians were constructed as the product of well defined creation and  annihilation operators extracted from  the set of elements of this algebra \cite{NAM-VH}. Then, a new generalization of the concept of vector coherent states to the matrix domain arose by substituting the characteristic complex parameter of the canonical Heisenberg-Weyl coherent states by a normal matrix and requiring  that the resulting states to verify the resolution of the identity \cite{TA-KT}.  The same successful substitution was done in the $SU(1,1)$ Gilmore-Perelomov and  Barut-Giraldello type coherent states.   An important consequence of this extension is  that  it led to the coherent state quantization of quaternions \cite{BM-KT}.  In this article we extend the definition of algebra eigenstates with eigenvalues in the complex domain, where by the way the eigenvectors are states in the Hilbert space of representation of the algebra, to consider eigenvalues on the complex matrix domain, that is, complex matrix eigenvalues, where now the associated  eigenvectors, that from now on  we will call vector algebra eigenstates,  are vectors whose components are by themselves  states  of the Hilbert space of representation of the algebra. We use this definition to compute  the vector algebra eigenstates of the  $h(1) \oplus su(2)$ Lie algebra, but it will be clear by regarding the subsequent development that it can be used  to compute the vector algebra eigenstates for  other Lie algebras.

 This article is organized as follows. In section \ref{sec-two} we introduce the $h(1) \oplus su(2)$  Lie algebra generators,  its  commutation relations and  its action on a chosen irreducible representation space of this algebra and establish the nomenclature to  describe the vector states.   In section  \ref{sec-three} we give the definition and compute the  vector algebra eigenstates of the   annihilation operator associated to the quantum harmonic oscillator and compare  the resulting expressions with the set of Heisenberg-Weyl vector coherent states on the matrix domain that we can find in the literature. Also, we choose the entries of the matrix eigenvalue as equal to the matrix elements of the elements of the $su(2) \oplus \hat{I}_j$ algebra in the $j$ irreducible representation. 
This alternative allows us, for a particular choice of the parameters, to generalize the quaternionic vector coherent states to an arbitrary $su(2)$ irreducible representation space. Moreover, we there we  show the connection of our general method with the coherent state quantization of quaternions.      In section \ref{sec-four},  we give the definition of vector algebra eigenstates  for the  $h(1) \oplus su(2)$ Lie algebra and compute these states for all possible linear combinations of generators of this algebra in the case when the matrix eigenvalue is  complex and normal.  In appendices \ref{appa} and  \ref{appb}, we recalculate  completely the $h(1) \oplus su(2)$ algebra eigenstates by combining  the exponential operator and the linear  differential  equations system methods and  prepare appropriately the expressions for using them in the main body of this article.

\section{ $h(1) \oplus su(2) $ Lie algebra and its generators}
\label{sec-two}
The $ h(1) \oplus su(2) $ Lie algebra is the direct sum of the Heisenberg Weyl algebra, whose generators are given by $ \hat{a},\hat{a}^{\dagger}$ and the identity $\hat{I}$, and the $su(2)$ algebra, generated by $\hat{J}_{-}, \hat{J}_{+} $ and $
\hat{J}_{3} $. These set of operators satisfy the commutation relations

 \begin{equation}
  \left[ \hat{a} ,\hat{a}^{\dagger}\right]= \hat{I},   \quad 
   \left[ \hat{I},\hat{a}  \right]=\hat{0}   \quad \mbox{and} \quad
    \left[ \hat{I} ,\hat{a}^{\dagger} \right]=  \hat{0},  \label{h1-Lie-algebra-commutators}
\end{equation}
where $\hat{0}$ is the null operator, and 
\begin{equation}
 \left[ \hat{J}_{+},\hat{J}_{-}\right]= 2  \hat{J}_{3} 
 \quad \mbox{and} \quad
    \left[\hat{J}_{3}, \hat{J}_{\pm}\right]=  \pm \hat{J}_{\pm}, \label{su2-Lie-algebra-commutators}
     \end{equation}
respectively.

The action of the former   on the orthonormal basis vectors $ \mid n \rangle , n=0,1,2,..$ spanning the infinite dimensional Hilbert space $ \mathcal{H} $  is given by
\begin{equation}
\hat{a}  \mid n \rangle  = \sqrt{n}  \mid n-1 \rangle, \quad  \hat{a}^{\dagger}  \mid n \rangle = \sqrt{n+1}  \mid n+1 \rangle \label{action-annihilation}
 \end{equation} 
 and 
\begin{equation}
 \hat{I} \mid n \rangle  = \mid n \rangle,
\end{equation} with $n=0,1,2...$ and the action of the latter on the  orthonormal basis vectors $\mid j,m \rangle$  of the irreducible representation $j,$  of dimension $2j+1,$ is given by
\begin{equation}
\hat{J}_{\pm} \mid j, m  \rangle =\sqrt{ ( j \mp m) ( j \pm m +1)}  \mid j, m \rangle \quad \mbox{and} \quad
  \quad  \hat{J}_{3} \mid j, m  \rangle = m \mid j, m \pm 1  \rangle,   \label{action-j-pm}
   \end{equation} where $j$ is a not negative integer or half-integer and  $m=-j,-j+1, ..., j-1,j.$

The orthonormal basis vectors of the $h(1) \oplus su(2) $ algebra, for fixed $j,$ can be written as the direct product \begin{equation}
\mid n \rangle \otimes \mid j , m\rangle = \mid n ; j,m\rangle, \quad  n=0,1,2...; \quad  m=-j, ....,j.
\end{equation}

Then, for fixed $j,$ a general state $\mid \psi \rangle^j  $ of the algebra $h(1) \oplus su(2)$
can be expressed in the form
\begin{equation}
\mid \psi \rangle^{j} = \sum_{n=0}^{\infty} \sum_{m=-j}^{j} C^{j}_{n,m} \mid n ; j , m \rangle, \label{big-psi}
\end{equation} 
where $ C^{j}_{n,m} $ are complex coefficients for all $n,m.$
\subsection{Realization of $h(1)$ generators on analytic functions space }
Another realization of the oscillator algebra can be found  in the  Fock -Bargman  repesentation space $\mathcal{H}$ of analytic functions $ f \left( \zeta  \right),  \zeta \in \mathbb{C},$  verifying the scalar product
\begin{equation}
\left( f_{1},f_{2}\right) = \int_{\mathbb{C}} 
f_{1}^{\ast} \left(\zeta \right) f_{2}\left( \zeta\right) e^{\vert \zeta \vert^{2}} \frac{d\zeta  d\zeta^{\ast}}{2 \pi i}, \zeta \in \mathbb{C}.
\end{equation} 
In this space, an arbitrary analytic function $f\left( \zeta \right)$ can be expressed as a linear combination of analytic elementary functions  $ \langle \zeta \mid n \rangle= \varphi_{n}(\zeta)= \frac{\zeta^{n}}{\sqrt{n!}}, n=0,1,2...,$ which span  $\mathcal{H}$ and verify the orthonormality property

\begin{equation}
\left(\varphi_{r}(\zeta),\varphi_{s}(\zeta) \right)= \delta_{rs}, \;  \forall r,s \in \mathbb{N}_{0}.
\end{equation}

In terms of these states the function $f(\zeta)$ writes
\begin{equation}
f\left( \zeta \right) = \sum_{n=0}^{\infty} c_n \varphi_n (\zeta) = \sum_{n=0}^{\infty} c_n \frac{\zeta^{n}}{\sqrt{n!}}.
\end{equation}

In this representation the $h(1)$ algebra generators take the form:
\begin{equation}
\hat{a} = \frac{d}{d\zeta}, \quad \hat{a}^{\dagger} = \zeta, \quad  \textrm{and} \; \hat{I}= 1. 
\label{action-annihilation-zeta}
\end{equation}

As we can see above, the  elementary analytic  states can be thought as  projections of  the number harmonic oscillator eigenstates $\mid n \rangle$ over the $\mid \zeta \rangle$ state. Using this fact, the projection of the general state (\ref{big-psi}) onto $\mid \zeta \rangle \otimes \hat{I}^{j}$ state, where $\hat{I}^{j}=\sum_{m=-j}^{j} \mid  j, m \rangle \langle j,m \mid $ is the identity operator in the $2j+1$ dimensional $j$ space, is given by
 
\begin{equation}
\mid \psi (\zeta) \rangle^j =  \hat{I}^j  \otimes \langle \zeta \mid \psi \rangle^j = \sum_{m=-j}^{j}  
\psi^{j}_{m}(\zeta) \otimes \mid j ,m \rangle, \label{general-state-expression}
\end{equation}
where
\begin{equation}
\psi^{j}_{m}(\zeta) \rangle = \sum_{n=0}^{\infty}  C^{j}_{n,m} \frac{\zeta^{n}}{\sqrt{n!}}, \quad m=-j, \cdots, j, \label{big-psi-zeta}
\end{equation}
are analytic states functions, which can  also be interpreted as a kind of coefficients of the expansion of the state $ \mid \psi (\zeta) \rangle^j $ in terms of  the basis states $\mid j ,m \rangle.$

\section{Quantum oscillator annihilation operator vector eigenstates  with matrix eigenvalues}
\label{sec-three} 

In this section we define and solve the matrix eigenvalue equation for the quantum harmonic oscillator annihilation operator.  We will use the basis vector  associated to the  $ h(1) \oplus su(2)$ algebra representation to  express  the states and we will see that the set of solutions of this  matrix eigenvalue equation contains the subset of vector coherent states over the matrix domain studied in the literature \cite{TA-KT}. We note that, in this particular case,  the use of the $su(2)$ sector representation space is optional, it could be replaced by another algebra representation space.  

Let us start with a preliminary  definition of the vector algebra eigenstates of  $\hat{a}$  as   those states that satisfy the matrix eigenvalue equation

\begin{equation}
\hat{a} \mid \Psi \rangle^{j}_K = \tilde{M} \mid \Psi \rangle^{j}_{K}, \label{eq-a-matrix-eigenvalues}
\end{equation}
   where  $\mid \Psi\rangle^{j}_K$  is the $K$ component vector state given by 
    \begin{equation} 
 \mid \Psi\rangle^{j}_K =\begin{pmatrix}
\mid \psi  \rangle^{j}_{[1]} \\
\mid \psi  \rangle^{j}_{[2]} \\
\vdots\\
\mid \psi \rangle^{j}_{[K-1]} \\
\mid \psi \rangle^{j}_{[K]} 
\end{pmatrix} ,
\end{equation}
where each state $\mid \psi \rangle^j_{[s]} =\sum_{n=0}^{\infty} \sum_{m=-j}^{j} C^{[s] j}_{n m} \mid n \rangle \otimes j , m \rangle, \;  s=1,\cdots,K$ is a unknown state  of the $h(1) \oplus su(2)$ representation space, to be determined,  and ${\tilde M}$ a diagonalizable complex  square matrix of  dimension $K \times K.$   
    
\subsection{When $\tilde{M}$ is a normal matrix}
\label{sec:normal-case}
In general, when  $\tilde{M}$ is normal, i.e., when $\tilde{M} \tilde{M}^\dagger = \tilde{M}^\dagger \tilde{M} ,$  a similarity transformation of $\tilde{M}$  by an unitary matrix $\tilde{U}$ can be done that leads ({\ref{eq-a-matrix-eigenvalues}}) to the simplified form  
\begin{equation}
\hat{a} \mid \tilde{\Psi} \rangle^{j}_K = \tilde{D} \mid \tilde{\Psi} \rangle^{j}_K, \label{eq-a-matrix-eigenvalues-diagonal}
\end{equation}
where  

\begin{equation} \tilde{D} = \tilde{U}^\dagger \tilde{M} \tilde{U}, \quad \mbox{and} \quad \mid \tilde{\Psi} \rangle^j_K = 
\tilde{U}^{\dagger} \mid \Psi \rangle^{j}_K,  \end{equation}    where $\tilde{D}$ is a diagonal matrix which entries are the eigenvalues of ${\tilde M}$  and $\mid \tilde{\Psi} \rangle^j_K   $ is the corresponding transformed state. Thus, each component of the matrix equation (\ref{eq-a-matrix-eigenvalues-diagonal}) verifies the eigenvalue equation

\begin{equation}
\hat{a} \mid \tilde{\psi} \rangle^j_{[s]} = \lambda^j_{[s]}  \mid  \tilde{\psi} \rangle^j_{[s]}, \quad  s=1,\cdots, K. 
\end{equation}
By solving this equations we realize that each  component  $\mid  \tilde{\psi} \rangle^j_{[s]}$  can be written as  a direct product between a canonical oscillator coherent state and a  generic state of the $j$ representation space of the $su(2)$  Lie algebra , i.e,

\begin{equation}
 \mid  \tilde{\psi}  \rangle^j_{[s]} = e^{\lambda^j_{[s]} \hat{a}^\dagger}  \mid  0  \rangle \otimes  \sum_{m=-j}^{j}  \tilde{\psi}^j_{[s]m} (0)  \mid j , m\rangle,   \quad   s =1, \cdots K,   \label{psi-componenents}
\end{equation}
then  we have  
 
 \begin{equation}
\mid \Psi (\tilde{M}) \rangle^j_K =  \tilde{U} \mid \tilde{\Psi} (\Lambda) \rangle^{j}_K = \mathcal{N}^{-1 /2} \tilde{U} e^{ \tilde{D} \hat{a}^\dagger} \mid \tilde{\Psi} (0) \rangle^{j}_K = \mathcal{N}^{-1 /2}  e^{ \tilde{M} \hat{a}^\dagger} \mid \Psi (0) \rangle^{j}_K
\label{VCS-expression-one} \end{equation} 
  where $\mathcal{N}$ is a normalization constant depending on $\lambda^j_{[s]}$ values but also on the choice of the $K \times 2j+1$ arbitrary constants $ \tilde{\psi}^j_{[s]m} (0),    \quad  s=1, \cdots,K, \quad m=-j,\cdots,j, $
and
\begin{equation}
 \mid \tilde{\Psi} (0) \rangle^{j}_K  = \begin{pmatrix}
 \mid 0  \rangle \otimes \mid \tilde{\psi} (0) \rangle^j_{[1]} \\  \mid 0  \rangle \otimes \mid \tilde{\psi}(0) \rangle^j_{[2]}  \\
    \vdots \\  \mid 0  \rangle \otimes \mid \tilde{\psi} (0) \rangle^j_{[K-1]} \\
   \mid 0  \rangle \otimes \mid \tilde{\psi} (0) \rangle^j_{[K]}
\end{pmatrix} , \label{zero-eigenvalue-state-tilde}
\end{equation} 
with  $\mid \tilde{\psi} (0) \rangle^j_{[s]} = \sum_{m=-j}^{j}  \tilde{\psi}^j_{[s]m} (0)  \mid j , m\rangle, \;     s=1, \cdots,K. $  

We notice that   (\ref{zero-eigenvalue-state-tilde}) represent a  set of $K \times 2j+1$  vector eigenstates of $\hat{a} $ with associated  eigenvalues equal to zero, just like the states  $\mid \Psi (0) \rangle^j_K = \tilde{U} \mid \tilde{\Psi} (0) \rangle^{j}_K .$ Then, by using the    Baker-Campbell-Hausdorff formula  $\exp(A + B)= \exp(A) \exp(B) \exp(- \frac{1}{2} [A,B]),$  for two operators $A$ and $B$ whose commutator commutes with both $A$ and $B,$    we get

\begin{equation}
\mid \Psi (\tilde{M}) \rangle^j_K =   \mathcal{N}^{-1 /2}   \tilde{U} \; e^{  \frac{1}{2}  \tilde{D} \tilde{D}^\dagger } e^{ (\tilde{D} \hat{a}^\dagger - 
\tilde{D}^\dagger \hat{a}) }  \mid \tilde{\Psi} (0) \rangle^{j}_K =\mathcal{N}^{-1 /2}   e^{  \frac{1}{2}  \tilde{M} \tilde{M}^\dagger } e^{ (\tilde{M} \hat{a}^\dagger - \tilde{M}^\dagger \hat{a}) }  \mid \Psi (0) \rangle^{j}_K.
\end{equation}

\subsubsection{A special choice of basis vectors}
The choice of the vector $\mid \tilde{\Psi} (0) \rangle^{j}_K $ has influence in the normalization constant $\mathcal{N}.$  
Choosing it as a vector which components form an orthonormal basis in the fixed $j$  representation subspace  makes it easier to  compute, choosing, for example, $\mid \tilde{\psi} (0) \rangle^j_{[s]} = \mid 0 \rangle \otimes  \mid j ,  m_{s} \rangle,    \;  s=1, \cdots, K,$ with $      \langle j ,  m_{\tilde{s}}    \mid j ,  m_{s} \rangle = \delta_{\tilde{s} s}, $ we get the normalized vector coherent states 

\begin{equation}
\mid \Psi (\Lambda) \rangle^j_K = \frac{1}{\sqrt{\sum_{s=1}^{K} e^{\|\lambda_{[s]}^{j}\|^2}}} \tilde{U} e^{\tilde{D} \hat{a}^\dagger} \mid \tilde{\Psi} (0) \rangle^{j}_K, 
\end{equation} 
which in the matrix form look like

\begin{eqnarray}
\mid \Psi (\Lambda) \rangle^j_K &=& \frac{1}{\sqrt{\sum_{s=1}^{K} e^{\|\lambda^{j}_{[s]}\|^2}}} \nonumber \\       
&\times& \tilde{U} \begin{pmatrix} e^{\lambda^j_{[1]} \hat{a}^\dagger} &0 &0& \cdots & \cdots&0\\0 &  e^{\lambda^j_{[2]} \hat{a}^\dagger} &0 & \dots& \cdots &0 \\ 0& 0& \ddots &0& \cdots &0 \\
 \vdots& \vdots & \vdots & \ddots & \vdots & \vdots \\ 
 0 & \cdots  &\cdots& 0 & e^{\lambda^j_{[K-1]} \hat{a}^\dagger}  &0 \\0 & \cdots& \cdots &0 & 0& e^{\lambda^j_{[K]} \hat{a}^\dagger}\end{pmatrix} 
 \begin{pmatrix}
 \mid 0  \rangle \otimes \mid  j, m_1 \rangle \\
 \mid 0  \rangle \otimes \mid j, m_2 \rangle\\
\vdots \\
 \mid 0  \rangle \otimes \mid  j ,  m_{k-1} \rangle \\
 \mid 0 \rangle  \otimes \mid j, m_{K} \rangle \\
 \end{pmatrix}. \label{vector-coherent-states-general-normal-case}
\end{eqnarray} 
On the other hand, if we choose these states in  such a way  that
\begin{equation}
 \mid \tilde{\Psi} (0) \rangle^{j}_K  = \tilde{U}^\dagger  \begin{pmatrix}
 \mid 0  \rangle \otimes \mid  j , m_1 \rangle\\
  \mid 0  \rangle \otimes \mid  j , -m_2 \rangle \\
  \vdots \\   \mid 0  \rangle \otimes \mid  j , m_{K-1} \rangle  \\
    \mid 0  \rangle \otimes \mid  j , m_{K} \rangle
\end{pmatrix} ,  \label{fundamental-state-choice-two}
\end{equation} 
the  corresponding vector coherent state in (\ref{VCS-expression-one}) writes

\begin{equation}
\mid \Psi (\tilde{M} ) \rangle^j  = \mathcal{N}^{-1 /2}  e^{ \tilde{M} \hat{a}^\dagger}  \begin{pmatrix}
 \mid 0  \rangle \otimes \mid  j , m_{1} \rangle\\
  \mid 0  \rangle \otimes \mid  j , m_{2} \rangle \\
  \vdots \\   \mid 0  \rangle \otimes \mid  j , m_{k-1} \rangle  \\
    \mid 0  \rangle \otimes \mid  j , m_{K} \rangle
\end{pmatrix} =  \mathcal{N}^{-1 /2}  \sum_{n=0}^{\infty}  \frac{\tilde{M}^n}{\sqrt{n!}}   \begin{pmatrix}
 \mid n  \rangle \otimes \mid  j , m_{1} \rangle\\
  \mid n  \rangle \otimes \mid  j , m_{2} \rangle \\
  \vdots \\   \mid n  \rangle \otimes \mid  j , m_{K -1} \rangle  \\
    \mid n \rangle \otimes \mid  j , m_{K} \rangle \label{VCS-expression-two}
\end{pmatrix},
\end{equation}
which, with a little difference in the choice  of the basis vectors, belong to the class of vector coherent states on the matrix domain  studied in \cite{TA-KT}. Now, the normalization constant $\mathcal{N} $ is more difficult to calculate, the general expression for it is:

\begin{equation}
\mathcal{N} =     {}^{j}\langle \Psi (0)  \mid  e^{ \tilde{M} \tilde{M}^\dagger }   \mid \Psi (0) \rangle^j. \label{factor-normalization-N}
\end{equation} 

 In this way,  the multiple possibilities of choosing the ground state vectors show us the  high level of degeneracy \cite{AT-FB}  of the energy eigenvalues of the oscillator harmonic Hamiltonian $H = \hat{a}^\dagger \hat{a}.$  Indeed, the construction of the energy eigenstates $\mid \mathbb{E}_n \rangle^j, n=0,1,\cdots, $ of $H,$ associated to the eigenvalue $n,$ which can be found in the usual way 
 
 \begin{equation}
 \mid \mathbb{E}_n \rangle^j = \frac{\hat{a}^n }{\sqrt{n!}}    \mid \tilde{\Psi} (0) \rangle^{j},
\end{equation}   
makes evidence of this phenomena.  In the remainder of this section, excepting   section \ref{quaternion-2}, in order to illustrate the theory,  we will use the choice made in equation   (\ref{vector-coherent-states-general-normal-case}) for the fundamental vector state. 

\subsubsection{ Vector coherent states linked to the matrix elements of the $su(2)$ algebra generators}

In this section we will study the special case when the matrix elements of $\tilde{M}$ are given by the matrix elements of the operator  $\beta \hat{I}^j - \beta_+ \hat{J}_- + \beta_- \hat{J}_+ + \beta_3 \hat{J}_3, $  where $ \beta, \beta_{\pm} $ and $\beta_3$ are complex numbers,  in the usual basis  spanning the $j$ irreducible  representation space  of the $su(2)$ algebra, i.e,
\begin{equation}
\tilde{M}_{m \ell} =   \langle j , m  \mid [\beta \hat{I}^j - \beta_+ \hat{J}_-  + \beta_-  \hat{J}_+ + \beta_3  \hat{J}_3 \mid j , \ell \rangle, \quad m, \ell = -j, -j+1, \cdots, j-1,j. 
\end{equation}
The explicit form of this matrix is given  by
  
\begin{equation}
M = 
 \left(
\begin{smallmatrix}
\beta + j \beta_{3} & - \sqrt{2j} \beta_{+} & 0 & 0 & \dots & 0 \\
- \sqrt{2j} \beta_{-} & \beta + (j -1) \beta_{3}  & - \sqrt{(2j-1) 2} \beta_{+}& 0 & \dots & 0 \\
0 & - \sqrt{(2j-1) 2} \beta_{-} & \beta + (j -2) \beta_{3} &   - \sqrt{(2j-2) 3} \beta_{+}
& \dots &0\\
\vdots  & \vdots & \ddots& \ddots & \ddots & \vdots \\
0& 0& - \sqrt{3 (2j-2) } \beta_{-} & \beta - (j -2) \beta_{3} & - \sqrt{2 (2j-1) } \beta_{+} &0 \\
0& 0&0&- \sqrt{2 (2j-1) } \beta_{-} & \beta - (j -1) \beta_{3}  & - \sqrt{2j} \beta_{+}\\
0& 0&0&0& - \sqrt{2j} \beta_{-}& \beta-j \beta_{3}
 \end{smallmatrix} \right)  \label{M-matrix-general-expression-part-annihilation}
 \end{equation}

This matrix becomes normal when the $\beta_{\pm}$ and $\beta_3$ parameters verify equation  (\ref{normality-conditions}), i.e, $\|\beta_-\| = \|\beta_+ \|$ and $\beta_3 {\beta_-}^\ast = \beta_{3}^\ast \beta_+. $  Under these conditions the  eigenvalues of this matrix $\lambda^j_{[m]} = \beta + m b, \; m=-j, \cdots,j,$  where $b=\sqrt{4\beta_+ \beta_- + \beta_3^2} \neq 0,$ are all different.Thus with the help of equation ( \ref{vector-coherent-states-general-normal-case}) and by spanning the vector  component states in the same $su (2)$ basis vector that we  used to compute the matrix elements of $\tilde{M} ,$   we can build  the vector coherent states associated to the $su(2)$ algebra:

\begin{eqnarray}
\mid \Psi (\Lambda) \rangle^j &=& \frac{1}{\sqrt{\sum_{m=-j}^{j} e^{\|(\beta - m b)     \|^2}}} \nonumber \\       
&\times& \quad \tilde{U}  \begin{pmatrix} e^{(\beta +j b) \hat{a}^\dagger} &0 &0& \cdots & \cdots&0\\0 &  e^{(\beta + (j-1) b) \hat{a}^\dagger} &0 & \dots& \cdots &0 \\ 0& 0& \ddots &0& \cdots &0 \\
 \vdots& \vdots & \vdots & \ddots & \vdots & \vdots \\ 
 0 & \cdots  &\cdots& 0 & e^{(\beta - (j-1)b) \hat{a}^\dagger}  &0 \\0 & \cdots& \cdots &0 & 0& e^{(\beta -j b) \hat{a}^\dagger}\end{pmatrix} 
 \begin{pmatrix}
 \mid 0 ; j, -j \rangle \\
 \mid 0 ; j, -j+1 \rangle\\
\vdots \\
 \mid 0 ; j, j -1 \rangle \\
 \mid 0 ; j, j \rangle \\
 \end{pmatrix}, \label{vector-coherent-states-general-normal-case-h1-su2}
\end{eqnarray}
where the entries of the unitary matrix  $\tilde{U}$ are given in appendix \ref{appb}, more precisely in equation (\ref{matrix-elements-T}):

\begin{equation}
U_{m \ell} = T^j_{m \ell}[\beta_+,\beta_- ,\beta_3].  
\end{equation}

\subsubsection{Vector coherent states in the  $j=\frac{1}{2}$ representation}
For example, in the special case when $j= \frac{1}{2},$ the unitary matrix $\tilde{U}$ is given by 

\begin{equation}
U= \begin{pmatrix}
\sqrt{\frac{b +\beta_3}{2b}}&  \frac{2 \beta_+}{\sqrt{2b (b+\beta_3)}}\\
\frac{- 2 \beta_-}{\sqrt{2b (b+\beta_3)}}&\sqrt{\frac{b +\beta_3}{2b}}
\end{pmatrix}\end{equation}
 and the  normal vector coherent states in  (\ref{vector-coherent-states-general-normal-case-h1-su2}) becomes

\begin{eqnarray}
\mid \Psi (\Lambda) \rangle^{\frac{1}{2}} = \frac{1}{\sqrt{e^{\|( \beta + \frac{b}{2} ) \|^2} + e^{\| (\beta - \frac{b}{2}) \|^2}}}
 \begin{pmatrix}
\sqrt{\frac{b +\beta_3}{2b}}&  \frac{2 \beta_+}{\sqrt{2b (b+\beta_3)}}\\ \times
\frac{- 2 \beta_-}{\sqrt{2b (b+\beta_3)}}&\sqrt{\frac{b +\beta_3}{2b}}
\end{pmatrix} \nonumber \\
\begin{pmatrix} e^{(\beta + b /2) \hat{a}^\dagger} &0  \\ 
 0 & e^{(\beta - b /2) \hat{a}^\dagger}
 \end{pmatrix} 
 \begin{pmatrix}
 \mid 0 ;  \frac{1}{2},  - \frac{1}{2} \rangle \\
  \mid 0 ;  \frac{1}{2} ,  + \frac{1}{2} \rangle \\
 \end{pmatrix}
\end{eqnarray}
where $b = \sqrt{4 \beta_{+} \beta_{-} + \beta_{3}^{2}},$
or better, after some few manipulations on the normalization constant:

 \begin{eqnarray}
\mid \Psi[\beta, b(\beta_{+},\beta_{-},\beta_{3}) ]\rangle^{\frac{1}{2}} =
\frac{e^{- \frac{1}{2}\left[ \|\beta\|^{2} + \frac{1}{4}    \|b\|^{2} \right]}}{\sqrt{2 \cosh \left[ \frac{1}{2}(\beta b^{\ast} + \beta^{\ast} b) \right]}} 
  \begin{pmatrix}
\sqrt{\frac{b +\beta_3}{2b}}&  \frac{2 \beta_+}{\sqrt{2b (b+\beta_3)}}\\ 
\frac{- 2 \beta_-}{\sqrt{2b (b+\beta_3)}}&\sqrt{\frac{b +\beta_3}{2b}}
\end{pmatrix} 
\nonumber \\ \times
\begin{pmatrix} 
e^{(\beta + \frac{1}{2} b ) 
\hat{a}^{\dagger}} & 0 \\
0 & e^{(\beta - \frac{1}{2} b) \hat{a}^{\dagger} }
\end{pmatrix} 
\begin{pmatrix}
\mid 0 ; \frac{1}{2}, -\frac{1}{2} \rangle \\
\mid  0 ; \frac{1}{2}, + \frac{1}{2} \rangle \end{pmatrix} \nonumber \\
\label{eq-general-vector-ch-j=1/2}
 \end{eqnarray} 
 
These last states  are  eigenstates of  $\hat{a}$  with matrix type  eigenvalues:
\begin{equation}
\tilde{M} = \begin{pmatrix} \beta - \frac{1}{2} \beta_3 & - \beta_+   \\ - \beta_- & \beta + \frac{1}{2} \beta_3 \end{pmatrix} , 
\label{M-tilde-j=1/2}
 \end{equation}
 where  $\beta_{\pm}$ and $\beta_3$ verify (\ref{normality-conditions}).

\subsubsection{Quaternionic  canonical  vector coherent states class from $su(2)$ algebra}
\label{quaternion-2}
For the special choice of parameters 

\begin{eqnarray}
\beta= r \cos \theta  &   \beta_{3} =2ir \sin \theta \cos \phi 
\nonumber \\
\beta_{+}= -i r \sin \theta \sin \phi e^{i \psi}  &  \beta_{-}= -i r \sin \theta \sin \phi e^{-i \psi}, \quad r \ne 0, \label{eq-quaternions-parameters}   
\end{eqnarray}
the parameter  $b= 2 i r \sin\theta$ and the $\tilde{M}$ matrix in (\ref{M-tilde-j=1/2})  becomes

\begin{equation}
\tilde{M}= r \cos \theta \; \sigma_{0} + i \; r \sin \theta 
\begin{pmatrix} \cos\phi & \sin \phi e^{i \psi} \\
\sin\phi e^{-i \psi}& - \cos \phi
\end{pmatrix}, \label{eq-quaternions-matrix}
\end{equation}
which is a complex representation of quaternions by $2 \times 2$ matrices in polar coordinates. 
Inserting the values of (\ref{eq-quaternions-parameters}) into equation (\ref{eq-general-vector-ch-j=1/2}), and simplifying the results we get the normalized vector coherent states

\begin{eqnarray}
\mid \Psi[r, \theta,\phi,\psi) ]\rangle^{\frac{1}{2}} &=&
 \begin{pmatrix}
\cos{\frac{\phi}{2}} & - \sin{\frac{\phi}{2}} \; e^{i \psi} \\
\sin{\frac{\phi}{2}} \;  e^{-i \psi}&  
\cos{\frac{\phi}{2}} 
\end{pmatrix} \nonumber \\
&\times& \frac{e^{- \frac{r^2}{2}} }{\sqrt{2}} \;
 \begin{pmatrix} 
e^{ (r e^{i\theta}) 
\hat{a}^{\dagger}} & 0 \\
0 & e^{ (r e^{-i \theta}) \hat{a}^{\dagger} }
\end{pmatrix} 
\begin{pmatrix}
\mid 0 ; \frac{1}{2}, -\frac{1}{2} \rangle \\
\mid  0 ; \frac{1}{2}, + \frac{1}{2} \rangle \end{pmatrix} 
\label{eq-quaternionic-vector-ch-j=1/2}
 \end{eqnarray} 
These states are eigenstates of the  annihilation operator associated to the  quantum harmonic oscillator with matrix  eigenvalues  given by $\tilde{M}$ of equation  (\ref{eq-quaternions-matrix}). These states are the so-called  quaternionic canonical coherent states  that we can find the literature \cite{TA-KT}. The only difference here is that our states, by  virtue of our particular  choice of the ground state,  mix the basis states of the $su(2)$ sector of the $h(1) \otimes su(2)$ algebra. Indeed, these states can be written in the form

\begin{equation}
\mid \Psi(r, \theta,\phi,\psi) \rangle^{\frac{1}{2}} = \frac{e^{- \frac{r^2}{2}} }{\sqrt{2}} \\
\begin{pmatrix} \cos(\frac{\phi}{2})  \mid r e^{i\theta} ; \frac{1}{2} , - \frac{1}{2} \rangle -  \sin(\frac{\phi}{2}) \; e^{i\psi} \mid r e^{-i\theta}; \frac{1}{2} , + \frac{1}{2} \rangle\\
\sin(\frac{\phi}{2})  \; e^{-i\psi} \mid r e^{- i\theta}; \frac{1}{2} , - \frac{1}{2} \rangle +  \cos(\frac{\phi}{2})  \mid r e^{-i\theta}; \frac{1}{2} , + \frac{1}{2}\rangle,
\end{pmatrix} \label{VCS-expression-one=quaternions}
\end{equation} 
that shows us explicitly that mixture.

On the other hand, if we  choose   the fundamental state in the form shown in   (\ref{fundamental-state-choice-two}) and  calculate the factor $\mathcal{N}$  with the help of  (\ref{factor-normalization-N}), as in this case $\tilde{M} \tilde{M}^\dagger = r^2 I, $    we get $\mathcal{N}= 2 \; e^{r^2}.$ Finally,  if we  insert this last result in equation (\ref{VCS-expression-two}), adapted to particular case that is  being studied here, we obtain  
\begin{equation}
\mid \Psi (\tilde{M} ) \rangle^{\frac{1}{2}}  =  \frac{e^{- \frac{r^2}{ 2} } }{\sqrt{2}}  \sum_{n=0}^{\infty}  \frac{\tilde{M}^n}{\sqrt{n!}}   \begin{pmatrix}
 \mid n  \rangle \otimes \mid  \frac{1}{2} , - \frac{1}{2}   \rangle\\
    \mid n  \rangle \otimes \mid  \frac{1}{2}  ,   + \frac{1}{2}  \rangle,  \label{VCS-expression-two=quaternions}
\end{pmatrix},
 \end{equation}
which  reproduces the results studied in \cite{TA-KT}, again with the little difference of the mixing of the states on the $su(2)$ sector, due to the particular choice of the fundamental state we are using here.

\subsection{Annihilation operator vector algebra eigenstates with non-normal but diagonalizable eigenvalue matrix}
When $\tilde{M}$ in  (\ref{M-matrix-general-expression-part-annihilation}) is diagonalizable but not normal, the process of obtaining the algebra eigenstates  associated to $\hat{a}$  is identical to  that  we followed when $\tilde{M}$ was normal, except for the fact that now the passing matrix $P$ that leads $\tilde{M}$ to the its diagonal form is not unitary. Indeed, the diagonalized form of $\tilde{M}$ is  reached by performing the similarity transformation  $\tilde{D} = P^{-1} \tilde{M}P,$ where $P^{-1}$ denotes  the inverse of $P.$    Then, following the same reasoning of section \ref{sec:normal-case}, we can establish that the vector eigenstates  are given by  

\begin{eqnarray}
\mid \Psi (\Lambda) \rangle^j &=& \tilde{\mathcal{N}}^{\frac{1}{2}} \nonumber \\       
&\times &P \begin{pmatrix} e^{\lambda^j_{[j]} \hat{a}^\dagger} &0 &0& \cdots & \cdots&0\\0 &  e^{\lambda^j_{[j-1]} \hat{a}^\dagger} &0 & \dots& \cdots &0 \\ 0& 0& \ddots &0& \cdots &0 \\
 \vdots& \vdots & \vdots & \ddots & \vdots & \vdots \\ 
 0 & \cdots  &\cdots& 0 & e^{\lambda^j_{[-j+1]} \hat{a}^\dagger}  &0 \\0 & \cdots& \cdots &0 & 0& e^{\lambda^j_{[-j]} \hat{a}^\dagger}\end{pmatrix} 
 \begin{pmatrix}
 \mid 0 ; j, -j \rangle \\
 \mid 0 ; j, -j+1 \rangle\\
\vdots \\
 \mid 0 ; j, j -1 \rangle \\
 \mid 0 ; j, j \rangle \\
 \end{pmatrix}, \label{vector-coherent-states-general-not-normal-case}
\end{eqnarray} 
where $\lambda^j_{[s]}, \; s=-j,\cdots,j, $ are the  eigenvalues of $\tilde{M}$ and   $\tilde{\mathcal{N}}$ is a normalization constant to be determined. 

As in general $P$ is not unitary, the computation of the normalization constant of these vector  states is now more difficult than the previous case when $\tilde{U}$ was unitary.  Indeed, the eigenvectors composing the columns of the  matrix $P$ are in general not orthogonal to each other. 

\subsubsection{Annihilation operator algebra eigenstates with a non-normal but diagonalizable  $su(2)$ eigenvalue matrix}\label{sec-VCS-not-normal}
Returning to the $su(2)$ eigenvalue matrix $\tilde{M}=M$  given by (\ref{M-matrix-general-expression-part-annihilation}). When that matrix is diagonalizable but when no special conditions on the  beta parameters are given, in certain cases, the matrix $P$  can still be computed with the formulas  of appendix \ref{appb}. Indeed, that is the case when $\beta_\pm \neq 0,$ whichever is the value of $\beta_3,$ provided that the $b$ parameter is different from zero, or  when $\beta_+$ or $\beta_-$ is equal to zero, but not both, and   $\beta_3 \neq 0.$  In all these cases   the matrix elements of  $P,$ for fixed $j$, can be extracted from the matrix elements of $\hat{T}$ given in  (\ref{matrix-elements-T}) in the following way  

\begin{eqnarray}
P_{m \ell} =  \begin{cases}   T^j_{m \ell}[\beta_+,\beta_- ,\beta_3]\; \mbox{when} \; \beta_+ \neq 0, \beta_- \neq 0  \;  \mbox{and} \;  \beta_3 \neq 0 \;  \mbox{with} \; b\neq 0  \\  
T^j_{m \ell}[0,\beta_- ,\beta_3]\; \mbox{when} \; \beta_+ = 0, \beta_- \neq 0  \;  \mbox{and} \;  \beta_3 \neq 0 \\
 T^j_{m \ell}[\beta_+,0 ,\beta_3]\; \mbox{when} \; \beta_+ \neq 0, \beta_- = 0  \;  \mbox{and} \;  \beta_3 \neq 0 \\
 T^j_{m \ell}[\beta_+, \beta_- ,0]\; \mbox{when} \; \beta_+ \neq 0, \beta_- \neq 0  \;  \mbox{and} \;  \beta_3 = 0
  \end{cases}  .\label{matrix-elements-P}
\end{eqnarray}

For example for $j=\frac{1}{2},$ when  $\beta_+ \neq 0, \beta_- \neq 0 $ and $ \beta_3 \neq 0,$  with  $b \neq 0,$ the normalized vector  states are given by

\begin{eqnarray}
\mid \Psi (\Lambda) \rangle^{\frac{1}{2}} = 
\frac{\sqrt{2 \|b\| \| b+ \beta_3\|}}{ \sqrt{ (4 \|\beta_-\|^2 +\| b+ \beta_3\|^2)    e^{\|( \beta + \frac{b}{2} ) \|^2} + (4 \|\beta_+\|^2 +\| b+ \beta_3\|^2)      e^{\| (\beta - \frac{b}{2}) \|^2}}}
  \nonumber \\
 \times
  \begin{pmatrix}
\sqrt{\frac{b +\beta_3}{2b}}&  \frac{2 \beta_+}{\sqrt{2b (b+\beta_3)}}\\ 
\frac{- 2 \beta_-}{\sqrt{2b (b+\beta_3)}}&\sqrt{\frac{b +\beta_3}{2b}}
\end{pmatrix} 
\nonumber  
\begin{pmatrix} 
e^{(\beta + \frac{1}{2} b ) 
\hat{a}^{\dagger}} & 0 \\
0 & e^{(\beta - \frac{1}{2} b) \hat{a}^{\dagger} }
\end{pmatrix} 
\begin{pmatrix}
\mid 0 ; \frac{1}{2}, -\frac{1}{2} \rangle \\
\mid  0 ; \frac{1}{2}, + \frac{1}{2} \rangle \end{pmatrix},
\label{eq-general-vch-j=1/2-no-unitary}
 \end{eqnarray}
which is an eigenstate of $\hat{a}$ with eigenvalues on the matrix domain given by (\ref{M-tilde-j=1/2}), but now without any constraint on the beta parameters regarding the normality of that matrix.

Furthermore, when for example,  $\beta_+ \neq 0,  \beta_- = 0 $  and  $ \beta_3 \neq 0,$ we get

\begin{eqnarray}
\mid \Psi (\Lambda) \rangle^{\frac{1}{2}} = 
\frac{\|\beta_3\|}{ \sqrt{ \|\beta_3\|^2    e^{\|( \beta + \frac{\beta_3}{2} ) \|^2} + (\|\beta_+\|^2 +\|\beta_3\|^2)      e^{\| (\beta - \frac{\beta_3}{2}) \|^2}}}
  \nonumber \\
 \times
  \begin{pmatrix}
1&   \frac{\beta_+}{\beta_3}\\ 
0&1
\end{pmatrix} 
\nonumber  
\begin{pmatrix} 
e^{(\beta + \frac{1}{2} \beta_3 ) 
\hat{a}^{\dagger}} & 0 \\
0 & e^{(\beta - \frac{1}{2} \beta_3) \hat{a}^{\dagger} }
\end{pmatrix} 
\begin{pmatrix}
\mid 0 ; \frac{1}{2}, -\frac{1}{2} \rangle \\
\mid  0 ; \frac{1}{2}, + \frac{1}{2} \rangle \end{pmatrix} \nonumber \\
\label{eq-general-vch-j=1/2-no-unitary-beta-0}
 \end{eqnarray}
which is an eigenstate of $\hat{a}$ with matrix eigenvalues 

\begin{equation}
\begin{pmatrix}
\beta + \frac{\beta_3}{2}& - \beta_+ \\
0 & \beta - \frac{\beta_3}{2}
\end{pmatrix},
\end{equation}
where no restrictions are assumed on the beta parameters with respect to the normality of this matrix.  

 \subsubsection{An example with a non-diagonalizable $su(2)$ eigenvalue matrix}
When $\tilde{M}$ in  (\ref{M-matrix-general-expression-part-annihilation}) is not diagonalizable, the process of obtaining the algebra eigenstates of $\hat{a}$  is, in general, more elaborate than the case when this matrix is diagonalizable, sometimes we can use the exponential form of the matrix  $M$ and the operator algebra or to solve the eigenvalue equation in a systematic way, component by component.   Let us illustrate the solving techniques with a simple example using  the $j=\frac{1}{2}$ basis vector representation for calculating the matrix elements of $\tilde{M}$ and leaving free choice for the dimension of the irreducible representation used to  span  the components of the vector states.    Let us take for this purpose the case when $\beta_+ \neq 0 $  and $\beta_- =\beta_3 =0,$ and consequently $b=0.$ Thus, the eigenvalue we have to solve is

\begin{equation}
\hat{a} \begin{pmatrix}   \mid  \psi \rangle^1 \\ \mid \psi \rangle^2 \end{pmatrix}  = \begin{pmatrix}   \beta  &  - \beta_+ \\  0 & 
\beta \end{pmatrix} \begin{pmatrix}   \mid  \psi \rangle^1\\ \mid \psi \rangle^2 \end{pmatrix} ,
\end{equation}
where the states  $ \mid \psi \rangle^k = \sum_{n=0}^\infty \sum_{m=-j}^j C^k_{n m} \mid n \rangle \otimes \mid j,m \rangle, \; k=1,2.$
By projecting  both sides of this equation on the analytic basis states $ \langle \zeta \mid$ and then using the realization (\ref{action-annihilation-zeta}) for the annihilation operator $\hat{a},$ we get a system of ordinary  linear differential equation for the components of  $ \psi^k (\zeta) = \langle \zeta  \mid \psi \rangle^k, \; k=1,2,$ that is,  

\begin{equation}
\frac{d}{d\zeta} \begin{pmatrix}   \psi^1_{m} (\zeta) \\  \psi^2_{m} (\zeta)   \end{pmatrix}  =  \begin{pmatrix}   \beta  &  - \beta_+ \\  0 & 
\beta \end{pmatrix}  \begin{pmatrix}   \psi^1_{m} (\zeta) \\  \psi^2_{m} (\zeta)   \end{pmatrix} , \quad m=-1, \cdots,j,
\end{equation} 
which explicitly corresponds to   
\begin{eqnarray}
\frac{d}{d\zeta}  \psi^1_{m} (\zeta) &=&   \beta  \psi^1_{m} (\zeta)   -  \beta_+  \psi^2_{m} (\zeta) \nonumber \\ 
 \frac{d}{d\zeta}  \psi^2_{m} (\zeta) &=  & \beta  \psi^2_{m} (\zeta)   , \quad m=-j, \cdots,j.  \label{AES-j-one-half-no-diagonalizable}
\end{eqnarray}   
The integration of  $\psi^2_{m} (\zeta)$ is direct, we get $\psi^2_{m} (\zeta)  = e^{\beta \zeta} \; \psi^2_{m} (0), \; m=-1, \cdots, j. $ Inserting this last result into the first of equations (\ref{AES-j-one-half-no-diagonalizable}) and then integrating the resulting 
non-homogeneous first order linear differential equation for each function $\psi^1_{m} (\zeta), \; m=-j, \cdots, j, $ we get $ \psi^1_{m} (\zeta) =  e^{\beta \zeta} \; \psi^1_{m} (0) - \beta_+ \zeta  e^{\beta \zeta} \; \psi^2_{m} (0),  \;  m=-j, \cdots, j.$ Finally, returning to the  Fock basis of the number eigenstates  $\mid n \rangle$ we obtain the vector component  state $$\mid \psi \rangle^1 =  e^{\beta \hat{a}^\dagger} \left[ \sum_{m=-j}^j   \psi^1_m (0) \mid 0 \rangle \otimes\mid j, m \rangle  - \beta_+ \hat{a}^{\dagger} \sum_{m=-j}^{j}  \psi^2_m (0) \mid 0 \rangle \otimes\mid  j,  m \rangle \right]  $$ and the vector component state $$\mid \psi \rangle^2 =  e^{\beta \hat{a}^\dagger}  \sum_{m=-j}^j   \psi^2_m (0) \mid 0 \rangle \otimes\mid j, m \rangle, $$ which represent the super-coherent states of the super-symmetric  Harmonic oscillator introduced by  Aragone and Zypman \cite{Ar-Zy}.  Let us recall that this same states also are included in the set of algebra eigenstates of the generalized annihilation operator $\hat{a} + \beta_+ \hat{J}_-, $ the difference is that here these states  arise as the components of a vector state, that is,  

\begin{equation}
\mid  \Psi \rangle^{\frac{1}{2}}_2 = \begin{pmatrix}   \mid  \psi \rangle^1 \\ \mid \psi \rangle^2 \end{pmatrix}  = e^{\beta \hat{a}^\dagger} \begin{pmatrix}   1  &  - \beta_+ \hat{a}^\dagger \\  0 & 
1  \end{pmatrix}   \begin{pmatrix}    \sum_{m=-j}^{j}  \psi^1_m (0) \mid 0 \rangle \otimes\mid j, m \rangle  \\ \sum_{m=-j}^{j}  \psi^2_m (0) \mid 0 \rangle \otimes\mid j, m \rangle  \end{pmatrix}  = e^{M \hat{a}^\dagger} \begin{pmatrix}    \sum_{m=-j}^{j}  \psi^1_m (0) \mid 0 \rangle \otimes\mid j, m \rangle  \\ \sum_{m=-j}^{j}  \psi^2_m (0) \mid 0 \rangle \otimes\mid j, m \rangle  \end{pmatrix},
\end{equation}
and are eigenstates  of the  annihilator $\hat{a}$ with matrix eigenvalues given by $   \left(\begin{smallmatrix}   \beta  &   - \beta_+ \\  0 & 
\beta \end{smallmatrix} \right).
$

\section{$h(1) \oplus su(2)$ vector algebra eigenstates with matrix eigenvalues}
\label{sec-four}
In this section we propose a generalization of the concept of vector coherent states associated to the  annihilation operator ${\hat a}$ to include  those states which are eigenvectors  of a class of  generalized $h(1) \oplus su(2)$ type annihilation operator. To start, let us consider the problem of finding the vector states that satisfy the following eigenvalue equation system:\footnote{In fact, we could define this equation as $[\alpha_-  \hat{a}  + \alpha_+ \hat{a}^\dagger + \alpha_3  \hat{I} + \beta_{-} \hat{J}_{+} + \beta_{+} \hat{J}_{-} + \beta_{3}  \hat{J}_{3} ]  \mid \Psi \rangle^j_{[K]} =  \tilde{M} \mid \Psi \rangle^j_{[K]},$ but  by performing a suitable squeezed state transformation and  then by absorbing  the identity operator coefficient into the $\beta$ parameter we get (\ref{vector-matrix-eigenvalue-total}), see  appendix  \ref{appa}. }   

\begin{equation}
\left[ \hat{a}  + \beta_{-} \hat{J}_{+} + \beta_{+} \hat{J}_{-} + \beta_{3}  \hat{J}_{3}\right] \mid \Psi \rangle^j_{[K]} =  \tilde{M} \mid \Psi \rangle^j_{[K]}, \label{vector-matrix-eigenvalue-total}
\end{equation} 
with $\tilde{M}$ being now a  $K \times K$ diagonalizable complex matrix and $\mid \Psi \rangle^{j}_{[K]}$ a vector state of $K$ components  having exactly the same characteristic they had in the last section, i.e.,  

\begin{equation} 
\mid \Psi\rangle^j_{[K]} =\begin{pmatrix}
\mid \psi  \rangle^j_{[1]} \\
\mid \psi  \rangle^j_{[2]} \\
\vdots\\
\mid \psi \rangle^j_{[K-1]} \\
\mid \psi \rangle^j_{[K]} 
\end{pmatrix} , \label{general-vector-solution-matrix-eigenvalues}
\end{equation}
where each state $\mid \psi \rangle^j_{[s]} = \sum_{m=-j}^{j} \sum_{n=0}^{\infty}    C^{[s] j}_{n  m} \mid n \rangle \otimes j , m \rangle \; 1 \leq s \leq K,$ is a unknown state of the $h(1) \oplus su(2)$ representation space to be determined.   

As before, if  $\tilde{M}$ is diagonalizable, then there is a $P$ matrix of dimension $K \times K$ such that $P^{-1} \tilde{M} P =\tilde{D} $ is a diagonal matrix which non-null entries are given by the eigenvalues of $\tilde{M}.$ Therefore, firstly, defining
\begin{equation}
\mid \Psi \rangle^j_{[K]} =  P \mid \tilde{\Psi}\rangle^j_{[K]}, \quad \rm{or \; equivalently} \quad \mid \tilde{\Psi}\rangle^j_{[K]} =  P^{-1} \mid \Psi\rangle^j_{[K]}, \label{transformed-vectors}
\end{equation}     
and  then by acting with $P^{-1}$ from the left on both sides of equation (\ref{vector-matrix-eigenvalue-total}), we get
\begin{equation}
\left[ \hat{a}  + \beta_{-} \hat{J}_{+} + \beta_{+} \hat{J}_{-} + \beta_{3}  \hat{J}_{3}\right] \mid \tilde{\Psi} \rangle^j_{[K]} =  \tilde{D} \mid \tilde{\Psi} \rangle^j_{[K]}. \label{vector-matrix-eigenvalue-total-diagonal-form}
\end{equation} 
If we denote  by $\tilde{\lambda}_{[s]}, \;  s=1,2,\cdots, K,$ the eigenvalues of $\tilde{M}$ which are distributed in this same order, from top to bottom,  on the diagonal principal of $\tilde{D},$ we obtain a system of $K$ independent $h(1) \oplus su(2)$ algebra eigenstates equations, that is

\begin{equation}
\left[ \hat{a}  + \beta_{-} \hat{J}_{+} + \beta_{+} \hat{J}_{-} + \beta_{3}  \hat{J}_{3}\right] \mid \tilde{\psi} \rangle^j_{[s]} =  \tilde{\lambda}_{[s]} 
\mid \tilde{\psi} \rangle^j_{[s]}, \quad s=1,2, \cdots, K.  \label{reduction-to-algebra-h1-su2-eigenstates}
\end{equation} 
As discussed in  section \ref{sec-VCS-not-normal}, the $h(1) \oplus su(2)$ algebra eigenstates can be obtained for all different values of the beta parameters. That means a wide variety of situations to be analyzed, each of which provides us with an interesting set of vector states verifying the original equation system. For example, when  $b\neq 0,$ the general expression for the  $h(1) \oplus su(2)$ algebra  eigenstates   $\mid \tilde{\psi} \rangle^j_{[s]}$ is given by equation (\ref{algebra-eigenstates-h1-su2-rewriten}), with  the corresponding $\tilde{\lambda}_{[s]} $ in place of the $\beta$  parameter, that is

\begin{eqnarray}
\mid \tilde{\psi} \rangle^j_{[s]} &=& N_{[s]}^{-1/2}  \nonumber \\ &\times& \begin{pmatrix}
\tilde{\psi}^j_{[s], -j} (0) e^{\lambda^{[s]j}_{j} \hat{a}^\dagger} & \tilde{\psi}^j_{[s],-j+1} (0) e^{\lambda^{[s] j}_{j-1} \hat{a}^\dagger}  & \cdots & \tilde{\psi}^j_{[s],j-1} (0) e^{\lambda^{[s]j}_{-j+1} \hat{a}^\dagger} & \tilde{\psi}^j_{[s],j} (0) e^{\lambda^{[s]j}_{-j} \hat{a}^\dagger}
 \end{pmatrix}  \nonumber \\ 
 &\times& V^{t} \;
 \begin{pmatrix}
 \mid 0 ; j, -j \rangle \\
 \mid 0 ; j, -j+1 \rangle\\
\vdots \\
 \mid 0 ; j, j -1 \rangle \\
 \mid 0 ; j, j \rangle \\
 \end{pmatrix}, \quad s=1,2, \cdots,K, \label{equation-components-h1-su2-algebra-eigenstates}
\end{eqnarray}
where
$\lambda^{[s]j}_{m}= \tilde{\lambda}_{[s]} + m b, \; s=1,2,\cdots, K $ and $V$ is an invertible   matrix whose matrix elements are given explicitly  in equation (\ref{matrix-elements-T}), then
\begin{equation}
V_{m \ell} = T^j_{m \ell}[\beta_+,\beta_- ,\beta_3], \quad b \neq 0.
\end{equation}   
  Finally  using (\ref{transformed-vectors}), we can return to the original vector states which are solutions of equation  (\ref{vector-matrix-eigenvalue-total}), thus  we get the $K$ components of the $h(1)\otimes su(2)$ vector state (\ref{general-vector-solution-matrix-eigenvalues}) that verifies (\ref{vector-matrix-eigenvalue-total}), they are: 
  \begin{equation}
 \mid  \psi \rangle^j_{[u]} = \sum_{s=1}^{K} P_{u s} \mid  \tilde{\psi} \rangle^j_{[s]}, \quad u=1,2, \cdots, K.  \label{vector-algebra-states-}
 \end{equation}
We recall that for each $s,$ equation (\ref{equation-components-h1-su2-algebra-eigenstates}) represent the general solution of equation (\ref{reduction-to-algebra-h1-su2-eigenstates}). It corresponds to a state formed from  a set of $2j+1$ linearly  independent states which are by themselves   solutions of (\ref{reduction-to-algebra-h1-su2-eigenstates}). Then, for each $s,$ we can choose an arbitrary combination of these states and then insert it in  equation (\ref{vector-algebra-states-}) to obtain the final structure of the vector states which verify equation   (\ref{vector-matrix-eigenvalue-total}). This freedom of choice  implies a very large variety of linearly independent solutions, more precisely   $K \times (2j+1)$ independent solutions. 

\pagebreak

\subsection{Generalized $h(1) \oplus su(2)$ vector coherent states on the matrix domain}
 In this article, we are interested in states that could be interpreted as vector coherent states of a given physics system. Then, instead of proceeding with the case-by-case analysis, let us select the parameters according to the necessary conditions  they have to fulfill  at the moment of  constructing   a suitable generalized annihilation operator extracted  from the linear combination of generators shown in equation (\ref{reduction-to-algebra-h1-su2-eigenstates}). Proceeding in this way, let us define the operator

\begin{equation}
\mathbb{A} = \hat{a}  + \beta_{-} \hat{J}_{+} + \beta_{+} \hat{J}_{-} + \beta_{3}  \hat{J}_{3}, \label{generalized-h1-su2-annihation-operator}
\end{equation} 
whose commutator with $\mathbb{A}^{\dagger} $ is given by

\begin{equation}
\left[ \mathbb{A}  ,   \mathbb{A}^\dagger \right]= \hat{I} + 2  \left( {\| \beta_- \|}^2 -  {\| \beta_+ \|}^2 \right) \hat{J}_3 +
\left( \beta_3 \beta_{+}^\ast - \beta_3^\ast \beta_- \right) \hat{J}_+ + \left( \beta_3^\ast \beta_+ - \beta_3  \beta_-^\ast \right) \hat{J}_-  .
\label{commutator-A-A-dagger}
\end{equation}
From this last equation we can see that when  $\|\beta_+\| = \|\beta_-\|$ and $\beta_3 \beta_+^\ast = \beta_3^\ast \beta_-,$which are the same conditions that make the matrix $M$ in (\ref{M-matrix-general-expression-part-annihilation}) normal,  including certainty here  the special case when $\beta_+ = \beta_-=0$ and $\beta_3 \neq 0,$  the commutator becomes  
\begin{equation}
\left[   \mathbb{A} , \mathbb{A}^\dagger \right] = \hat{I},   \label{commutator-A-A-dagger=I}
\end{equation}
i.e., the corresponding set of vector states which are  solutions of equation(\ref{vector-matrix-eigenvalue-total}) can be interpreted as  vector coherent states in the context of the  physical system whose Hamiltonian is given by $\mathbb{H} = \mathbb{A}^\dagger  \mathbb{A} .$ If we write the  beta  parameters in the form $\beta_{\pm}= R e^{i \theta_{\pm}}$ and $\beta_3 = R_3 e^{i \theta_3 },$ with $R$ and  $R_3 $ being no negative real numbers,  the new operator thus defined takes the form
\begin{equation}
\mathbb{A} = \hat{a} + R [ e^{i \theta_-} \hat{J}_+ + e^{i \theta_+} \hat{J}_{-} ] + R_3  e^{i \frac{(\theta_+ + \theta_- )}{2}} \hat{J}_3,  
\end{equation}
and then the above newly defined Hamiltonian writes
\begin{eqnarray}
\mathbb{H} &=&  \hat{a}^{\dagger} \hat{a} + R \hat{a}^\dagger [e^{i \theta_{-} } \hat{J}_{+}  + e^{i \theta_{+}} \hat{J}_{-} ] +  R  [e^{- i \theta_{-} } \hat{J}_{-} + e^{- i \theta_{+} } \hat{J}_{+} ] \hat{a} +
 R_3   [ e^{i \frac{(\theta_+ + \theta_- )}{2}}\hat{a}^\dagger + e^{- i \frac{(\theta_+ + \theta_- )}{2}} \hat{a}] 
 \hat{J}_3  \nonumber \\  &+&  R^2 [ \hat{J}_{-}  \hat{J}_{+} + \hat{J}_{+} \hat{J}_{-} ] + R^2 [  e^{i (\theta_{+} - \theta_{-} )}  \hat{J}_{-}^{2} +  e^{- i (\theta_{+} - \theta_{-} )} \hat{J}_{+}^2 ]  + R_{3}^{2}   \hat{J}_{3}^{2} \nonumber \\
 &+&  R R_3 e^{- i \frac{(\theta_{+} - \theta_{-} )}{2}} [ \hat{J}_{+}  \hat{J}_{3}  +  \hat{J}_{3}  \hat{J}_{+} ] +  R R_{3} e^{i \frac{(\theta_+ - \theta_- )}{2}} [ \hat{J}_{-}  \hat{J}_{3}  +  \hat{J}_{3}  \hat{J}_{-} ]. \label{generalized-oscillator-hamiltonian}
 \end{eqnarray} 

We can see that, by the way this last Hamiltonian was constructed  and in accordance with the commutation relation (\ref{commutator-A-A-dagger=I}),  this Hamiltonian is isospectral with the standard harmonic oscillator Hamiltonian although  its eigenstates  are very different. Nevertheless, if we define $\mid  \tilde{0}\rangle$ as being a suitable ground state of this system such that $\mathbb{A} \mid \tilde{0} \rangle = 0, $ we can be sure that the associated vector coherent states, which are eigenstates of $\mathbb{A}$ with the entries of    $\tilde{M}$  as its eigenvalues, are given by $  e^{ [ \tilde{M} \mathbb{A}^\dagger - \tilde{M}^\dagger \mathbb{A} ]} \mid \tilde{0} \rangle,$   as long as  $\tilde{M}$ is normal.  We will probe now  in general  this last assertion  and show  explicitly the technique used in doing that  at the end of the following section with an example  for the the special case $K=2$ and $j=\frac{1}{2}.$  

\subsubsection{Generalized matrix displacement  operator}
\label{sec-canonical}
Using the operator algebra described in appendix \ref{appb}, we can show that a generic algebra eigenstate $\mid \tilde{\psi} \rangle^j_{[s] m }$ verifying  (\ref{reduction-to-algebra-h1-su2-eigenstates}), in the case that $b \ne 0,$  can be written in the form
\begin{equation}
\mid \tilde{\psi} \rangle^j_{[s] m} = \mid \tilde{\lambda}_{[s]}  - m  b \rangle \otimes {\hat T} \mid j , m \rangle = \hat{D} (\tilde{\lambda}_{[s]} - m b )   \mid 0  \rangle \otimes {\hat T} \mid j , m \rangle,
\end{equation}
where $\hat{T}$ is the exponential operator defined in  (\ref{master-operator-T}),  which parameters are defined in  (\ref{phases-conditions}). We notice that when $\beta_+ = \beta_-=0$ and $\beta_3 \neq 0,$ the operator $T=I.$ 

Using the displacement operator property $\hat{D} (z_1 + z_2) = \hat{D} (z_1 )   \hat{D} ( z_2) \exp{ \frac{1}{2} [z_1^\ast z_2 - z_1 z_2^\ast  ]},  $ we get

\begin{equation}
\mid \tilde{\psi} \rangle^j_{[s] m} = \hat{D} (\tilde{\lambda}_{[s]}) \exp{\frac{m}{2} [     \tilde{\lambda}_{[s]}  b^\ast-\tilde{\lambda}^{\ast}_{[s]}  b   ] }  \mid  - m  b \rangle \otimes {\hat T} \mid j , m \rangle.\label{solution-tilde-oscillator-generalized}
\end{equation}
 Defining $\vec{\beta} \cdot \vec{\hat{J}} = \beta_+ \hat{J}_{-}  + \beta_-  \hat{J}_{+}  +  \beta_3 \hat{J}_{3}$ and using the fact that in this particular case
 
 \begin{equation}
 (\vec{\beta} \cdot  \vec{\hat{J}} )  \hat{T}  \mid j , m \rangle = m b  \hat{T}  \mid j , m \rangle, \quad \mbox{and} \quad (\vec{\beta} \cdot  \vec{\hat{J}})^{\dagger}   \hat{T}  \mid j , m \rangle = m b^\ast  \hat{T}  \mid j , m \rangle,
 \end{equation} 
equation (\ref{solution-tilde-oscillator-generalized}) becomes
 
 \begin{equation}
 \mid \tilde{\psi} \rangle^j_{[s] m} = \hat{D} (\tilde{\lambda}_{[s]}) \exp{\left(\frac{1}{2} \left[     \tilde{\lambda}_{[s]}  (\vec{\beta} \cdot \vec{\hat{J}})^{\dagger}    -\tilde{\lambda}^{\ast}_{[s]}  \vec{\beta} \cdot \vec{\hat{J}}   \right] \right) }  \mid  - m  b \rangle \otimes {\hat T} \mid j , m \rangle. \label{factorized-solution-tilde}
\end{equation}
Thus the normalized general solution of  (\ref{reduction-to-algebra-h1-su2-eigenstates}) is given by

\begin{equation}
\mid \tilde{\psi} \rangle^j_{[s]} =   \frac{ \sum_{m=-j}^{j} \varphi^j_{[s] m } (0)   \mid \tilde{\psi} \rangle^j_{[s] m}}{\sqrt{\sum_{m=-j}^{j} \| \varphi^j_{[s] m } (0)   \|^2} } ,
\end{equation}
or more explicitly by 
\begin{equation}
\mid \tilde{\psi} \rangle^j_{[s]} =    \hat{D} (\tilde{\lambda}_{[s]} ) \exp{\left(\frac{1}{2} \left[     \tilde{\lambda}_{[s]}  (\vec{\beta} \cdot \vec{\hat{J}})^{\dagger}    -\tilde{\lambda}^{\ast}_{[s]}  \vec{\beta} \cdot \vec{\hat{J}}   \right] \right) } \frac{\sum_{m=-j}^{j} \varphi^j_{[s] m } (0)     \mid  - m  b \rangle \otimes {\hat T} \mid j , m \rangle}{{\sqrt{\sum_{m=-j}^{j} \| \varphi^j_{[s] m } (0)   \|^2} }  }.
\end{equation}
 Inserting this last result into equation (\ref{vector-algebra-states-}), which give us   the component of the general vector coherent states verifying  (\ref{vector-matrix-eigenvalue-total}),  we obtain the matrix realization of these states:

\begin{eqnarray} 
\mid \Psi \rangle^j_{[K]} &=&  \tilde{U}   \begin{pmatrix}  \hat{D} (\tilde{\lambda}_{[1]})  & 0& \cdots & 0 &  0 \\
0&  \hat{D} (\tilde{\lambda}_{[2]}) & 0 & \cdots  & 0\\
\vdots& 0& \ddots & 0 & \vdots  \\
0& \cdots&0&  \hat{D} (\tilde{\lambda}_{[K-1]}) &0 \\
0&0& \cdots&0 & \hat{D} (\tilde{\lambda}_{[K]})
 \end{pmatrix}  \nonumber \\  &\times& \begin{pmatrix}  \hat{B} (\tilde{\lambda}_{[1]})  & 0& \cdots & 0 &  0 \\
0&  \hat{B} (\tilde{\lambda}_{[2]}) & 0 & \cdots  & 0\\
\vdots& 0& \ddots & 0 & \vdots  \\
0& \cdots&0&  \hat{B} (\tilde{\lambda}_{[K-1]}) &0 \\
0&0& \cdots&0 & \hat{B} (\tilde{\lambda}_{[K]})
 \end{pmatrix} 
 \begin{pmatrix}
   \frac{\sum_{m=-j}^{j}   \tilde{\varphi}^j_{[1] m} (0)  \mid  - m  b \rangle \otimes {\hat T} \mid j , m \rangle } {\sum_{m=-j}^{j} \| \varphi^j_{[1] m } (0)   \|^2}  \\
  \frac{\sum_{m=-j}^{j}  \tilde{\varphi}^j_{[2] m} (0)    \mid  - m  b \rangle \otimes {\hat T} \mid j , m \rangle }{\sum_{m=-j}^{j} \| \varphi^j_{[2] m } (0)   \|^2}  \\
 \vdots\\
  \frac{\sum_{m=-j}^{j}   \tilde{\varphi}^j_{[K] m} (0)    \mid  - m  b \rangle \otimes {\hat T} \mid j , m \rangle}{\sum_{m=-j}^{j} \| \varphi^j_{[K] m } (0)   \|^2} 
 \end{pmatrix}, \label{general-set-VCS-matrix-form}
\end{eqnarray}
where  $\hat{B} (\tilde{\lambda}_{[s]}) = \exp{\left(\frac{1}{2} \left[     \tilde{\lambda}_{[s]}  (\vec{\beta} \cdot \vec{\hat{J}})^{\dagger}    -\tilde{\lambda}^{\ast}_{[s]}  \vec{\beta} \cdot \vec{\hat{J}}   \right] \right) }, \; s=1,\cdots, K.$ Equation (\ref{general-set-VCS-matrix-form}) represent a set of $ (2j+1)^K$  linearly independent  vector coherent states associated to the generalized harmonic oscillator system here defined. 

 By setting the integration parameters appropriately we can change the form of these states to a more familiar one, for example, by choosing

\begin{equation}
\varphi^j_{[s] m} (0) =  \sum_{u=1}^{K} \sum_{r=1}^{K} \tilde{U}^\dagger_{[s][u]} \exp\left(  \frac{m}{2}  \left[ \tilde{M} b^\ast - \tilde{M^\dagger} b    \right) \right]_{[u][r]}  \delta_{m_{[r]} m}, 
\end{equation}
 where $\delta_{m_{[r]} m}$ is the Kronecker delta, (\ref{general-set-VCS-matrix-form}) writes
 
  \begin{eqnarray} 
\mid \Psi \rangle^j_{[K]} &=&  \tilde{U}  \;  \begin{pmatrix}  \hat{D} (\tilde{\lambda}_{[1]})  & 0& \cdots & 0 &  0 \\
0&  \hat{D} (\tilde{\lambda}_{[2]}) & 0 & \cdots  & 0\\
\vdots& 0& \ddots & 0 & \vdots  \\
0& \cdots&0&  \hat{D} (\tilde{\lambda}_{[K-1]}) &0 \\
0&0& \cdots&0 & \hat{D} (\tilde{\lambda}_{[K]})
 \end{pmatrix} 
\nonumber \\ & \times& 
 \begin{pmatrix}  \hat{B} (\tilde{\lambda}_{[1]})  & 0& \cdots & 0 &  0 \\
0&  \hat{B} (\tilde{\lambda}_{[2]}) & 0 & \cdots  & 0\\
\vdots& 0& \ddots & 0 & \vdots  \\
0& \cdots&0&  \hat{B} (\tilde{\lambda}_{[K-1]}) &0 \\
0&0& \cdots&0 & \hat{B} (\tilde{\lambda}_{[K]})
 \end{pmatrix}  \;  \tilde{U}^\dagger \nonumber \\ &\times&   \exp\left[ \frac{1}{2}  \left( \tilde{M} (\vec{\beta} \cdot \vec{\hat{J}} )^\dagger  - \tilde{M^\dagger}     (\vec{\beta} \cdot \vec{\hat{J}})  \right) \right]       
 \begin{pmatrix}
   \mid  - m_1  b \rangle \otimes {\hat T} \mid j , m_1 \rangle  \\
    \mid  - m_2  b \rangle \otimes {\hat T} \mid j , m_2 \rangle \\
 \vdots\\
  \mid  - m_{K}  b \rangle \otimes {\hat T} \mid j , m_{K} \rangle
 \end{pmatrix} \label{general-set-VCS-matrix-form-before-end}.
\end{eqnarray}
Finally, using the fact that the exponential of a diagonal square matrix is equal to a diagonal square matrix which elements are given by the exponential of the  diagonal elements of the starting matrix and vice versa,   and  that  $\tilde{U} \tilde{D} \tilde{U}^\dagger = \tilde{M},$ we can put  (\ref{general-set-VCS-matrix-form-before-end}) in the compact form
\begin{equation}
\mid \Psi \rangle^j_{[K]} = \exp\left[  \tilde{M } \mathbb{A}^\dagger - \tilde{M}^\dagger \mathbb{A}   \right]
\begin{pmatrix}
   \mid  - m_1  b \rangle \otimes {\hat T} \mid j , m_1 \rangle  \\
    \mid  - m_2  b \rangle \otimes {\hat T} \mid j , m_2 \rangle \\
 \vdots\\
  \mid  - m_{K}  b \rangle \otimes {\hat T} \mid j , m_{K} \rangle
 \end{pmatrix}, \label{general-set-VCS-matrix-form-end}.
\end{equation}
This result can evidently be obtained in a more heuristic way. Indeed, as  here  $[ \hat{\mathbb{A}} , \hat{\mathbb{A}}^\dagger ] = \hat{I},$ we can try a solution of  (\ref{reduction-to-algebra-h1-su2-eigenstates}) in the  exponential form $\mid \Psi \rangle = \exp\left(\tilde{M}  \hat{\mathbb{A}} \right) \mid \Omega \rangle .$ Then with the help of  the relation  $\hat{\mathbb{A}} \exp\left( \tilde{M}  \hat{\mathbb{A}} \right) = \exp\left(\tilde{M}  \hat{\mathbb{A}} \right)  \hat{\mathbb{A}} + \tilde{M}   \exp \left(\tilde{M}  \hat{\mathbb{A}} \right),  $ we can reduce the problem to simply calculate the fundamental state, that is, the  eigenstate of $\hat{\mathbb{A}} $ with matrix eigenvalue equal to zero,     $\hat{\mathbb{A}} \mid \Omega \rangle =0.$

This set of these vector coherent states verify all properties of the standard  harmonic oscillator coherent states with respect to completeness,    resolution of the identity  and minimum uncertainty states. They are obtained by  acting with a  generalized matrix displacement operator on a zero energy eigenstate.

\subsubsection{Vector coherent states of two components}
When $K=2,$  the $\tilde{M}$ matrix has the general form

\begin{equation}
\tilde{M}= \begin{pmatrix}
\tilde{m}_{11} & \tilde{m}_{12} \\
 \tilde{m}_{21} & \tilde{m}_{22} 
\end{pmatrix} ,  \label{tilde-matrix-K=2}
\end{equation}
and its associated eigenvalues are
\begin{equation}
\tilde{\lambda}_{[\pm]} = \frac{1}{2} (\tilde{m}_{11} + \tilde{m}_{22}) \pm \tilde{b}, \quad
\mbox{where} \quad  \tilde{b} = \sqrt{4 \tilde{m}_{12} \tilde{m}_{21} + {(\tilde{m}_{11} - \tilde{m}_{22})}^2}.
\end{equation}
 Assuming that $\tilde{b} \neq 0,$ we can show that  the  passing matrices take the form
 \begin{equation}
 P = \begin{pmatrix}
2 \tilde{m}_{12} & \tilde{m}_{11} - \tilde{m}_{22} - \tilde{b} \\
 \tilde{m}_{22} - \tilde{m}_{11} + \tilde{b} & 2 \tilde{m}_{21}\end{pmatrix}, \quad  
 P^{-1} = \frac{ \begin{pmatrix}
2 \tilde{m}_{21} & \tilde{m}_{22} - \tilde{m}_{11} + \tilde{b} \\
 \tilde{m}_{11} - \tilde{m}_{22} - \tilde{b} & 2 \tilde{m}_{12}\end{pmatrix}}{2 \tilde{b} (1+\tilde{m}_{22} - \tilde{m}_{11} )}  , \label{matrix-P-K=2}
\end{equation}  
 and we can verify that
 
 \begin{equation}
 P^{-1} \tilde{M} P =\tilde{D} =   \begin{pmatrix}
\tilde{\lambda}_{+} & 0 \\
 0 & \tilde{\lambda}_{-}
\end{pmatrix}.
 \end{equation}
Then the two component vector coherent state writes 

\begin{equation}
\mid \Psi\rangle^j_{[2]} = {\mathcal{N}}^{-1/2} P \mid \tilde{\Psi} \rangle^j_{[2]} = {\mathcal{N}}^{-1/2} \begin{pmatrix}
2 \tilde{m}_{12} & \tilde{m}_{11} - \tilde{m}_{22} - \tilde{b} \\
 \tilde{m}_{22} - \tilde{m}_{11} + \tilde{b} & 2 \tilde{m}_{21}\end{pmatrix}
 \begin{pmatrix}  \mid \tilde{\psi}\rangle^j_{[1]} \\ 
 \mid \tilde{\psi}\rangle^j_{[2]}  \end{pmatrix}  \label{vector-coherent-state-K=2}
\end{equation}
where $\mathcal{N} $ is a normalization constant to be determined by the condition $  {}_{[2]}^j \langle \Psi \mid \Psi\rangle_{[2]}^j =1,$
or more explicitly 

\begin{equation}
\mid \Psi\rangle^j_{[2]} = {\mathcal{N}}^{-1/2} 
\begin{pmatrix} 
2 \tilde{m}_{12}  \mid \tilde{\psi}\rangle^j_{[1]} +  [\tilde{m}_{11} - \tilde{m}_{22} - \tilde{b}] \mid \tilde{\psi}\rangle^j_{[2]}  \\
 [\tilde{m}_{22} - \tilde{m}_{11} + \tilde{b}]  \mid \tilde{\psi}\rangle^j_{[1]} +  2 \tilde{m}_{21} \mid \tilde{\psi}\rangle^j_{[2]}
\end{pmatrix}. \label{vector-coherent-state-K=2-explicitly}
\end{equation}
When $\tilde{M}$ in (\ref{tilde-matrix-K=2}) is normal, i.e., when its entries verify 
$ \tilde{m}_{12}(\tilde{m}_{11} - \tilde{m}_{22})^{\ast}  =  \tilde{m}_{21}^\ast (\tilde{m}_{11} - \tilde{m}_{22}) $ and  $\|\tilde{m}_{12}\| = \|\tilde{m}_{21}\|, $ the $P$ matrix in 
(\ref{matrix-P-K=2}) becomes  unitary, which we  name $\tilde{U}$, then normalization constant in (\ref{vector-coherent-state-K=2}) is  easier  
to compute.
\subsubsection{Generalized quaternionic vector coherent states}
When we choose the entries of $\tilde{M}$ in   equation (\ref{tilde-matrix-K=2}) as

\begin{eqnarray}
\tilde{m}_{11} &=& r[ \cos\theta + i \sin\theta \cos\phi], \quad \tilde{m}_{12} = i r \sin\theta \sin\phi e^{i \psi} \nonumber \\
\tilde{m}_{21} &=& i r \sin\theta \sin\phi e^{- i \psi}  \quad  \mbox{and} \quad \tilde{m}_{22} = r[ \cos\theta - i \sin\theta \cos\phi], 
\end{eqnarray}
then $\tilde{b}= 2ir\sin\theta,$ $\tilde{\lambda}_{[\pm]} = r e^{\pm i \theta}$ and $\tilde{M}$ becomes normal and takes the form given in equation (\ref{eq-quaternions-matrix}), which corresponds to a complex representation of quaternions by square matrices of dimension $2.$ Also, the matrix $P$ becomes unitary and the states in equation  (\ref{vector-coherent-state-K=2-explicitly}) turn into the following generalized quaternionic vector coherent states:

\begin{equation}
\mid \Psi\rangle^j_{[2]} = {\mathcal{N}}^{-1/2} 
\begin{pmatrix} 
\cos\left(\frac{\phi}{2}\right) e^{i \psi} \mid \tilde{\psi}\rangle^j_{[1]} -  \sin\left(\frac{\phi}{2}\right) \mid \tilde{\psi}\rangle^j_{[2]}  \\
  \sin\left(\frac{\phi}{2}\right)  \mid \tilde{\psi}\rangle^j_{[1]} +  \cos\left(\frac{\phi}{2}\right) e^{-i\psi}  \mid \tilde{\psi}\rangle^j_{[2]}
\end{pmatrix},  \label{vector-coherent-states-all-j-K=2} 
\end{equation}
where some multiplicative factors have been absorbed by the normalization constant $\mathcal{N}.$

For example, in the special case $j=\frac{1}{2},$ the states

\begin{equation}
\mid \tilde{\psi} \rangle^{\frac{1}{2}}_{[s]} = N_{[s]}^{-1/2}\; \begin{pmatrix}
\tilde{\psi}^{\frac{1}{2}}_{[s]  \frac{-1}{2}}(0) \; e^{\lambda^{[s] \frac{1}{2} }_{\frac{1}{2}} \hat{a}^\dagger} & \tilde{\psi}^{\frac{1}{2}}_{[s]  \frac{+ 1}{2}}(0) \; e^{\lambda^{[s] \frac{1}{2}}_{- \frac{1}{2}} \hat{a}^\dagger}
 \end{pmatrix} \;
 V^{t} \;
 \begin{pmatrix}
 \mid 0 ; \frac{1}{2}, - \frac{1}{2} \rangle \\
 \mid 0 ; \frac{1}{2}, + \frac{1}{2} \rangle\\
 \end{pmatrix}, \quad s=1,2,
\end{equation}
where $\lambda^{[s] \frac{1}{2} }_{\frac{1}{2}} = \tilde{\lambda}_{[s]} + \frac{1}{2} b, s=1,2$ and  $\lambda^{[s] \frac{1}{2} }_{-\frac{1}{2}} =\tilde{\lambda}_{[s]} - \frac{1}{2} b, s=1,2,$ with $b=\sqrt{4 \beta_+ \beta_- + \beta_3^2},$ and 

\begin{equation}
V= \begin{pmatrix}
\sqrt{\frac{b +\beta_3}{2b}}&  \frac{2 \beta_+}{\sqrt{2b (b+\beta_3)}}\\
\frac{- 2 \beta_-}{\sqrt{2b (b+\beta_3)}}&\sqrt{\frac{b +\beta_3}{2b}}
\end{pmatrix},\end{equation} 
where $\beta_{\pm}$ and $\beta_3$ satisfy the necessary conditions for $[\mathbb{A},\mathbb{A}^\dagger]=\mathbb{I}.$ Explicitly, the eigenvalues of the $h(1) \otimes su(2)$ sector, depending on the eigenvalues of the $\tilde{M}$ matrix, are given by 
\begin{eqnarray}
\lambda^{[1] \frac{1}{2} }_{+ \frac{1}{2}} = r e^{i \theta} + \frac{b}{2}, \quad 
\lambda^{[1] \frac{1}{2} }_{- \frac{1}{2}} = r e^{i \theta} - \frac{b}{2}, \nonumber \\ 
\lambda^{[2] \frac{1}{2} }_{+\frac{1}{2}} = r e^{-i \theta} + \frac{b}{2}, \quad  
\lambda^{[2] \frac{1}{2} }_{- \frac{1}{2}} = r e^{- i \theta} - \frac{b}{2}, 
\end{eqnarray}
with $b= e^{i \theta_3} \sqrt{4 R^2 + R_3^2},$ where $e^{i \theta_3} = e^{\frac{i}{2} (\theta+ + \theta_-)},$
and the unitary matrix $V$ reads

\begin{eqnarray}
V= \sqrt{\frac{R_3 +\sqrt{4 R^2 + R_3^2}}{2 \sqrt{ 4 R^2 + R_3^2 }}} \begin{pmatrix}
1 & \frac{2R \; e^{\frac{i}{2} (\theta_+ - \theta_-)}}{R_3 +\sqrt{4 R^2 + R_3^2} } \\
\frac{ - 2R \; e^{\frac{- i}{2} (\theta_+ - \theta_-)}}{R_3 +\sqrt{4 R^2 + R_3^2} } & 1
 \end{pmatrix} .
 \end{eqnarray}  
Finally using (\ref{equation-components-h1-su2-algebra-eigenstates}) for the particular case being treated  here we get the set of states
\begin{eqnarray}
\mid \tilde{\psi}\rangle^{\frac{1}{2}}_{[1]} &=& \sqrt{\frac{R_3 +\sqrt{4 R^2 + R_3^2}}{2 \sqrt{ 4 R^2 + R_3^2 }}}  \nonumber \\  
&\times &  \left\lbrace   \tilde{\psi}^{\frac{1}{2}}_{[1]  \frac{-1}{2}}(0)  \left[ 
 \mid r e^{i \theta} + \frac{b}{2} ; + \frac{1}{2} , -  \frac{1}{2} \rangle  -  \frac{  2R \; e^{\frac{- i}{2} (\theta_+ - \theta_-)}}{R_3 +\sqrt{4 R^2 + R_3^2} } \mid r e^{i \theta} + \frac{b}{2} ; + \frac{1}{2} , +  \frac{1}{2} \rangle \right]   \right.  \nonumber \\
&+&\left.  \tilde{\psi}^{\frac{1}{2}}_{[1]  \frac{+1}{2}}(0)   \left[
\frac{2R \; e^{\frac{i}{2} (\theta_+ - \theta_-)}}{R_3 +\sqrt{4 R^2 + R_3^2} } \mid r e^{i \theta} - \frac{b}{2}  ; + \frac{1}{2} , -  \frac{1}{2}  \rangle  +  \mid r e^{i \theta} - \frac{b}{2} ; + \frac{1}{2} , +  \frac{1}{2} \rangle \right]  \right\rbrace
\end{eqnarray}
and
 \begin{eqnarray}
\mid \tilde{\psi}\rangle^{\frac{1}{2}}_{[2]} &=& \sqrt{\frac{R_3 +\sqrt{4 R^2 + R_3^2}}{2 \sqrt{ 4 R^2 + R_3^2 }}}  \nonumber \\  
&\times &  \left\lbrace   \tilde{\psi}^{\frac{1}{2}}_{[2]  \frac{-1}{2}}(0)  \left[ 
 \mid r e^{- i \theta} + \frac{b}{2} ; + \frac{1}{2} , -  \frac{1}{2} \rangle  -  \frac{  2R \; e^{\frac{- i}{2} (\theta_+ - \theta_-)}}{R_3 +\sqrt{4 R^2 + R_3^2} } \mid r e^{-i \theta} + \frac{b}{2} ; + \frac{1}{2} , +  \frac{1}{2} \rangle \right]   \right.  \nonumber \\
&+&\left.  \tilde{\psi}^{\frac{1}{2}}_{[2]  \frac{+1}{2}}(0)   \left[
\frac{2R \; e^{\frac{i}{2} (\theta_+ - \theta_-)}}{R_3 +\sqrt{4 R^2 + R_3^2} } \mid r e^{-i \theta} - \frac{b}{2}  ; + \frac{1}{2} , -  \frac{1}{2}  \rangle  +  \mid r e^{-i \theta} - \frac{b}{2} ; + \frac{1}{2} , +  \frac{1}{2} \rangle \right]  \right\rbrace .
\end{eqnarray}

Thus, if we insert these last results into equation (\ref{vector-coherent-state-K=2-explicitly}) for fixed $j = 1/2,$ with the corresponding unitary matrix $\tilde{U}$ given by
\begin{equation}
\tilde{U}=  \begin{pmatrix} 
\cos\left( \frac{\phi}{2}\right)  e^{i\psi}   &  -  \sin\left(\frac{\phi}{2}\right)  \\
  \sin\left(\frac{\phi}{2}\right)  &  \cos\left(\frac{\phi}{2}\right) e^{-i\psi} 
\end{pmatrix},
\end{equation} we finally obtain the generalized quaternionic vector coherent states based on the $h(1) \oplus su(2) $ algebra.

\begin{equation}
\mid \Psi\rangle^{\frac{1}{2}}_{[2]} = \mathcal{N}^{\frac{- 1}{2}}  \tilde{U}
\begin{pmatrix}\tilde{\psi}\rangle^{\frac{1}{2}}_{[1]} \\
\tilde{\psi}\rangle^{\frac{1}{2}}_{[2]}\end{pmatrix}
=  \mathcal{N}^{\frac{- 1}{2}} 
\begin{pmatrix} 
\cos\left( \frac{\phi}{2}\right) e^{i \psi} \mid \tilde{\psi}\rangle^{\frac{1}{2}}_{[1]} -  \sin\left(\frac{\phi}{2}\right) \mid \tilde{\psi}\rangle^{\frac{1}{2}}_{[2]}  \\
  \sin\left(\frac{\phi}{2}\right)  \mid \tilde{\psi}\rangle^{\frac{1}{2}}_{[1]} +  \cos\left(\frac{\phi}{2}\right) e^{-i\psi}  \mid \tilde{\psi}\rangle^{\frac{1}{2}}_{[2]}
\end{pmatrix}.  \label{VCS-j=1/2}
\end{equation}

These last states are then eigenstates of the generalized annihilation operator $\mathbb{A}$ with eigenvalues  on a $2 \times 2 $ complex matrix representing the quaternions. We emphasize  that equation {\ref{VCS-j=1/2}} represent a set of four linearly independent vector coherent states  associated to the quantum system described by the Hamiltonian (\ref{generalized-oscillator-hamiltonian}), in the $j= \frac{1}{2}$ representation.  Also we note that in this case  the vector coherent states  (\ref{VCS-j=1/2}) can be written in the normalized form
\begin{eqnarray}
\mid \Psi\rangle^{\frac{1}{2}}_{[2]} &=&  
\begin{pmatrix} 
\cos\left( \frac{\phi}{2}\right)  e^{i\psi}   &  -  \sin\left(\frac{\phi}{2}\right)  \\
  \sin\left(\frac{\phi}{2}\right)  &  \cos\left(\frac{\phi}{2}\right) e^{-i\psi} 
\end{pmatrix}
\begin{pmatrix}
e^{(r e^{i \theta} \hat{a}^\dagger-  r e^{-i \theta}\hat{a}) } & 0 \\
0 & e^{(r e^{-i \theta} \hat{a}^\dagger    -  r e^{i \theta} \hat{a}) } 
\end{pmatrix}  \nonumber \\  &\times&  \begin{pmatrix}
e^{ \frac{1}{2} (r e^{i \theta} (\mathbb{A}- \hat{a})^\dagger  -  r e^{-i \theta}   (\mathbb{A} - \hat{a})    ) } & 0 \\
0 &  e^{ \frac{1}{2}  (r e^{-i \theta} (\mathbb{A}- \hat{a})^\dagger  -  r e^{i \theta}   (\mathbb{A} - \hat{a})    )  } 
\end{pmatrix}   \nonumber \\  &\times& 
 \sqrt{\frac{R_3 +\sqrt{4 R^2 + R_3^2}}{4 \sqrt{ 4 R^2 + R_3^2 }}}
\begin{pmatrix}
\frac{\tilde{\psi}^{\frac{1}{2}}_{[1]  \frac{-1}{2}}(0)}{ \sqrt{\|\tilde{\psi}^{\frac{1}{2}}_{[1]  \frac{-1}{2}}(0)\|^2 + \|\tilde{\psi}^{\frac{1}{2}}_{[1]  \frac{+1}{2}}(0) \|^2 } }      & \frac{\tilde{\psi}^{\frac{1}{2}}_{[1]  \frac{+1}{2}}(0)}{ \sqrt{\|\tilde{\psi}^{\frac{1}{2}}_{[1]  \frac{-1}{2}}(0)\|^2 + \|\tilde{\psi}^{\frac{1}{2}}_{[1]  \frac{+1}{2}}(0) \|^2 } }  \\
\frac{ \tilde{\psi}^{\frac{1}{2}}_{[2]\frac{-1}{2}}(0)}{ \sqrt{\|\tilde{\psi}^{\frac{1}{2}}_{[2]  \frac{-1}{2}}(0)\|^2 + \|\tilde{\psi}^{\frac{1}{2}}_{[2]  \frac{+1}{2}}(0) \|^2 } }   & \frac{ \tilde{\psi}^{\frac{1}{2}}_{[2]  \frac{+1}{2}}(0)}{ \sqrt{\|\tilde{\psi}^{\frac{1}{2}}_{[2]  \frac{-1}{2}}(0)\|^2 + \|\tilde{\psi}^{\frac{1}{2}}_{[2]  \frac{+1}{2}}(0) \|^2 } } 
\end{pmatrix}  \nonumber \\  \label{VCS-j=1/2-first-transformation}
&\times&
 \begin{pmatrix} 
 \mid  \frac{b}{2} ; + \frac{1}{2} , -  \frac{1}{2} \rangle  -  \frac{  2R \; e^{\frac{- i}{2} (\theta_+ - \theta_-)}}{R_3 +\sqrt{4 R^2 + R_3^2} } \mid  \frac{b}{2} ; + \frac{1}{2} , +  \frac{1}{2} \rangle 
 \\
\frac{2R \; e^{\frac{i}{2} (\theta_+ - \theta_-)}}{R_3 +\sqrt{4 R^2 + R_3^2} } \mid  - \frac{b}{2}  ; + \frac{1}{2} , -  \frac{1}{2}  \rangle  +  \mid  - \frac{b}{2} ; + \frac{1}{2} , +  \frac{1}{2} \rangle 
 \end{pmatrix}.
\end{eqnarray} 
Choosing now the integration constants as follows 
\begin{equation}
 \begin{pmatrix}
\frac{\tilde{\psi}^{\frac{1}{2}}_{[1]  \frac{-1}{2}}(0)}{ \sqrt{\|\tilde{\psi}^{\frac{1}{2}}_{[1]  \frac{-1}{2}}(0)\|^2 + \|\tilde{\psi}^{\frac{1}{2}}_{[1]  \frac{+1}{2}}(0) \|^2 } }      & \frac{\tilde{\psi}^{\frac{1}{2}}_{[1]  \frac{+1}{2}}(0)}{ \sqrt{\|\tilde{\psi}^{\frac{1}{2}}_{[1]  \frac{-1}{2}}(0)\|^2 + \|\tilde{\psi}^{\frac{1}{2}}_{[1]  \frac{+1}{2}}(0) \|^2 } }  \\
\frac{ \tilde{\psi}^{\frac{1}{2}}_{[2]\frac{-1}{2}}(0)}{ \sqrt{\|\tilde{\psi}^{\frac{1}{2}}_{[2]  \frac{-1}{2}}(0)\|^2 + \|\tilde{\psi}^{\frac{1}{2}}_{[2]  \frac{+1}{2}}(0) \|^2 } }   & \frac{ \tilde{\psi}^{\frac{1}{2}}_{[2]  \frac{+1}{2}}(0)}{ \sqrt{\|\tilde{\psi}^{\frac{1}{2}}_{[2]  \frac{-1}{2}}(0)\|^2 + \|\tilde{\psi}^{\frac{1}{2}}_{[2]  \frac{+1}{2}}(0) \|^2 } } 
\end{pmatrix}
 = \; \tilde{U}^{\dagger}  \;  \exp \left[ -  \frac{1}{2} \left(  \tilde{M}  \frac{b^\ast}{2} - \tilde{M}^\dagger \frac{b}{2} \right) \sigma_3 \right]   
\end{equation}
and using the Baker–Campbell–Hausdorff formula for splitting the displacement operators and the fact that $\tilde{U} \exp \left(\begin{smallmatrix} r e^{i \theta}& 0 \\ 0 & r e^{-i \theta}\end{smallmatrix} \right) \tilde{U}^\dagger = \exp{\tilde{M}} ,$ and then recomposing the resulting  expression following the inverse process , we can put  (\ref{VCS-j=1/2-first-transformation}) in the compact form

\begin{eqnarray}
\mid \Psi (\tilde M) \rangle^{\frac{1}{2}}_{[2]} &=&   e^{(\tilde{M} \mathbb{A}^\dagger - \tilde{M}^\dagger \mathbb{A} )}
 \sqrt{\frac{R_3 +\sqrt{4 R^2 + R_3^2}}{ 4 \sqrt{ 4 R^2 + R_3^2 }}}
 \begin{pmatrix} 
 \mid  \frac{b}{2} ; + \frac{1}{2} , -  \frac{1}{2} \rangle  -  \frac{  2R \; e^{\frac{- i}{2} (\theta_+ - \theta_-)}}{R_3 +\sqrt{4 R^2 + R_3^2} } \mid  \frac{b}{2} ; + \frac{1}{2} , +  \frac{1}{2} \rangle 
 \\
\frac{2R \; e^{\frac{i}{2} (\theta_+ - \theta_-)}}{R_3 +\sqrt{4 R^2 + R_3^2} } \mid  - \frac{b}{2}  ; + \frac{1}{2} , -  \frac{1}{2}  \rangle  +  \mid  - \frac{b}{2} ; + \frac{1}{2} , +  \frac{1}{2} \rangle 
 \end{pmatrix}, \label{VCS-j=1/2-second-transformation}
\end{eqnarray} 
from where we can extract one version of the fundamental state of the system governed by the Hamiltonian (\ref{generalized-oscillator-hamiltonian}), that is
\begin{eqnarray}
\mid \Psi (0) \rangle^{\frac{1}{2}}_{[2]} =  
 \sqrt{\frac{R_3 +\sqrt{4 R^2 + R_3^2}}{ 4 \sqrt{ 4 R^2 + R_3^2 }}}
 \begin{pmatrix} 
 \mid  \frac{b}{2} ; + \frac{1}{2} , -  \frac{1}{2} \rangle  -  \frac{  2R \; e^{\frac{- i}{2} (\theta_+ - \theta_-)}}{R_3 +\sqrt{4 R^2 + R_3^2} } \mid  \frac{b}{2} ; + \frac{1}{2} , +  \frac{1}{2} \rangle 
 \\
\frac{2R \; e^{\frac{i}{2} (\theta_+ - \theta_-)}}{R_3 +\sqrt{4 R^2 + R_3^2} } \mid  - \frac{b}{2}  ; + \frac{1}{2} , -  \frac{1}{2}  \rangle  +  \mid  - \frac{b}{2} ; + \frac{1}{2} , +  \frac{1}{2} \rangle \end{pmatrix}.\label{fundamental-state-j=1/2-quaternions}
\end{eqnarray}

To conclude this section, let us say a few words about the system we are dealing with. The fundamental state   (\ref{fundamental-state-j=1/2-quaternions}) can be used for finding the energy eigenstates of the hamiltonian  (\ref{generalized-oscillator-hamiltonian}). By the way it latter was conceived, we know that it is isospectral with the harmonic oscillator Hamiltonian. The normalized energy eigenstates of this system are given by

\begin{equation}
 \mid \tilde{n} \rangle^{\frac{1}{2}}= \frac{(\hat{\mathbb{A}}^\dagger)^n}{ \sqrt{n!}} \mid \Psi (0) \rangle^{\frac{1}{2}}_{[2]} .
\end{equation}  
 The vector coherent states (\ref{VCS-j=1/2-second-transformation}) based on the $h(1) \oplus su(2)$ algebra constitute an interesting generalization of the quaternionic vector  coherent states studied in the literature. It depends certainly on the choice of the integration constants that remains us the highly degenerate  system we are studying.

\subsection{ Extended canonical commutation relation and intelligent states}
Let us now choose the beta parameters in such a way that the commutation relation (\ref{commutator-A-A-dagger})  verify

\begin{equation}
\left[ \mathbb{A}  ,   \mathbb{A}^\dagger \right]= \hat{I}  + 2  x \hat{J}_3,
\label{commutator-A-A-dagger-x-J-3}
\end{equation}
where $x \in \mathbb{R} $ such that  $x  > 0 $ or $x < 0.$  Here four scenarios are possible,  all of them requiring $\beta_3=0:$

\begin{equation}
    \begin{aligned}\setcounter{mysubequations}{0}
      \mysubnumber   \quad x  > 0;  & \quad  \beta_{+} = \sqrt{x} \sinh \alpha e^{i \theta_+} , \quad \beta_{-} =   \sqrt{x}\cosh \alpha e^{i \theta_-}; \quad \alpha \neq 0  \Rightarrow b \neq 0\\
      \mysubnumber \quad x> 0; & \quad \beta_{+} = 0, \quad \beta_{-} =   \sqrt{x} e^{i \theta_-};  \; \Rightarrow \; b=0\\
      \mysubnumber \quad x< 0; & \quad \beta_{+} = \sqrt{-x} \cosh \alpha e^{i \theta_+} , \quad \beta_{-} = \sqrt{-x}  \cosh \alpha e^{i \theta_-}; \quad \alpha \neq 0 ; \; \Rightarrow b \neq 0 \\ 
      \mysubnumber  \quad  x < 0; & \quad \beta_{+} = \sqrt{-x} e^{i \theta_+} , \quad \beta_{-} = 0;   \Rightarrow b = 0 
    \end{aligned} .
   \end{equation}

 In the  cases listed as (i) and (iii) the $b$ parameter is different from zero and hence, in the process of solving the eigenvalue  equation (\ref{reduction-to-algebra-h1-su2-eigenstates} ), which is necessary to solve  the matrix eigenvalue equation (\ref{vector-matrix-eigenvalue-total}),  we find that  the matrix  formed from the  matrix elements  of the linear combination of the generators of the  $ su(2)$ algebra  is diagonalizable but  the passing matrix $V$ is not unitary.  However, the  technique to obtain the generalized vector states  has no difference with the one followed in the above section except for the fact that the computation of the normalization constants is now  more difficult. We will not deal with this case now, we will leave it for last, we will treat it together with all similar cases. 
  The cases listed as (ii) and \setcounter{mysubequations}{3} \mysubnumber   are very special because in both the parameter $b$ is equal to zero and the aforementioned matrix diagonalization process does not apply. In appendix \ref{appb} we  show  explicitly the  method to solve the corresponding  $h(1) \oplus su(2)$  eigenvalue type  equations  appearing after of performing the  diagonalization of the $\tilde{M}$ matrix in  (\ref{vector-matrix-eigenvalue-total}),  in the special case where   
$\beta_-=\beta_3=0$ and $\beta_+ \neq 0,$  and  also we adapt the results of these latter for obtaining those in the case $\beta_+=\beta_3=0$ and $\beta_- \neq 0.$   

 Thus, in the case listed as \setcounter{mysubequations}{3}  \mysubnumber,  inserting  (\ref{algebra-eigenstates-beta3=0-beta-=0}) into (\ref{general-solution-beta3=0,beta-=0}) and this latter into    (\ref{vector-algebra-states-}), but not without first substituting the  $\beta$ parameter by the corresponding  eigenvalues of $\tilde{M},$ i.e.,   $\tilde{\lambda}_{[s]}, s=1,2, \cdots, K,$   we get the components of the vector state  verifying   (\ref{vector-matrix-eigenvalue-total}):

\begin{equation}             
\mid  \psi \rangle^j_{[u]} = \sum_{s=1}^{K} P_{u s}    \mid \Psi [\tilde{\lambda}_{[s]}, \beta_+ ] \rangle^j
=  \sum_{s=1}^{K}  \sum_{m=-j}^{j}  P_{u s}  \tilde{\varphi}^j_{[s] m} (0) \mid \psi  [\tilde{\lambda}_{[s]} ,\beta_+]  \rangle^j_m  \quad u=1,2, \cdots, K,  \label{vector-algebra-eigenstates-beta3=0-beta-=0}
\end{equation}   
 or better using \ref{algebra-eigenstates-beta3=0-beta-=0-D}
\begin{equation}             
\mid  \psi \rangle^j_{[u]} =  \sum_{s=1}^{K}  \sum_{m=-j}^{j}  P_{u s}       \tilde{\varphi}^j_{[s] m} (0)   \hat{D} (\tilde{\lambda}_{[s]})  \mid \tilde{\psi}  [\tilde{\lambda}_{[s]} ,\beta_+]  \rangle^j_m  \quad u=1,2, \cdots, K,  \label{vector-algebra-eigenstates-beta3=0-beta-=0-D}
\end{equation} 
which in the matrix form look like
\begin{equation}
\mid  \Psi \rangle^j_{[K]} = P \begin{pmatrix}  \hat{D} (\tilde{\lambda}_{[1]})  & 0& \cdots & 0 &  0 \\
0&  \hat{D} (\tilde{\lambda}_{[2]}) & 0 & \cdots  & 0\\
\vdots& 0& \ddots & 0 & \vdots  \\
0& \cdots&0&  \hat{D} (\tilde{\lambda}_{[K-1]}) &0 \\
0&0& \cdots&0 & \hat{D} (\tilde{\lambda}_{[K]})
 \end{pmatrix} 
  \begin{pmatrix}
 \sum_{m=-j}^{j}   \tilde{\varphi}^j_{[1] m} (0) \mid \tilde{\psi}   [\tilde{\lambda}_{[1]} ,\beta_+]  \rangle^j_m \\
  \sum_{m=-j}^{j}  \tilde{\varphi}^j_{[2] m} (0) \mid \tilde{\psi}  [\tilde{\lambda}_{[2]} ,\beta_+]  \rangle^j_m \\
 \vdots\\
  \sum_{m=-j}^{j}   \tilde{\varphi}^j_{[K] m} (0) \mid \tilde{\psi}  [\tilde{\lambda}_{[K]} ,\beta_+]  \rangle^j_m
 \end{pmatrix}. \label{vector-algebra-eigenstates-beta3=0-beta-=0-D-matrix-form}
\end{equation}
When  $\tilde{M}$ is normal   $P=\tilde{U}$ is unitary, then the vector states  (\ref{vector-algebra-eigenstates-beta3=0-beta-=0-D-matrix-form}) can be written   in the suggested form

\begin{equation}
\mid  \Psi \rangle^j_{[K]} =  \mathcal{N}^{- \frac{1}{2}} \exp\left[  \tilde{M} \hat{a}^\dagger  - \tilde{M}^{\dagger}  \hat{a}     \right]  \tilde{U}
  \begin{pmatrix}
 \sum_{m=-j}^{j}   \tilde{\varphi}^j_{[1] m} (0) \mid \tilde{\psi}   [\tilde{\lambda}_{[1]} ,\beta_+]  \rangle^j_m \\
  \sum_{m=-j}^{j}  \tilde{\varphi}^j_{[2] m} (0) \mid  \tilde{\psi}  [\tilde{\lambda}_{[2]} ,\beta_+]  \rangle^j_m \\
 \vdots\\
  \sum_{m=-j}^{j}   \tilde{\varphi}^j_{[K] m} (0) \mid \tilde{\psi}   [\tilde{\lambda}_{[K]} ,\beta_+]  \rangle^j_m
 \end{pmatrix}. \label{vector-algebra-eigenstates-beta3=0-beta-=0-D-matrix-form-normal-case}
\end{equation}
 We note that we cannot do the same with the vector states  (\ref{vector-algebra-eigenstates-beta3=0-beta-=0-D-matrix-form}) due to the fact that in this case $\tilde{M}$ is not normal. Indeed, in such a case we can use the  Baker–Campbell–Hausdorff formula to separate the harmonic oscillator type displacement operators but we cannot reconstitute it in a closed unitary exponential form by the inverse process because $\tilde{M}$ does not commute with $\tilde{M}^\dagger.$

We remark that 

\begin{equation}
 \begin{pmatrix}
 \sum_{m=-j}^{j}   \tilde{\varphi}^j_{[1] m} (0) \mid \tilde{\psi}  [\tilde{\lambda}_{[1]} ,\beta_+]  \rangle^j_m \\
  \sum_{m=-j}^{j}  \tilde{\varphi}^j_{[2] m} (0) \mid  \tilde{\psi}   [\tilde{\lambda}_{[2]} ,\beta_+]  \rangle^j_m \\
 \vdots\\
  \sum_{m=-j}^{j}   \tilde{\varphi}^j_{[K] m} (0) \mid   \tilde{\psi}   [\tilde{\lambda}_{[K]} ,\beta_+]  \rangle^j_m
 \end{pmatrix}
\end{equation}
is an eigenstate of $\hat{a} + \beta_+ \hat{J}_-$ with matrix eigenvalue equals to $0.$ Then using the displacement operator property
$  \exp\left[ -(  \tilde{M} \hat{a}^\dagger  - \tilde{M}^{\dagger)}  \hat{a}     \right] (\hat{a} + \beta_+ \hat{J}_-)  \exp\left[  \tilde{M} \hat{a}^\dagger  - \tilde{M}^{\dagger}  \hat{a}     \right]  = \hat{a} + \beta_+ \hat{J}_- +\tilde{M},$ we can also show  that (\ref{vector-algebra-eigenstates-beta3=0-beta-=0-D-matrix-form-normal-case}) is effectively  an eigenstate of  $\hat{a} + \beta_+ \hat{J}_-$ with matrix eigenvalue equals to $\tilde{M}.$ 

The vector states in (\ref{vector-algebra-eigenstates-beta3=0-beta-=0-D-matrix-form-normal-case})   represent a large set of different states depending on the choice of the integration constants  $\tilde{\varphi}^j_{[s] m} (0). $ In general, they cannot be considered as vector coherent states  associated to the Hamiltonian $\mathbb{H}=\mathbb{A}^\dagger \mathbb{A} = \hat{a}^\dagger \hat{a} + \beta_+  \hat{a}^\dagger \hat{J}_+ + \beta^\ast_{+} \hat{a} \hat{J}_- +  \|\beta_+\|^2  \hat{J}_- \hat{J}_+ ,$ but as intelligent states in the sense that they minimize the generalized Schr\"odinger-Robertson uncertainty relation $(\Delta \hat{\mathcal{X}})^2 (\Delta \hat{\mathcal{P}})^2 \geq  \frac{1}{4} \left[   \langle\hat{C} \rangle^2     +     \langle \hat{F} \rangle^2 \right],$ where $i \hat{C} = [\hat{\mathcal{X}},\hat{\mathcal{P}}] $ and $\langle \hat{F} \rangle=  \langle \{ \hat{\mathcal{X}}, \hat{\mathcal{X}}  \} \rangle - 2 \langle \hat{\mathcal{X}} \rangle  \; \langle \hat{\mathcal{P}} \rangle, $  with  $\hat{\mathcal{X}} =  \frac{\mathbb{A} + \mathbb{A}^\dagger }{\sqrt{2}}$ and  $\hat{\mathcal{P}} =  \frac{\mathbb{A}  -  \mathbb{A}^\dagger }{\sqrt{2} i}.$

\subsubsection{Heisenberg-Weyl  vector coherent states on the matrix domain}

 A special choice of the integration constants  in (\ref{vector-algebra-eigenstates-beta3=0-beta-=0-D-matrix-form-normal-case})  is given by

\begin{equation}
\tilde{\varphi}^j_{[s] m} (0) = \delta^{[s] }_{-j, m } \sum_{u=1}^{K} \tilde{U}^{\dagger}_{s u }       , \quad s=1, \cdots, K; \quad m= -j, \cdots, j,
\end{equation}
where  $\delta^{[s] }_{-j,  m } $ is the Kronecker delta which is equal to $1$ if $m=-j$ and $0$ otherwise. With this choice the states  (\ref{vector-algebra-eigenstates-beta3=0-beta-=0-D-matrix-form-normal-case}) become
\begin{equation}
\mid  \Psi [\tilde{M} ] \rangle^j_{[K]} =  \frac{1}{\sqrt{K}}  \exp\left[  \tilde{M} \hat{a}^\dagger  - \tilde{M}^{\dagger}  \hat{a}     \right] 
  \begin{pmatrix}
\mid 0 \rangle \otimes \mid j , - j  \rangle  \\
  \mid 0 \rangle \otimes \mid j , - j  \rangle  \\
 \vdots\\
  \mid 0 \rangle \otimes \mid j , - j  \rangle  
 \end{pmatrix},\label{standard-VCS+j-j}
\end{equation}
which are standard Heisenberg- Weyl vector coherent states  on the matrix domain, which have been  obtained  here by applying the  matrix  extension  of the harmonic oscillator displacement operator on the simpler vector eigenstate of $\hat{a} + \beta_{+} \hat{J}_-$ with zero matrix eigenvalue.   

\subsubsection{Vector algebra eigenstates  associated to the standard  $h(1) \oplus su(2)$ annihilator operator }

Looking at equation (\ref{vector-algebra-eigenstates-beta3=0-beta-=0-D-matrix-form-normal-case})  we can see that we can go even further in the change of structure  of these states and express  them in a more compact form, where the contribution of each sub algebra is more easily visualized.   To do this, we can insert equation   (\ref{algebra-eigenstates-beta3=0-beta-=0-D-J-} ) into  (\ref{vector-algebra-eigenstates-beta3=0-beta-=0-D-matrix-form-normal-case})  and proceed with the inverse diagonalization process  in the same way   we have done in previous subsections, doing all that we get

\begin{equation}
\mid  \Psi \rangle^j_{[K]} =  \mathcal{N}^{- \frac{1}{2}} \exp\left[  \tilde{M} \hat{a}^\dagger  - \tilde{M}^{\dagger}  \hat{a}     \right]  
  \exp\left[  - ( \hat{a}^\dagger  I + \tilde{M}^\dagger  ) \beta_+ \hat{J}_-     \right]  \tilde{U} \begin{pmatrix}
 \sum_{m=-j}^{j}   \frac{\tilde{\varphi}^j_{[1] m} (0)}{\sqrt{\tilde{\mathcal{N}}^j_m  [\tilde{\lambda}^j_{[1]}, \beta_+]}} \mid  0 \rangle \otimes \mid  j,  m \rangle\\
  \sum_{m=-j}^{j}   \frac{\tilde{\varphi}^j_{[2] m} (0)}{\sqrt{\tilde{\mathcal{N}}^j_m  [\tilde{\lambda}^j_{[2]}, \beta_+]}} \mid  0 \rangle \otimes \mid  j,  m \rangle  \\
 \vdots\\
  \sum_{m=-j}^{j}    \frac{\tilde{\varphi}^j_{[K] m} (0)}{\sqrt{  \tilde{\mathcal{N}}^j_m  [\tilde{\lambda}^j_{[K]}, \beta_+]}}  \mid  0 \rangle \otimes \mid  j,  m \rangle
 \end{pmatrix}. \label{vector-algebra-HW-BG-one}
\end{equation}
Choosing now the set of arbitrary  parameters in the form 

\begin{equation}
\frac{\varphi^j_{[s] m} (0)}{ \sqrt{\tilde{\mathcal{N}}^j_m  [\tilde{\lambda}^j_{[s]}, \beta_+]}}    =  \sum_{r=1}^{K} \tilde{U}^\dagger_{[s][r]}  \delta_{m_{[r]} m}, 
\end{equation}
we arrive to the special set of ${(2j+1)}^K$  linearly independent vector states labeled by $m1, \cdots, m_K,$ that is.  

\begin{equation}
\mid  \Psi \rangle^j_{[K]} =  \mathcal{N}^{- \frac{1}{2}} \exp\left[  \tilde{M} \hat{a}^\dagger  - \tilde{M}^{\dagger}  \hat{a}     \right]  
  \exp\left[  - ( \hat{a}^\dagger  I + \tilde{M}^\dagger  ) \beta_+ \hat{J}_-     \right]   \begin{pmatrix}
 \mid  0 \rangle \otimes \mid  j,  m_1 \rangle\\
 \mid  0 \rangle \otimes\mid  j,  m_2 \rangle  \\
 \vdots\\
 \mid  0 \rangle \otimes \mid  j,  m_K \rangle
 \end{pmatrix}. \label{vector-algebra-HW-BG-two}
\end{equation}
Thus, for example, if  $m_1=m_2 = \cdots=m_K=-j,$  we recuperate the states  (\ref{standard-VCS+j-j}). On the other hand, if we take
$m_1=m_2 = \cdots=m_K=j ,$   and   use the standard disentangled formula  of the  exponential  operator $$\exp{\left[- \frac{\tilde{\theta}}{2} \left( e^{-i \tilde{\phi}} \hat{J}_+  -  e^{i \tilde{\phi}} \hat{J}_-  \right)  \right]}= \exp{ \left[ \tan\left( {\frac{\tilde{\theta}}{2} } \right)  e^{i \tilde{\phi}} \hat{J}_-  \right]}  \exp{ \left[ -  \ln  \left( {\sec^2 ( \frac{\tilde{\theta}}{2} ) } \right)   \hat{J}_3  \right]}      \exp{ \left[ - \tan\left( {\frac{\tilde{\theta}}{2} } \right)  e^{- i \tilde{\phi}} \hat{J}_+  \right]} ,  $$   we get formally the symmetric form

\begin{eqnarray}
\mid  \Psi \rangle^j_{[K]} &=&   \mathcal{N}^{- \frac{1}{2}} \exp\left[   \tilde{M} \hat{a}^\dagger  - \tilde{M}^{\dagger}  \hat{a}     \right]  
  \exp\left[  -  I  \hat{a}^\dagger \beta_+ \hat{J}_-    \right]     \nonumber \\ &\times&  \exp\left[  \frac{\arctan \left(  \|\beta_+ \|   \tilde{M}  \tilde{M}^\dagger \right)}{  \sqrt{\|\beta_+ \| \tilde{M}  \tilde{M}^\dagger}}     \left(  \tilde{M}  \beta^\ast_+  \hat{J}_+          -   \tilde{M}^\dagger  \beta_+ \hat{J}_-     \right)\right]   \begin{pmatrix}
 \mid  0 \rangle \otimes \mid  j, j \rangle\\
 \mid  0 \rangle \otimes \mid  j,  j \rangle  \\
 \vdots\\
 \mid  0 \rangle \otimes \mid  j,  j \rangle
 \end{pmatrix}, \label{vector-algebra-HW-BG-three}
\end{eqnarray}
which extrapolate between the Heisenberg-Weyl  vector coherent states and the vector eigenstates  of the generalized annihilation operator $\hat{a} + \beta_{+} \hat{J}_-. $

\subsection{Non-canonical commutation relation}
Let us now choose the beta parameters in such a way that the commutation relation (\ref{commutator-A-A-dagger})  verify

\begin{equation}
\left[ \mathbb{A}  ,   \mathbb{A}^\dagger \right] = \hat{I}  +  \rho e^{i \nu } \hat{J}_+  + \rho e^{ - i \nu} \hat{J}_- 
\label{commutator-A-A-dagger-rho-nu-J-J}
\end{equation}
where $\rho \in \mathbb{R} $   such that  $\rho  > 0$ and $0 \leq \nu \leq 2 \pi .$  Here two scenarios are possible depending of the value of  $b$, both  requiring 
$\|\beta_+ \|= \| \beta_- \| = R  > 0,$ and $\beta_3 \neq 0,$ with $\theta_3 \neq \frac{\theta_+ + \theta_- }{2} + k \pi, k=0,1, \cdots , $ they are

\begin{equation}
    \begin{aligned}\setcounter{mysubequations}{0}
      \mysubnumber   \quad   & \beta_{+} = R  e^{i \theta_+} , \quad \beta_{-} =   R e^{i \theta_-}; \quad \beta_3 = R_3 e^{i \theta_3}  \\
\quad & R_3  \ne  2R  \quad \mbox{or} \quad     \theta_3 \neq \frac{\theta_+ + \theta_- }{2} + ( k + \frac{1}{2}) \pi, \;  k=0,1, \cdots ,   \Rightarrow \; b \neq 0, \\
 \quad & \rho e^{i \nu } = 2i R R_3   \sin{\left(     \theta_3 -  \frac{\theta_+ + \theta_- }{2}           \right)} \;  e^{-  \frac{i}{2} \left( \theta_+ - \theta_-  \right)},   \\
 \mysubnumber  \quad & \beta_{+} = R  e^{i \theta_+} , \quad \beta_{-} =   R e^{i \theta_-}; \quad \beta_3 =   R_3 e^{i \theta_3}    \\
      \quad & R_3  =  2R  \quad \mbox{ and} \quad     \theta_3 = \frac{\theta_+ + \theta_- }{2} + ( k + \frac{1}{2}) \pi, \;  k=0,1, \cdots ,   \Rightarrow \; b = 0,\\
      \quad & \rho e^{i \nu } = 4 R^2  e^{ - \frac{i}{2} (\theta_+ - \theta_-)   }  \; e^{i ( k + \frac{1}{2}) \pi},  \; k=1,2, \cdots. 
      \end{aligned} .
   \end{equation}
As before, we will leave the case listed in (i), where $b\neq 0,$ for the last and we will analyze  here the case listed in (ii), where $b=0.$ 
 Then, if we insert  the algebra eigenstates   (\ref{general-solution-AES-b=0-remaining-non-zero}) into (\ref{vector-algebra-states-}), which gives us the vector components of the state that represents the solutions of the matrix eigenvalue equation (\ref{vector-matrix-eigenvalue-total}),  and supposing that $\tilde{M}$ is normal, we get   
 \begin{equation}
 \mid  \psi \rangle^j_{[u]} =  \mathcal{N}_{[u]}^{-\frac{1}{2}}   \sum_{s=1}^{K} \tilde{U}_{u s}  
\sum_{m=-j}^{j}  \tilde{\varphi}^j_{m} (0) \hat{D} (\tilde{\lambda}_{[s]}) \mid \tilde{\psi} [   \tilde{\lambda}_{[s]}     ,\beta_+, \beta_-, \beta_3] \rangle^j_m
     \quad u=1,2, \cdots, K,  \label{vector-algebra-states-components-non-canonical-case}
 \end{equation}
with the states $\mid \tilde{\psi} [   \tilde{\lambda}_{[s]}     ,\beta_+, \beta_-, \beta_3] \rangle^j_m $  given by equation  (\ref{AES-b=0-all-different-zero-operator-form}):  

 \begin{eqnarray}
\mid \tilde{\psi} [     \tilde{\lambda}_{[s]}           ,\beta_+, \beta_-, \beta_3] \rangle^j_m &=& \frac{1}{\sqrt{ \tilde{\mathcal{N}}^{j}_{m}  [\tilde{\lambda}_{[s]} , \beta_+, \beta_-, \beta_3 ]  }}  \sum_{n=0}^{j+m}  \quad \sum_{\ell=0}^{j+n-m}  \; \frac{(\vartheta \hat{J}_+ )^\ell}{\ell !}   \nonumber \\
  &\times&   (-1)^n \;     \frac{\left( (\hat{a}^\dagger +   \tilde{\lambda}^\ast_{[s]}               )  \beta_+ \hat{J_-} \right)^n}{n!} 
   \mid 0  \rangle \otimes 
\mid  j, m  \rangle, \quad m=-j, \cdots,j,  \label{AES-b=0-all-different-zero-operator-form-lambda-s} 
\end{eqnarray}
where in this special case $\vartheta = \frac{\beta_3}{2 \beta_+} = -  \frac{2 \beta_- }{\beta_3} = e^{i (\theta_3 - \theta_+ )} = {(-1)}^k  i e^{- \frac{i}{2} ( \theta_+ - \theta_-)}, k=1,2.$ 
In a first step we can write  equation (\ref{vector-algebra-states-components-non-canonical-case}) in matrix form, in the same way  we did in the last section, 

\begin{equation}
\mid  \Psi \rangle^j_{[K]} =  \mathcal{N}^{- \frac{1}{2}} \exp\left[  \tilde{M} \hat{a}^\dagger  - \tilde{M}^{\dagger}  \hat{a}     \right]  
    \tilde{U} \begin{pmatrix}
 \sum_{m=-j}^{j}   \tilde{\varphi}^j_{[1] m} (0)    \mid \tilde{\psi} [     \tilde{\lambda}_{[1]}           ,\beta_+, \beta_-, \beta_3] \rangle^j_m                  \\
  \sum_{m=-j}^{j}   \tilde{\varphi}^j_{[2] m} (0)    \mid \tilde{\psi} [     \tilde{\lambda}_{[2]}           ,\beta_+, \beta_-, \beta_3] \rangle^j_m                   \\
 \vdots\\
  \sum_{m=-j}^{j}   \tilde{\varphi}^j_{[K] m} (0)    \mid \tilde{\psi} [     \tilde{\lambda}_{[K]}           ,\beta_+, \beta_-, \beta_3] \rangle^j_m                 
 \end{pmatrix}. \label{vector-algebra-non-canonical-first-step}
\end{equation}
We could  do better but the states (\ref{AES-b=0-all-different-zero-operator-form-lambda-s}) are not easy to disentangle and write as a product of  exponential factors acting on the basis states, except for the case $m=-j,$ which we will study in the next subsection. To achieve this last goal, we will try another method in the subsequent subsection.   

\subsubsection{Vector coherent states from non-canonical commutation relations}     
When $m=-j,$ from  (\ref{AES-b=0-all-different-zero-operator-form-lambda-s}) we obtain the normalized state

 \begin{eqnarray}
\mid \tilde{\psi} [ \tilde{\lambda}_{[s]} ,\beta_+, \beta_-, \beta_3] \rangle^j_{-j} &=& \frac{ \sum_{\ell=0}^{2j}  \; \frac{(\vartheta \hat{J}_+ )^\ell}{\ell !} }{\sqrt{ \tilde{\mathcal{N}}^{j}_{-j}  [\tilde{\lambda}_{[s]} , \beta_+, \beta_-, \beta_3 ]  }}    
   \mid 0  \rangle \otimes
\mid  j, -j  \rangle  \nonumber \\  &=& \frac{\exp[ \vartheta \hat{J}_+ ]  }{\sqrt{ \tilde{\mathcal{N}}^{j}_{-j}  [\tilde{\lambda}_{[s]} , \beta_+, \beta_-, \beta_3 ]  }}    \mid 0  \rangle \otimes 
\mid  j, -j  \rangle  \nonumber \\ &=&  \exp\left[ {(-1)}^k  \frac{i \pi}{4} \left(e^{\frac{- i}{2} (\theta_+ - \theta_-)} \hat{J}_+  +  e^{\frac{i}{2} (\theta_+ - \theta_-)} \hat{J}_- \right) \right]   \mid 0  \rangle \otimes 
\mid  j, -j  \rangle, \quad k=0,1,
\end{eqnarray}
which is independent on $\tilde{\lambda}_{[s]}.$

Thus choosing  (\ref{vector-algebra-non-canonical-first-step}) the integration constants as follows: 
\begin{equation}
\varphi^j_{[s] m} (0)  =  \sum_{r=1}^{K} \tilde{U}^\dagger_{[s][r]}  \delta_{m_{[r]} m}, \quad m_{[r]} =-j, \forall r=1,\cdots, K,
\end{equation}
we finally reach the vector coherent states
\begin{eqnarray}
\mid  \Psi \rangle^j_{[K]} &=&  \frac{1}{\sqrt{K}}\exp\left[  \tilde{M} \hat{a}^\dagger  - \tilde{M}^{\dagger}  \hat{a}  \right]  \nonumber \\
   &\times & \exp\left[  \frac{i \pi}{4}  {(-1)}^k  \left(e^{\frac{- i}{2} (\theta_+ - \theta_-)} \hat{J}_+  +  e^{\frac{i}{2} (\theta_+ - \theta_-)}  \hat{J}_- \right)  \right]  \nonumber \\  &\times&  \begin{pmatrix}
 \mid 0  \rangle \otimes 
\mid  j, -j  \rangle   \\        \mid 0  \rangle \otimes 
\mid  j, -j  \rangle        \\
 \vdots\\
   \mid 0  \rangle \otimes 
\mid  j, -j  \rangle
   \end{pmatrix}, \quad k=0,1. \label{vector-coherent-state-from-non-canonical-relation}
\end{eqnarray}
These last states, certainty,  are eigenstates of the annihilation operator 
\begin{equation}
\mathbb{A} = a + R e^{i \theta_+} \hat{J}_-  +  R  e^{i \theta_-} \hat{J}_+  +  2 i R {(-1)}^k e^{ \frac{i}{2} (\theta_+ + \theta_-)} , \quad k=0,1,
\label{A-operator-non-canonical}
\end{equation}
with matrix eigenvalue equal to $\tilde{M}.$

\subsubsection{Disentangling through a unitary transformation}
\label{disentangling}
The commutator (\ref{commutator-A-A-dagger-rho-nu-J-J})  can be written in the form 

\begin{equation}
\left[ \mathbb{A}  ,   \mathbb{A}^\dagger \right] = \hat{I}  -  2  \rho  \hat{\mathbb{J}}_3,
\label{commutator-A-A-dagger-rho-cal-J}
\end{equation}
where 
\begin{equation}
  \hat{\mathbb{J}}_3 =  \frac{-1}{2} \left[e^{i \nu } \hat{J}_+  +   e^{ - i \nu} \hat{J}_- \right]= {(-1)}^{k+1}  \frac{i}{2}  \left[e^{ - \frac{i}{2} (\theta_+ -  \theta_-) }  \hat{J}_+  -  e^{  \frac{i}{2} (\theta_+ -  \theta_-) }  \hat{J}_- \right], \quad k=0,1. \label{j3-transformation}
\end{equation}
On the other hand, the operator $\mathbb{A}$ in (\ref{A-operator-non-canonical}) also can be rewritten in a different way, that is:

\begin{equation}
\mathbb{A} = a + \mathcal{B}_+  \hat{\mathbb{J}}_- ,  \label{annihilator-transformed}
\end{equation}
with
\begin{equation}
 \mathcal{B}_+ = 2R  e^{\frac{i}{2} (\theta_+ + \theta_-) } \quad \mbox{and}   \quad \hat{\mathbb{J}}_- = 
 \frac{1}{2} \left[e^{  \frac{i}{2} (\theta_+ -  \theta_-) }  \hat{J}_-  +  e^{  - \frac{i}{2} (\theta_+ -  \theta_-) }  \hat{J}_+ \right] +  {(-1)}^k i \hat{J}_3, \quad k=0,1.
 \end{equation}
  
  Thus, by defining $\hat{\mathbb{J}}_+ = \hat{\mathbb{J}}_-^{\dagger} $ we observe that the set  generators $\hat{\mathbb{J}}_+ , \hat{\mathbb{J}}_- $ and $\hat{\mathbb{J}}_3$ verify the $su(2)$ Lie algebra commutation relations (\ref{su2-Lie-algebra-commutators} ), for all $k=0,1.$  Hence, the problem of finding the algebra eigenstates of  (\ref{A-operator-non-canonical})  reduces to the same already solved in the last section, the only difference is a change of the $su(2)$ representation basis vectors at the moment of spanning the states. We will denote this new basis vectors as $\mid \tilde{j} , \tilde{m} \rangle, \; \tilde{m}=-\tilde{j}, \cdots \tilde{j}.$ Here, it is important to  mention that the Casimir operator is invariant under the transformation, then the label $\tilde{j}$ denoting the representation coincides with the label $j.$  Using then equation   (\ref{vector-algebra-HW-BG-two}), but taking into account the new operators and basis states, the solution of  the eigenvalue equation (\ref{vector-matrix-eigenvalue-total}), in this particular  case, is given by

\begin{equation}
\mid  \Psi \rangle^j_{[K]} =  \mathcal{N}^{- \frac{1}{2}} \exp\left[  \tilde{M} \hat{a}^\dagger  - \tilde{M}^{\dagger}  \hat{a}     \right]  
  \exp\left[  - ( \hat{a}^\dagger  I + \tilde{M}^\dagger  ) \mathcal{B}_+ \hat{\mathbb{J}}_-     \right]   \begin{pmatrix}
 \mid  0 \rangle \otimes \mid  \tilde{j},  \tilde{m}_1 \rangle\\
 \mid  0 \rangle \otimes\mid  \tilde{j},  \tilde{m}_2 \rangle  \\
 \vdots\\
 \mid  0 \rangle \otimes \mid  \tilde{j},  \tilde{m}_K \rangle
 \end{pmatrix}, \label{vector-algebra-HW-BG-two-transformed}
\end{equation}
  where $ \mid \tilde{j} ,  \tilde{m}_{r}  \rangle, \; r=1,\cdots,k$ are eigenstates of $ \hat{\mathbb{J}}_3$ associated to the eigenvalue $\tilde{m}_{r},$ respectively.
 
 To know the connection of these last states with the regular states $|j ,m\rangle, $ we need to solve  the  following $su(2)$ algebra eigenstate equation for  $ \hat{\mathbb{J}}_3$ given in (\ref{j3-transformation}):
  \begin{equation}
 [ \alpha_-  \hat{J}_+  +  \alpha_+  \hat{J}_- ]  \mid \tilde{j} ,  \tilde{m}   \rangle =  \tilde{m}  \;   \mid \tilde{j} ,  \tilde{m}   \rangle, \end{equation} 
 where $\alpha_- =    - \frac{1}{2}    e^{ - \frac{i}{2} (\theta_+ -  \theta_-) } e^{i \left( k +\frac{1}{2}    \right) \pi } , k=0,1, $ $\alpha_+ =  -  \;  \frac{1}{2}    e^{ \frac{i}{2} (\theta_+ -  \theta_-) }       e^{- i \left( k +\frac{1}{2}    \right) \pi }, \; k=0,1, $ and $\alpha_3 =0.$  Then using the results of appendix  \ref{appb}, in particular, equations 
  (\ref{master-operator-T}) and (\ref{phases-conditions}), we find the parameter $b= \sqrt{4 \alpha_+ \alpha_- + \alpha_3^2 } = 1, $ which implies $ \tilde{m}= m b =m,$ 
 \begin{equation} 
\frac{\tilde{\theta}}{2} =  \arctan \left(\sqrt{\frac{b- \alpha_3}{ b + \alpha_3}}\right) = \arctan( 1) = \frac{\pi}{4},  \quad
\mbox{and} \quad e^{i \tilde{\phi}} = \sqrt{\frac{\alpha_+}{ \alpha_-}} =  e^{\frac{i}{2} (\theta_+ - \theta_-)}   e^{- i \left( k +\frac{1}{2}    \right) \pi }   , \quad k=0,1,
\end{equation}   
 from where we extract the unitary operator 
 
 \begin{equation}
 T = \exp{ \left[ \frac{i \pi}{4} {(-1)}^k \left(  e^{\frac{- i}{2} (\theta_+ - \theta_-)}  \hat{J}_+ +     e^{\frac{i}{2} (\theta_+ - \theta_-)}   \hat{J}_-                 \right) \right] }, \quad k=0,1.  \label{T-operator-special-non-canonical-case}
\end{equation} 
Thus, following the reasoning of appendix \ref{appb},  we finally conclude that the transformed states $|\tilde{j}, \tilde{m} \rangle = T \mid j , m \rangle, \; m=-j, \cdots, j, $ are eigenstates of   $ \hat{\mathbb{J}}_3,$ corresponding to the eigenvalue $\tilde{m} =m.$ 

At this point, before concluding the section, let us return briefly to the case $ m=-j.$  As  $ \hat{\mathbb{J}}_-  \mid \tilde{j} , -j \rangle =0,$ then  when  in (\ref{vector-algebra-HW-BG-two-transformed}) we choose $\tilde{m}_r = - \tilde{j} = -j, \forall r =1,\cdots,K$ and we use the  results we have just obtained, we regain the vector coherent states (\ref{vector-coherent-state-from-non-canonical-relation}).

 Continuing now with the line of argument,  using the recently obtained results,   we can return to the original $su(2)$ representation  basis states and  write  (\ref{vector-algebra-HW-BG-two-transformed}) in the disentangled form

\begin{eqnarray}
\mid  \Psi \rangle^j_{[K]} =  \mathcal{N}^{- \frac{1}{2}} \exp\left[  \tilde{M} \hat{a}^\dagger  - \tilde{M}^{\dagger}  \hat{a}     \right]  
  \exp\left[  - ( \hat{a}^\dagger  I + \tilde{M}^\dagger  ) \mathcal{B}_+ \hat{\mathbb{J}}_-     \right]  \nonumber \\  \exp{ \left[ \frac{i \pi}{4} \left(  e^{\frac{- i}{2} (\theta_+ - \theta_-)}  \hat{J}_+ +     e^{\frac{i}{2} (\theta_+ - \theta_-)}   \hat{J}_-                 \right) I \right]}  \begin{pmatrix}
 \mid  0 \rangle \otimes \mid  \tilde{j},  m_1 \rangle\\
 \mid  0 \rangle \otimes\mid  \tilde{j},  m_2 \rangle  \\
 \vdots\\
 \mid  0 \rangle \otimes \mid  \tilde{j},  m_K \rangle
 \end{pmatrix}. \label{vector-algebra-HW-BG-two-transformed-final}
\end{eqnarray}

\subsubsection{Disentangled form of the vector states}
On the other hand, when we choose $\tilde{m}_r =  \tilde{j} = j, \forall r =1,\cdots,K$ the vector states   in (\ref{vector-algebra-HW-BG-two-transformed}) can be written in the form

 \begin{eqnarray}
\mid  \Psi \rangle^j_{[K]} &=&   \mathcal{N}^{- \frac{1}{2}} \exp\left[   \tilde{M} \hat{a}^\dagger  - \tilde{M}^{\dagger}  \hat{a}     \right]  
 \nonumber \\ &\times& \exp\left[  -  I  \hat{a}^\dagger     \mathcal{B}_+  \hat{\mathbb{J}}_-  \right]       \exp\left[  \frac{\arctan \left( \|  \mathcal{B}_+           \|    \sqrt{ \tilde{M}  \tilde{M}^\dagger } \right)}{ \|  \mathcal{B}_+  \|  \sqrt{ \tilde{M}  \tilde{M}^\dagger}}     \left(  \tilde{M}   \mathcal{B}^\ast_+            \hat{\mathbb{J}}_+    -   \tilde{M}^\dagger   \mathcal{B}_+    \hat{\mathbb{J}}_-   \right)\right]  \nonumber \\ 
   &\times&    \exp{ \left[ \frac{i \pi}{4} {(-1)}^k \left(  e^{\frac{- i}{2} (\theta_+ - \theta_-)}  \hat{J}_+ +     e^{\frac{i}{2} (\theta_+ - \theta_-)}   \hat{J}_-                 \right) \right] }    \begin{pmatrix}
 \mid  0 \rangle \otimes \mid  j, j \rangle\\
 \mid  0 \rangle \otimes \mid  j,  j \rangle  \\
 \vdots\\
 \mid  0 \rangle \otimes \mid  j,  j \rangle
 \end{pmatrix},  \quad k=0,1, \label{vector-algebra-HW-BG-generalized}
\end{eqnarray}
which extrapolates between the vector coherent states over the matrix domain associated to the  quantum harmonic oscillator   and the intelligent vector states associated to  the generalized annihilation operator $ \hat{a} + {\mathcal{B}}_+     \hat{\mathbb{J}}_- ,$ whose matrix eigenvalues are given by $\tilde{M}.$

 \subsection{More general sets of vector algebra eigenstates}
When we are interested in physics systems which Hamiltonian is formed from the suitable product of  the generalized annihilation operator (\ref{generalized-h1-su2-annihation-operator}) and its adjoint, both verifying the commutation relation (\ref{commutator-A-A-dagger}), and we want that all  coefficients of the left side of this last expression be different from $0,$ again we need to distinguish between several possibilities, all of them requiring  $\beta_3 \neq 0.$ Indeed, in this case we need $ \|\beta_-\|^2 - \| \beta_+ \|^2 \neq 0 ,$ and $\beta_3 \beta^\ast_+ - \beta^\ast_3 \beta_- \neq 0.$ Thus, we have

\begin{equation}
    \begin{aligned}\setcounter{mysubequations}{0}
      \mysubnumber   \quad \beta_+  \neq 0,  & \quad  \beta_{-}  =  0, \quad \beta_{3}  \neq 0,    \quad  \Rightarrow b \neq 0\\
      \mysubnumber   \quad \beta_+  = 0,  & \quad  \beta_{-}  \neq  0, \quad \beta_{3}  \neq 0,    \quad  \Rightarrow b \neq 0  \\
      \mysubnumber     \quad  \beta_+ =  R_+ e^{i \theta_+} \neq 0,  & \quad  \beta_{-} =  R_-  e^{i \theta_-}  \neq  0,     \quad                                                                                        \beta_3= R_3  e^{i \theta_3} \neq 0, \quad   R_+ \neq R_- \\
\quad \mbox{and} \quad R_3 \neq 2 \sqrt{ R_+ R_-}   & \quad \mbox{or} \quad \theta_3 \neq \frac{ \theta_+ + \theta_- }{2} + \left( k + \frac{1}{2}\right) \pi, \quad k=0,1,   \quad  \Rightarrow b \neq 0  \\ 
      \mysubnumber    \quad   \beta_+ = R_+ e^{i \theta_+} \neq 0,  & \quad  \beta_{-} =  R_-  e^{i \theta_-}  \neq  0,     \quad                                                                                        \beta_3= R_3  e^{i \theta_3}   \neq 0, \quad   R_+ \neq R_- \\
\quad R_3 = 2 \sqrt{ R_+ R_- }  & \quad \mbox{and} \quad \theta_3 = \frac{ \theta_+ + \theta_- }{2} + \left( k + \frac{1}{2}\right) \pi, \quad k=0,1,   \quad  \Rightarrow b = 0\\
\quad \rho e^{i \nu} = & R_3 (R_+ +R_-) e^{- \frac{i}{2}( \theta_+  - \theta_- )} \; e^{i \left( k + \frac{1}{2}  \right) \pi}
, \quad k=0,1.
    \end{aligned} .
   \end{equation}
The cases listed in (i,ii,iii),  where parameter $b \neq 0,$  will be treated in the next subsection together with their similar ones  mentioned above. The case listed in $(iv),$ where $b=0,$ will be solved here using the technique of the $su(2)$  operator algebra transformations used in the last subsection.

As we have done in the in section \ref{disentangling}, the  commutator (\ref{commutator-A-A-dagger-rho-nu-J-J})  can be written in the form 

\begin{equation}
\left[ \mathbb{A}  ,   \mathbb{A}^\dagger \right] = \hat{I}  -  2  \rho \frac{R_+ + R_-}{R_3}   \hat{\mathbb{J}}_3,
\label{commutator-A-A-dagger-rho-cal-J-second}
\end{equation}
where now
\begin{eqnarray}
  \hat{\mathbb{J}}_3 &=& R_3  \frac{(R_+ - R_-)}{\rho}  \hat{J}_3 -   \frac{R_3}{2(R_+ + R_-)}  \left[e^{i \nu } \hat{J}_+  +   e^{ - i \nu} \hat{J}_- \right] \nonumber \\&=&  \frac{(R_+ - R_-)}{(R_+ + R_-)}  \hat{J}_3  -   \frac{\sqrt{R_+ R_-}}{ (R_+ + R_-)}    \left[e^{ - \frac{i}{2} (\theta_+ -  \theta_-) }  e^{i \left( k + \frac{1}{2}  \right) \pi} \hat{J}_+  +  e^{  \frac{i}{2} (\theta_+ -  \theta_-) }  e^{- i \left( k + \frac{1}{2}  \right) \pi} \hat{J}_- \right], \quad k=0,1. \label{j3-second-transformation}
\end{eqnarray}
On the other hand, the operator $\mathbb{A}$ in (\ref{A-operator-non-canonical}) again can be written in the form:

\begin{equation}
\mathbb{A} = a + \mathcal{B}_+  \hat{\mathbb{J}}_- ,  \label{annihilator-second-transformation}
\end{equation}
where now
\begin{equation}
 \mathcal{B}_+ =  (R_+ + R_-)  e^{\frac{i}{2} (\theta_+ + \theta_-) } \end{equation}
 and
 \begin{equation} \hat{\mathbb{J}}_- = 
 \frac{1}{(R_+ + R_- )} \left[R_+ \;  e^{ \frac{i}{2} (\theta_+ -  \theta_-) }  \hat{J}_-  +  R_- \;  e^{  - \frac{i}{2} (\theta_+ -  \theta_-) }  \hat{J}_+  +  2 \sqrt{R_+ R_-}     \;  e^{i \left( k + \frac{1}{2}  \right) \pi}     \hat{J}_3\right] , \quad k=0,1.
 \end{equation}
 Thus, as before, by defining $\hat{\mathbb{J}}_+ = \hat{\mathbb{J}}^\dagger_- ,$ we realize that the transformed operators verify the $su(2)$ algebra. Then the problem of computing the algebra eigenstates reduces to that of the  previous sections.

%%%%%%%%%%%%%%%%%%%%%%%%%%%%%%%%%%%%%%%%%%%%%%%%%%%%%%%%%
 Using again equation   (\ref{vector-algebra-HW-BG-two}), but taking into account the new operators and basis states, the solution of  the eigenvalue equation (\ref{vector-matrix-eigenvalue-total}), in this particular  case, is given by

\begin{equation}
\mid  \Psi \rangle^j_{[K]} =  \mathcal{N}^{- \frac{1}{2}} \exp\left[  \tilde{M} \hat{a}^\dagger  - \tilde{M}^{\dagger}  \hat{a}     \right]  
  \exp\left[  - ( \hat{a}^\dagger  I + \tilde{M}^\dagger  ) \mathcal{B}_+ \hat{\mathbb{J}}_-     \right]   \begin{pmatrix}
 \mid  0 \rangle \otimes \mid  \tilde{j},  \tilde{m}_1 \rangle\\
 \mid  0 \rangle \otimes\mid  \tilde{j},  \tilde{m}_2 \rangle  \\
 \vdots\\
 \mid  0 \rangle \otimes \mid  \tilde{j},  \tilde{m}_K \rangle
 \end{pmatrix}, \label{vector-algebra-HW-BG-two-second-transformation}
\end{equation}
  where $ \mid \tilde{j} ,  \tilde{m}_{r}  \rangle, \; r=1,\cdots,k$ are eigenstates of $ \hat{\mathbb{J}}_3$ associated to the eigenvalue $\tilde{m}_{r},$ respectively.
 
 To know the connection of these last states with the original states $|j ,m\rangle, $ we have to solve  the  following $su(2)$ algebra eigenstate equation for  $ \hat{\mathbb{J}}_3$ given in (\ref{j3-second-transformation}):
  \begin{equation}
 [ \alpha_-  \hat{J}_+  +  \alpha_+  \hat{J}_-  + \alpha_3 \hat{J}_3  ]  \mid \tilde{j} ,  \tilde{m}   \rangle =  \tilde{m}  \;   \mid \tilde{j} ,  \tilde{m}   \rangle,\end{equation} 
 where \begin{equation}
 \alpha_- =    -   \;  \frac{\sqrt{R_+ R_-}}{ (R_+ + R_-)}  e^{ - \frac{i}{2} (\theta_+ -  \theta_-) } e^{i \left( k +\frac{1}{2}    \right) \pi }, \quad \alpha_+ =  -  \;    \frac{\sqrt{R_+ R_-}}{ (R_+ + R_-)}     e^{ \frac{i}{2} (\theta_+ -  \theta_-) }       e^{- i \left( k +\frac{1}{2}    \right) \pi },  \quad  k=0,1,   \end{equation}
 and  \begin{equation} \alpha_3 = \frac{R_+ - R_-}{ R_+ +  R_-}, \quad  k=0,1. \end{equation}  Then using the results of appendix  \ref{appb}, in particular, equations 
  (\ref{master-operator-T}) and (\ref{phases-conditions}), we find 
  \begin{equation}b= \sqrt{4 \alpha_+ \alpha_- + \alpha_3^2 } =1, \quad \mbox{ which implies} \quad  \tilde{m}= m, \end{equation}
 \begin{equation} 
\frac{\tilde{\theta}}{2} =  \arctan \left(\sqrt{\frac{b- \alpha_3}{ b + \alpha_3}}\right) = \arctan\left( \sqrt{\frac{R_-}{R_+}}\right) \quad
\mbox{and} \quad e^{i \tilde{\phi}} = \sqrt{\frac{\alpha_+}{ \alpha_-}} =  e^{\frac{i}{2} (\theta_+ - \theta_-)}   e^{- i \left( k +\frac{1}{2}    \right) \pi }   , \quad k=0,1,
\end{equation}   
 from where we extract the unitary operator 
 
 \begin{equation}
 T = \exp{ \left[ i    \arctan\left( \sqrt{\frac{R_-}{R_+}}\right)    {(-1)}^k \left(  e^{\frac{- i}{2} (\theta_+ - \theta_-)}  \hat{J}_+ +     e^{\frac{i}{2} (\theta_+ - \theta_-)}   \hat{J}_-                 \right) \right] }, \quad k=0,1.  \label{T-operator-special-second-non-canonical-case}
\end{equation} 
Thus, following the reasoning of appendix \ref{appb},  we finally conclude that the transformed  states $|\tilde{j}, \tilde{m} \rangle = T \mid j , m \rangle, \; m=-j, \cdots, j, $ are eigenstates of   $ \hat{\mathbb{J}}_3,$ corresponding to the eigenvalue $\tilde{m} =m.$ 

Finally, inserting these last states into (\ref{vector-algebra-HW-BG-two-second-transformation}) and using   (\ref{T-operator-special-second-non-canonical-case}) for $T,$ we arrive to the generalized vector states

\begin{eqnarray}
\mid  \Psi \rangle^j_{[K]} &=&  \mathcal{N}^{- \frac{1}{2}} \exp\left[  \tilde{M} \hat{a}^\dagger  - \tilde{M}^{\dagger}  \hat{a}     \right]  
  \exp\left[  - ( \hat{a}^\dagger  I + \tilde{M}^\dagger  ) \mathcal{B}_+ \hat{\mathbb{J}}_-     \right]   \nonumber \\  &\times& \exp{ \left[ i    \arctan\left( \sqrt{\frac{R_-}{R_+}}\right)    {(-1)}^k \left(  e^{\frac{- i}{2} (\theta_+ - \theta_-)}  \hat{J}_+ +     e^{\frac{i}{2} (\theta_+ - \theta_-)}   \hat{J}_-                 \right) \; I\right] } \begin{pmatrix}
 \mid  0 \rangle \otimes \mid  j,  m_1 \rangle\\
 \mid  0 \rangle \otimes\mid  j,  m_2 \rangle  \\
 \vdots\\
 \mid  0 \rangle \otimes \mid  j,  m_K \rangle
 \end{pmatrix},  \; \; k=0,1,\label{vector-algebra-HW-BG-two-second-transformation-final}
\end{eqnarray}
which constitute a set of $(2j+1)^K$ of linearly independent vector states.

\subsubsection{Generalized disentangled form of new  vector coherent states}
If we choose $\tilde{m}_r =  - \tilde{j} = -j, \forall r =1,\cdots,K$ the vector states   in (\ref{vector-algebra-HW-BG-two-second-transformation}) can be written in the form

  \begin{eqnarray}
\mid  \Psi \rangle^j_{[K]} &=&  \frac{1}{\sqrt{K}}\exp\left[  \tilde{M} \hat{a}^\dagger  - \tilde{M}^{\dagger}  \hat{a}  \right]  
    \nonumber \\ &\times &  \exp{ \left[ i   \arctan\left( \sqrt{\frac{R_-}{R_+}}\right)        {(-1)}^k \left(  e^{\frac{- i}{2} (\theta_+ - \theta_-)}  \hat{J}_+ +     e^{\frac{i}{2} (\theta_+ - \theta_-)}   \hat{J}_-                 \right)  \;  I  \right] }    \begin{pmatrix}
 \mid 0  \rangle \otimes 
\mid  j, -j  \rangle   \\        \mid 0  \rangle \otimes 
\mid  j, -j  \rangle        \\
 \vdots\\
   \mid 0  \rangle \otimes 
\mid  j, -j  \rangle
   \end{pmatrix}, \quad k=0,1, \label{vector-coherent-state-from-non-canonical-relation-second}
\end{eqnarray}
that generalize   (\ref{vector-coherent-state-from-non-canonical-relation}), and represent the  product of the canonical harmonic oscillator coherent states and the Perelomov type $su(2)$ coherent states, in the matrix domain.  

\subsubsection{Generalized disentangled form of the vector states II}
On the other hand, if we choose $\tilde{m}_r =  \tilde{j} = j, \forall r =1,\cdots,K$ the vector states   in (\ref{vector-algebra-HW-BG-two-second-transformation}) can be written in the form

 \begin{eqnarray}
\mid  \Psi \rangle^j_{[K]} &=&   \mathcal{N}^{- \frac{1}{2}} \exp\left[   \tilde{M} \hat{a}^\dagger  - \tilde{M}^{\dagger}  \hat{a}     \right]  \nonumber \\ &\times&
  \exp\left[  -  I  \hat{a}^\dagger     \mathcal{B}_+  \hat{\mathbb{J}}_-  \right]       \exp\left[  \frac{\arctan \left(  \|  \mathcal{B}_+      \|  \sqrt{ \tilde{M}  \tilde{M}^\dagger} \right)}{ \|  \mathcal{B}_+  \| \sqrt{ \tilde{M}  \tilde{M}^\dagger}}     \left(  \tilde{M}   \mathcal{B}^\ast_+            \hat{\mathbb{J}}_+    -   \tilde{M}^\dagger   \mathcal{B}_+    \hat{\mathbb{J}}_-   \right)\right]  \nonumber \\ 
   &\times&    \exp{ \left[ i   \arctan\left( \sqrt{\frac{R_-}{R_+}}\right)        {(-1)}^k \left(  e^{\frac{- i}{2} (\theta_+ - \theta_-)}  \hat{J}_+ +     e^{\frac{i}{2} (\theta_+ - \theta_-)}   \hat{J}_-                 \right) \right] }    \begin{pmatrix}
 \mid  0 \rangle \otimes \mid  j, j \rangle\\
 \mid  0 \rangle \otimes \mid  j,  j \rangle  \\
 \vdots\\
 \mid  0 \rangle \otimes \mid  j,  j \rangle
 \end{pmatrix},  \quad k=0,1, \label{vector-algebra-HW-BG-generalized-trasformation}
\end{eqnarray}
which extrapolates between the vector coherent states over the matrix domain associated to the  harmonic oscillator  annihilation operator $ \hat{a},$ 		and the algebra eigenstates of the generalized operator $ \hat{a} + {\mathcal{B}}_+     \hat{\mathbb{J}}_- ,$ which associated eigenvalues are given by $\tilde{M}.$  

\subsection{The cases with $ b\neq 0$}
In this section we are going to compute the vector algebra eigenstates for the set of generators from the $su(2)$ sector whose parameter b is different from zero,  except for the generators that give rise to the canonical case whose vector coherent states have already been calculated  in section \ref{sec-canonical}. In all the remaining non-canonical cases,   the operator $\hat{T}$ that  act on the basis states spanning the complete space  of the $j$ irreducible representation of the $su(2)$ algebra is not unitary.   In any case, this fact does not greatly affect the procedure for obtaining the desired states. Actually, the process of obtaining such states is identical to the one used in the canonical case.  Furthermore, despite the non-unitary character of the $T$ operators, there are some situations that depend on the choice of integration constants in which the states can be written in the form of a unitary operator acting on the ground state, in other words, it is still possible to construct some vector coherent states. On the other hand, the non-unitary attribute of $\hat{T}$ provides us with an adequate metric that can be used to construct  some linear and quadratics $su(2)$ pseudo-Hermitian Hamiltonians \cite{PEGA}  in the context of this article.    

According to the results of appendices \ref{appa} and \ref{appb}, the   solutions of the eigenvalue equation

\begin{equation}
[\hat{a} + \beta \cdot \vec{\hat{J}} ] \mid \tilde{\psi} \rangle^j_{[s]} = \tilde{\lambda}_{[s]} \mid  \tilde{\psi} \rangle^j_{[s]} \label{eigen-value-equation-non-unitary}
\end{equation}
are given by

\begin{equation}
\mid \tilde{\psi} \rangle^j_{[s]} =    \mid \tilde{\lambda}_{[s]} - m  b \rangle \otimes  \hat{T} \mid j, m \rangle, \quad m=-j, \cdots, j.
\end{equation}
Using  the fact that $\hat{D} (\tilde{\lambda}_{[s]} - m b)  =   \hat{D} (\tilde{\lambda}_{[s]} )    \;   \hat{D} (-mb)   \; \exp{\left[ \frac{1}{2}  \left(\tilde{\lambda}_{[s]} b^\ast - \tilde{\lambda}_{[s]}^\ast b  \right) m \right]},    $  we can see that this last equation can be written in the form 
 
\begin{equation} 
 \mid \tilde{\psi} \rangle^j_{[s]}  =  \hat{D} (\tilde{\lambda}_{[s]} )    \;  \hat{T}  \;  \exp{\left[ \frac{1}{2}  \left(\tilde{\lambda}_{[s]} b^\ast - \tilde{\lambda}_{[s]}^\ast b  \right) \hat{J}_3\right]}   \mid  - m  b \rangle \otimes  \mid j, m \rangle, \quad m=-j, \cdots, j.
 \end{equation}
 Thus, the general solution of the eigenvalue equation (\ref{eigen-value-equation-non-unitary}) is
 
 \begin{equation}
 \mid \tilde{\Psi} \rangle^j_{[s]} = \hat{D} (\tilde{\lambda}_{[s]} )    \;  \hat{T}  \;  \exp{\left[ \frac{1}{2}  \left(\tilde{\lambda}_{[s]} b^\ast - \tilde{\lambda}_{[s]}^\ast b  \right) \hat{J}_3\right]} \sum_{m=-j}^{j}  \tilde{\varphi}^j_{[s] m} (0)  \mid  - m  b \rangle \otimes  \mid j, m \rangle
 \end{equation}
 and by consequent the $ (2 j + 1)^K$ solutions of equation  (\ref{vector-matrix-eigenvalue-total}), depending on the $\tilde{\varphi}^j_{[s] m}$  parameter values are given by
 \begin{equation} 
 \mid \Psi \rangle^j_{[K]} = \mathcal{N}^{-\frac{1}{2}} \; \hat{D} (\tilde{M})    \;  \hat{T}  \;  \exp{\left[ \frac{1}{2}  \left(\tilde{M} b^\ast - \tilde{M}^\dagger b  \right) \hat{J}_3\right]} \;  \tilde{U}  \;   \begin{pmatrix} \sum_{m=-j}^{j}  \tilde{\varphi}^j_{[1] m} (0)  \mid  - m  b \rangle \otimes  \mid j, m \rangle \\
  \sum_{m=-j}^{j}  \tilde{\varphi}^j_{[2] m} (0)  \mid  - m  b \rangle \otimes  \mid j, m \rangle \\
  \vdots \\
   \sum_{m=-j}^{j}  \tilde{\varphi}^j_{[K-1] m} (0)  \mid  - m  b \rangle \otimes  \mid j, m \rangle \\
    \sum_{m=-j}^{j}  \tilde{\varphi}^j_{[K] m} (0)  \mid  - m  b \rangle \otimes  \mid j, m \rangle
  \end{pmatrix}. \label{general-vector-eigenstates-no-unitary-b-neq-0}
 \end{equation}
 By choosing the integration constant in the form
 
\begin{equation}
\tilde{\varphi}^j_{[s] m} = \sum_{r =1}^{K}\tilde{U}^\dagger_{[s] [r]} \delta_{m_{[r]} m } \label{choice-varphi-s}
\end{equation}
we get the elementary set  of solutions 

\begin{equation}
 \mid \Psi \rangle^j_{[K]} = \mathcal{N}^{-\frac{1}{2}} \; \hat{D} (\tilde{M})    \;  \hat{T}  \;  \exp{\left[ \frac{1}{2}  \left(\tilde{M} b^\ast - \tilde{M}^\dagger b  \right) \hat{J}_3\right]}    \begin{pmatrix}   \mid  - m_1  b \rangle \otimes  \mid j, m_1 \rangle \\
   \mid  - m_2  b \rangle \otimes  \mid j, m_2 \rangle \\
  \vdots \\
    \mid  - m_{K-1}  b \rangle \otimes  \mid j, m_{K-1} \rangle \\
    \mid  - m_K  b \rangle \otimes  \mid j, m_K \rangle
  \end{pmatrix}. \label{master-vector-states-non-unitary}
\end{equation}

 \subsubsection{The case $[\hat{\mathbb{A}}, \hat{\mathbb{A}}^\dagger ] = \hat{I} + 2 x \; \hat{J}_3 $}
   
 -  When $x >  0$ and with the choice of parameters  $\beta_+ = \sqrt{x} \sinh{\alpha} e^{i \theta_+},$  $\beta_- = \sqrt{x} \cosh{\alpha} e^{i \theta_-}$ and $\beta_3=0$, we have  $b= e^{\frac{i}{2}(\theta_+ - \theta_-) }\sqrt{ 2 x \cosh{(2 \alpha)}}.$ Thus,  from equations (\ref{master-operator-T})   and (\ref{phases-conditions}) we deduce
 \begin{equation}
 \hat{T} = \exp{\left[- \frac{\pi}{4} \left(  \sqrt{\coth(\alpha)}  e^{- \frac{i}{2} (\theta_+ - \theta_-)} \hat{J}_+ -      \sqrt{\tanh(\alpha)}  e^{ \frac{i}{2} (\theta_+ - \theta_-)} \hat{J}_-              \right)\right]}.
 \end{equation}
Inserting this last result into equation  (\ref{master-vector-states-non-unitary}),  we obtain the vector algebra eigenstates of the element  $\hat{\mathbb{A}}= a + \sqrt{x} \sinh{\alpha} e^{i \theta_+} \hat{J}_- +  \sqrt{x} \cosh{\alpha} e^{i \theta_-} \hat{J}_+,$ that is,

\begin{eqnarray}
 \mid \Psi \rangle^j_{[K]} &=& \mathcal{N}^{-\frac{1}{2}} \; \hat{D} (\tilde{M})    \exp{\left[- \frac{\pi}{4} \left(  \sqrt{\coth(\alpha)}  e^{- \frac{i}{2} (\theta_+ - \theta_-)} \hat{J}_+ -      \sqrt{\tanh(\alpha)}  e^{ \frac{i}{2} (\theta_+ - \theta_-)} \hat{J}_-              \right)\right]}    \nonumber \\  &\times&  \exp{\left[ \frac{1}{2}  \left(\tilde{M} b^\ast - \tilde{M}^\dagger b  \right) \hat{J}_3\right]}    \begin{pmatrix}   \mid  - m_1  b \rangle \otimes  \mid j, m_1 \rangle \\
   \mid  - m_2  b \rangle \otimes  \mid j, m_2 \rangle \\
  \vdots \\
    \mid  - m_{K-1}  b \rangle \otimes  \mid j, m_{K-1} \rangle \\
    \mid  - m_K  b \rangle \otimes  \mid j, m_K \rangle
  \end{pmatrix}.
\end{eqnarray}

From this set of vector states we can extract two classes of vector coherent states. Indeed, if we choose $m_1 = m_2 = \cdots = m_K = -j,$  and suitably express  the exponential operator $\hat{T}$ as a product of three exponential operators depending of $\hat{J}_+,$ $\hat{J}_3$ and $\hat{J}_-,$ respectively and then perform the action of the ensemble of the exponential operators on the state $\mid j , -j \rangle,$   we obtain the equivalent exponential  operator action on this last state: $\exp\left[- \sqrt{\coth(\alpha)}  e^{- \frac{i}{2} (\theta_+ - \theta_-)} \hat{J}_+ \right] \mid j, -j  \rangle.$ Finally, by replacing this last non-unitary exponential operator by its unitary equivalent, we obtain the normalized vector coherent states

\begin{equation}
 \mid \Psi \rangle^j_{[K]} =  \frac{1}{\sqrt{K}}  \hat{D} (\tilde{M})    \exp{\left[- \arctan\left( \sqrt{\coth(\alpha)}\right)   \left( e^{- \frac{i}{2} (\theta_+ - \theta_-)} \hat{J}_+ -   e^{ \frac{i}{2} (\theta_+ - \theta_-)} \hat{J}_-  \right)\right]}        
  \begin{pmatrix}   \mid  j  b \rangle \otimes  \mid j, -j \rangle \\
   \mid  j b \rangle \otimes  \mid j, -j\rangle \\
  \vdots \\
    \mid  j b \rangle \otimes  \mid j, -j \rangle \\
    \mid  j b \rangle \otimes  \mid j, -j \rangle
  \end{pmatrix}.
\end{equation}

In the same way, if we choose  $m_1=m_2= \cdots= m_K = j,$ and perform the same process that we just described, we get a second set of vector coherent states, that is, 

\begin{equation}
 \mid \Psi \rangle^j_{[K]} =   \frac{1}{\sqrt{K}} \hat{D} (\tilde{M})    \exp{\left[- \arctan\left( \sqrt{\tanh(\alpha)} \right) \left( e^{- \frac{i}{2} (\theta_+ - \theta_-)} \hat{J}_+ -   e^{ \frac{i}{2} (\theta_+ - \theta_-)} \hat{J}_-  \right)\right]}         
 \begin{pmatrix}   \mid  - j  b \rangle \otimes  \mid j, j \rangle \\
   \mid  - j b \rangle \otimes  \mid j, j\rangle \\
  \vdots \\
    \mid  - j b \rangle \otimes  \mid j, j \rangle \\
    \mid  - j b \rangle \otimes  \mid j, j \rangle
  \end{pmatrix}.
\end{equation}

-  When $x <  0$ and  with the choice of parameters  $\beta_+ = \sqrt{ - x} \cosh{\alpha} e^{i \theta_+},$  $\beta_- = \sqrt{ - x} \sinh{\alpha} e^{i \theta_-}$ and $\beta_3=0$,  we have  $b= e^{\frac{i}{2}(\theta_+ - \theta_-) }\sqrt{ 2 \mid x \mid \cosh{(2 \alpha)}}.$ The expressions for the corresponding algebra eigenstates in this case can be extracted from the expressions of the  previous one by interchanging $\sinh{\alpha}$ by $\cosh(\alpha)$ and vice-versa. Doing that, we get

\begin{eqnarray}
 \mid \Psi \rangle^j_{[K]} &=& \mathcal{N}^{-\frac{1}{2}} \; \hat{D} (\tilde{M})    \exp{\left[- \frac{\pi}{4} \left(  \sqrt{\tanh(\alpha)}  e^{- \frac{i}{2} (\theta_+ - \theta_-)} \hat{J}_+ -      \sqrt{\coth(\alpha)}  e^{ \frac{i}{2} (\theta_+ - \theta_-)} \hat{J}_-              \right)\right]}    \nonumber \\  &\times&  \exp{\left[ \frac{1}{2}  \left(\tilde{M} b^\ast - \tilde{M}^\dagger b  \right) \hat{J}_3\right]}    \begin{pmatrix}   \mid  - m_1  b \rangle \otimes  \mid j, m_1 \rangle \\
   \mid  - m_2  b \rangle \otimes  \mid j, m_2 \rangle \\
  \vdots \\
    \mid  - m_{K-1}  b \rangle \otimes  \mid j, m_{K-1} \rangle \\
    \mid  - m_K  b \rangle \otimes  \mid j, m_K \rangle
  \end{pmatrix}.
\end{eqnarray}
    
From this last expression, as before,  we can also extract two classes of vector coherent states. Indeed, if we choose $m_1 = m_2 = \cdots = m_K = -j,$  following the same procedure as above, we get

\begin{equation}
 \mid \Psi \rangle^j_{[K]} =  \frac{1}{\sqrt{K}} \hat{D} (\tilde{M})    \exp{\left[- \arctan\left( \sqrt{\tanh(\alpha)} \right)  \left( e^{- \frac{i}{2} (\theta_+ - \theta_-)} \hat{J}_+ -   e^{ \frac{i}{2} (\theta_+ - \theta_-)} \hat{J}_-  \right)\right]}         
 \begin{pmatrix}   \mid   j  b \rangle \otimes  \mid j, -j \rangle \\
   \mid  j b \rangle \otimes  \mid j, - j\rangle \\
  \vdots \\
    \mid   j b \rangle \otimes  \mid j, -j \rangle \\
    \mid   j b \rangle \otimes  \mid j, -j \rangle
  \end{pmatrix}.
\end{equation}
On the other hand, if we choose  $m_1 = m_2 = \cdots = m_K = j,$ we get

\begin{equation}
 \mid \Psi \rangle^j_{[K]} = \frac{1}{\sqrt{K}}  \hat{D} (\tilde{M})    \exp{\left[- \arctan\left( \sqrt{\coth(\alpha)} \right)  \left( e^{- \frac{i}{2} (\theta_+ - \theta_-)} \hat{J}_+ -   e^{ \frac{i}{2} (\theta_+ - \theta_-)} \hat{J}_-  \right)\right]}         
 \begin{pmatrix}   \mid   - j  b \rangle \otimes  \mid j, j \rangle \\
   \mid  - j b \rangle \otimes  \mid j, j\rangle \\
  \vdots \\
    \mid  - j b \rangle \otimes  \mid j, j \rangle \\
    \mid  - j b \rangle \otimes  \mid j, j \rangle
  \end{pmatrix}.
\end{equation}

    \subsubsection{The case $[\hat{\mathbb{A}}, \hat{\mathbb{A}}^\dagger ] = \hat{I} + \rho e^{i \nu} \hat{J}_+ + \rho e^{- i \nu} \hat{J}_-  $}
In case that  $\beta_+ = R e^{i \theta_+},$     $\beta_- = R e^{i \theta_-}$ and $\beta_3 = R_3 e^{i \theta_3},$ with $\theta_3 \neq \frac{1}{2} (\theta_+ + \theta_-) + k  \pi, \; k=1,2,\cdots $ and   $R_3 \neq 2R$  or $\theta_3 \neq \frac{1}{2} (\theta_+ + \theta_-) + (k + \frac{1}{2}) \pi,$  then $ b  =  2 R  \; e^{ \frac{i}{2} (\theta_+ + \theta_-) } \sqrt{1  +\left[ \frac{R_3}{2R}  e^{ i   \left(\theta_3 - \frac{1}{2} (\theta_+ + \theta_-)\right)  }\right]^2 } \neq 0.$ Thus, from equation (\ref{phases-conditions}) we can compute the parameters $\tilde{\theta}$ and $\tilde{\phi},$ serving to construct the operator $\hat{T},$ we get

\begin{equation}
\tan\left(\frac{\tilde{\theta}}{2} \right)= \sqrt{\frac{ b - \beta_3}{b + \beta_3}} = \sqrt{1 + \gamma^2} - \gamma, \quad e^{i \tilde{\phi}} = e^{\frac{i}{2} (\theta_+ - \theta_-)}, \label{tilde-theta-complex}
\end{equation}
where $\gamma = \frac{R_3}{2R}  e^{ i   \left(\theta_3 - \frac{1}{2} (\theta_+ + \theta_-)\right)}.   $
From these last relations we can see that $\tilde{\phi}$ is a real parameters whereas $\tilde{\theta}$ is, in general, a complex one. Indeed, if we write  $\gamma = \varrho e^{i \varphi},$ where $\varrho= \frac{R_3}{2R}$ and $\varphi = \theta_3 - \frac{1}{2} (\theta_+ + \theta_-) \neq k \pi, \;  k=1,2,\cdots, $ and insert it into (\ref{tilde-theta-complex}), after some manipulations we obtain

\begin{eqnarray}
\tan\left(\frac{\tilde{\theta}}{2} \right)&=& \left[ \sqrt{ \frac{
\sqrt{1 + \varrho^4 + 2 \varrho^2 \cos(2 \varphi)} + \left(1 + \varrho^2 \cos(2 \varphi)\right)} 
{2}} - \varrho \cos(\varphi) \right] \nonumber \\
&+& i \left[ \sqrt{ \frac{
\sqrt{1 + \varrho^4 + 2 \varrho^2 \cos(2 \varphi)} - \left(1 + \varrho^2 \cos(2 \varphi)\right)} 
{2}} - \varrho \sin(\varphi) \right] . \label{complex-tan-tilde-theta}
\end{eqnarray}
 
We note that if we write $\tan \left(\frac{\tilde{\theta}}{2} \right) = \mathcal{Z} =   \mathcal{Z}_{R} + i \mathcal{Z}_{Im} , $ where $ \mathcal{Z}_{Re}  $ and $\mathcal{Z}_{Im}$ are the real and imaginary parts of the complex number $\mathcal{Z},$ respectively, then formally we have

\begin{equation}
\frac{\tilde{\theta}}{2} =\frac{i}{2} \ln \left[ \frac{1 - i \mathcal{Z}}{ 1 + i \mathcal{Z}}  \right]= - \frac{\delta}{2} + \frac{i}{2} \ln \left[   \frac{\sqrt{\left(1- \| \mathcal{Z}\|^2 \right)^2 + 4 \mathcal{Z}^2_{Re} } }{1 -2 \mathcal{Z}_{Im} + \| \mathcal{Z}\|^2    }  \right] , \label{complex-tilde-theta-dos}
\end{equation}
where $ \delta = \arctan \left [  \frac{2 \mathcal{Z}_{Re}} {\| \mathcal{Z}\|^2 -1}  \right] $ when $ \|\mathcal{Z}\| \neq 1$ and 
$\delta= - \frac{\pi}{2}$ if $\| \mathcal{Z}\| =1 $ and $ \mathcal{Z}_{Re } > 0$ and  $\delta=  \frac{\pi}{2}$ if $\| \mathcal{Z}\| =1 $ and $ \mathcal{Z}_{Re }  < 0,$ where $\mathcal{Z}_{R} $  and $ \mathcal{Z}_{Im}  $ represent the real and imaginary parts of the right side of equation  (\ref{complex-tan-tilde-theta}), respectively.  Thus, we can show that the general structure of the operator $\hat{T}$ in this case is given by the product of two exponential operators, one of them unitary and the other one Hermitian:

\begin{eqnarray}
\hat{T} &=&   \exp\left[ \frac{\delta}{2}  \left(  e^{- \frac{i}{2} (\theta_+ - \theta_-)  }  \hat{J}_+  -   e^{ \frac{i}{2} (\theta_+ - \theta_-)  }  \hat{J}_-    \right) \right] \nonumber \\ &\times&   \exp\left[ - \frac{i}{2}      \ln \left(   \frac{\sqrt{\left(1- \| \mathcal{Z}\|^2 \right)^2 + 4 \mathcal{Z}^2_{Re} } }{1 -2 \mathcal{Z}_{Im} + \| \mathcal{Z}\|^2    }  \right)              \left(  e^{- \frac{i}{2} (\theta_+ - \theta_-)  }  \hat{J}_+  -   e^{ \frac{i}{2} (\theta_+ - \theta_-)  }  \hat{J}_-    \right) \right]. 
\end{eqnarray}
By inserting this operator into the general expression (\ref{general-vector-eigenstates-no-unitary-b-neq-0}) or into (\ref{master-vector-states-non-unitary}) we finally obtain the vector algebra eigenstates of the operator $\hat{\mathbb{A}}=\hat{a} + R e^{i \theta_+} \hat{J}_- + R e^{i \theta_-} \hat{J}_+  + R_3 e^{i \theta_3}  \hat{J}_3 , $ which parameter satisfy the condition given in this subsection:
 
 \begin{eqnarray}
 \mid \Psi \rangle^j_{[K]} &=& \mathcal{N}^{-\frac{1}{2}} \; \hat{D} (\tilde{M})   \exp\left[ \frac{\delta}{2}  \left(  e^{- \frac{i}{2} (\theta_+ - \theta_-)  }  \hat{J}_+  -   e^{ \frac{i}{2} (\theta_+ - \theta_-)  }  \hat{J}_-    \right) \right]  \nonumber \\  &\times& \exp\left[ - \frac{i}{2}      \ln \left(   \frac{\sqrt{\left(1- \| \mathcal{Z}\|^2 \right)^2 + 4 \mathcal{Z}^2_{Re} } }{1 -2 \mathcal{Z}_{Im} + \| \mathcal{Z}\|^2    }  \right)              \left(  e^{- \frac{i}{2} (\theta_+ - \theta_-)  }  \hat{J}_+  -   e^{ \frac{i}{2} (\theta_+ - \theta_-)  }  \hat{J}_-    \right) \right]    \nonumber \\ 
 &\times&
   \exp{\left[ \frac{1}{2}  \left(\tilde{M} b^\ast - \tilde{M}^\dagger b  \right) \hat{J}_3\right]}    \begin{pmatrix}   \mid  - m_1  b \rangle \otimes  \mid j, m_1 \rangle \\
   \mid  - m_2  b \rangle \otimes  \mid j, m_2 \rangle \\
  \vdots \\
    \mid  - m_{K-1}  b \rangle \otimes  \mid j, m_{K-1} \rangle \\
    \mid  - m_K  b \rangle \otimes  \mid j, m_K \rangle
  \end{pmatrix}. \label{AES-case-J+-J-}
\end{eqnarray}
  If  in (\ref{AES-case-J+-J-} ) we choose $m_1=m_2= \cdots= m_K=-j,  $ and proceed as in the previous sections, we get the set of  normalized vector coherent states:
  
  \begin{eqnarray}
 \mid \Psi \rangle^j_{[K]} &=& \frac{1}{\sqrt{K}} \hat{D} (\tilde{M})   \\  &\times& 
 \exp\left[ -     \arctan \left( \| \mathcal{Z} \|\right)   \left( e^{-  i  \left(  \frac{\theta_+ - \theta_-}{2} - \arctan{ \left( \frac{ \mathcal{Z}_{Im}}{\mathcal{Z}_{Re}} \right)}   \right)}   \hat{J}_+  -   e^{ i \left( \frac{\theta_+ - \theta_-}{2} - \arctan{ \left(\frac{ \mathcal{Z}_{Im}}{\mathcal{Z}_{Re}} \right)}    \right)}     \hat{J}_-    \right)  \right]  \nonumber \\ &\times&   
 \begin{pmatrix}   \mid  j  b \rangle \otimes  \mid j, -j \rangle \\
   \mid  j  b \rangle \otimes  \mid j, -j\rangle \\
  \vdots \\
    \mid  j b \rangle \otimes  \mid j, -j\rangle \\
    \mid  j  b \rangle \otimes  \mid j, -j\rangle
 \end{pmatrix}, \label{AES-case-J+-J-j-menos}
\end{eqnarray}
where  $ \arctan{\left( \frac{ \mathcal{Z}_{Im}}{\mathcal{Z}_{Re}} \right)}$ is replaced by  $\frac{\pi}{2}$ when $\mathcal{Z}_{Re}=0.$

On the other hand, if we choose  $m_1=m_2= \cdots= m_K=j,  $ and again we proceed as in the previous sections, we get a second set  of  normalized vector coherent states:
   \begin{eqnarray}
 \mid \Psi \rangle^j_{[K]} &=& \frac{1}{\sqrt{K}} \hat{D} (\tilde{M})   \\  &\times& 
 \exp\left[ -     \arctan \left( \| \mathcal{Z} \|\right)   \left( e^{-  i  \left(  \frac{\theta_+ - \theta_-}{2} + \arctan{ \left( \frac{ \mathcal{Z}_{Im}}{\mathcal{Z}_{Re}} \right)}   \right)}   \hat{J}_+  -   e^{ i \left( \frac{\theta_+ - \theta_-}{2} + \arctan{ \left(\frac{ \mathcal{Z}_{Im}}{\mathcal{Z}_{Re}} \right)}    \right)}     \hat{J}_-    \right)  \right]  \nonumber \\ &\times&   
 \begin{pmatrix}   \mid - j  b \rangle \otimes  \mid j, j \rangle \\
   \mid  - j  b \rangle \otimes  \mid j, j\rangle \\
  \vdots \\
    \mid  - j b \rangle \otimes  \mid j, j\rangle \\
    \mid  - j  b \rangle \otimes  \mid j, j\rangle
 \end{pmatrix}. 
 \label{AES-case-J+-J-j-plus}
\end{eqnarray}
 
 \paragraph{ Some special cases}
For example, in the special case where $\varphi = \frac{\pi}{2}$ and $\varrho > 1,$  we have  $\mathcal{Z}= i ( \sqrt{\varrho^2-1} -  \varrho ),$ that implies $\delta=0,$ and by consequent $ \frac{\tilde{\theta}}{2} = - \frac{i}{2} \ln\left( \sqrt{\frac{\varrho +1}{\varrho-1}} \right) = - \; \frac{i}{2} \; \ln\left(\sqrt{\frac{R_3 + 2R}{R_3 - 2R}} \right)   $ and $b=i \; e^{\frac{i}{2} (\theta_+ + \theta_-)} \sqrt{R_3^2 - 4 R^2} .$ Thus, the vector eigenstates of   $\mathbb{A} = \hat{a} + R e^{i \theta_+} \hat{J}_-  + 
 R e^{i \theta_-}  \hat{J}_+ +  i  R_3  e^{\frac{i}{2} (\theta_+ + \theta_-)} \hat{J}_3 $ with matrix eigenvalue $\tilde{M}$ are given by
 
  \begin{eqnarray}
 \mid \Psi \rangle^j_{[K]} &=& \mathcal{N}^{-\frac{1}{2}} \; \hat{D} (\tilde{M})    \exp\left[  \frac{i}{2}        \ln\left( \sqrt{\frac{R_3 + 2R}{R_3 - 2R}} \right)          \left(  e^{- \frac{i}{2} (\theta_+ - \theta_-)  }  \hat{J}_+  -   e^{ \frac{i}{2} (\theta_+ - \theta_-)  }  \hat{J}_-    \right) \right]    \nonumber \\ 
 &\times&
   \exp{\left[ \frac{1}{2}  \left(\tilde{M} b^\ast - \tilde{M}^\dagger b  \right) \hat{J}_3\right]}    \begin{pmatrix}   \mid  - m_1  b \rangle \otimes  \mid j, m_1 \rangle \\
   \mid  - m_2  b \rangle \otimes  \mid j, m_2 \rangle \\
  \vdots \\
    \mid  - m_{K-1}  b \rangle \otimes  \mid j, m_{K-1} \rangle \\
    \mid  - m_K  b \rangle \otimes  \mid j, m_K \rangle
  \end{pmatrix}. \label{AES-case-J+-J-special-1}
\end{eqnarray}

On the other hand, when $\varphi = \frac{\pi}{2}$ and $ 0 < \varrho < 1,$  we have $\mathcal{Z}=  \sqrt{1-\varrho^2} - i \varrho ,$  that implies $\delta = - \frac{\pi}{2},$ and by consequent  $ \frac{\tilde{\theta}}{2} = \frac{\pi}{4}  - \frac{i}{2} \ln\left( \sqrt{\frac{1+ \varrho}{1-\varrho}} \right) = \frac{\pi}{4} - \frac{i}{2} \ln\left( \sqrt{\frac{2R+R_3}{2R-R_3}} \right)   $ and $b=e^{\frac{i}{2} (\theta_+ + \theta_-)} \sqrt{4 R^2 - R_3^2} .$ Thus, the vector eigenstates of $\mathbb{A} = \hat{a} + R e^{i \theta_+} \hat{J}_-  + 
 R e^{i \theta_-}  \hat{J}_+ +  i  R_3  e^{\frac{i}{2} (\theta_+ + \theta_-)} \hat{J}_3 $ with matrix eigenvalue $\tilde{M}$ are given by
 
  \begin{eqnarray}
 \mid \Psi \rangle^j_{[K]} &=& \mathcal{N}^{-\frac{1}{2}} \; \hat{D} (\tilde{M})    \exp\left[ - \frac{\pi}{4}  \left(  e^{- \frac{i}{2} (\theta_+ - \theta_-)  }  \hat{J}_+  -   e^{ \frac{i}{2} (\theta_+ - \theta_-)  }  \hat{J}_-    \right) \right]  \nonumber \\  &\times& \exp\left[  \frac{i}{2}        \ln\left( \sqrt{\frac{2R+R_3}{2R-R_3}} \right)          \left(  e^{- \frac{i}{2} (\theta_+ - \theta_-)  }  \hat{J}_+  -   e^{ \frac{i}{2} (\theta_+ - \theta_-)  }  \hat{J}_-    \right) \right]    \nonumber \\ 
 &\times&
   \exp{\left[ \frac{1}{2}  \left(\tilde{M} b^\ast - \tilde{M}^\dagger b  \right) \hat{J}_3\right]}    \begin{pmatrix}   \mid  - m_1  b \rangle \otimes  \mid j, m_1 \rangle \\
   \mid  - m_2  b \rangle \otimes  \mid j, m_2 \rangle \\
  \vdots \\
    \mid  - m_{K-1}  b \rangle \otimes  \mid j, m_{K-1} \rangle \\
    \mid  - m_K  b \rangle \otimes  \mid j, m_K \rangle
  \end{pmatrix}. \label{AES-case-J+-J-special-2}
\end{eqnarray}
From these last two expression we can also extract the corresponding set of vector coherent states. Indeed, putting $m_1=m_2=\cdots=m_K= \mp \, j$   we can obtain the desired vector states: 
\begin{eqnarray}
 \mid \Psi \rangle^j_{[K]} &=& \frac{1}{\sqrt{K}} \hat{D} (\tilde{M})   \\  &\times& 
 \exp\left[ \mp  i      \arctan \left(   \frac{R_3 - \sqrt{R_3^2 -4 R^2}}{2R}  \right)   \left(  e^{-  \frac{i}{2}  (\theta_+ - \theta_-) }  \hat{J}_+  + 
  e^{  \frac{i}{2} (\theta_+ - \theta_- )}    \hat{J}_-    \right)  \right]  \nonumber \\ &\times&   
 \begin{pmatrix}   \mid  \pm j  b \rangle \otimes  \mid j, \mp j \rangle \\
   \mid  \pm j  b \rangle \otimes  \mid j, \mp\rangle \\
  \vdots \\
    \mid  \pm j b \rangle \otimes  \mid j, \mp\rangle \\
    \mid  \pm j  b \rangle \otimes  \mid j, \mp \rangle
 \end{pmatrix}, \label{VCS-case-J+-J-j-menos}
\end{eqnarray}
when $R_3 > 2R,$
and

\begin{eqnarray}
 \mid \Psi \rangle^j_{[K]} &=& \frac{1}{\sqrt{K}} \hat{D} (\tilde{M})   \\  &\times& 
 \exp\left[ -    \frac{\pi}{4}   \left( e^{-  i  \left(  \frac{\theta_+ - \theta_-}{2} \pm \arctan{ \left( \frac{ R_3}{\sqrt{4 R^2 - R_3^2} } \right)}   \right)}   \hat{J}_+  -   e^{ i \left( \frac{\theta_+ - \theta_-}{2} \pm \arctan{ \left(\frac{ R_3}{\sqrt{4 R^2 - R_3^2} } \right)}    \right)}     \hat{J}_-    \right)  \right]  \nonumber \\ &\times&   
 \begin{pmatrix}   \mid  \pm j  b \rangle \otimes  \mid j, \mp j \rangle \\
   \mid  \pm j  b \rangle \otimes  \mid j, \mp j\rangle \\
  \vdots \\
    \mid  \pm j b \rangle \otimes  \mid j, \mp j\rangle \\
    \mid  \pm j  b \rangle \otimes  \mid j, \mp j\rangle
 \end{pmatrix}, 
\end{eqnarray}
when $0 < R_3 < 2 R. $

\subsubsection{The case $[\hat{\mathbb{A}}, \hat{\mathbb{A}}^\dagger ] = \hat{I} + 2  x \; \hat{J}_3 + \rho e^{i \nu} \hat{J}_+ + \rho e^{- i \nu} \hat{J}_- ,  \quad  x \neq 0 $  and $ \rho > 0    $ }    
In this subsection we will study the situation where the  commutator  between $\hat{\mathbb{A}}$ and $\hat{\mathbb{A}}^\dagger$ is equal to the identity plus a linear combination of the $su(2)$ algebra generators with non-zero coefficients.  Here, three cases are possible, they are:

 -  When $\beta_- = R_- e^{i \theta_-} \neq 0,$ $\beta_- =0$ and $\beta_3 = R_3 e^{i \theta_3} \neq 0,$ the parameter $b= \beta_3  $ $x = R_-^2 $ and $\rho e^{i \nu} = - R_3 R_-  e^{- \frac{i}{2}  (\theta_3 - \theta_-) }. $ Then, with help of equation (\ref{AES-j3-j--normalized-operator-factor}) and  the general structure (\ref{general-vector-eigenstates-no-unitary-b-neq-0}) we can construct the vector states verifying (\ref{vector-matrix-eigenvalue-total}), they are given by
 \begin{equation} 
 \mid \Psi \rangle^j_{[K]} = \mathcal{N}^{-\frac{1}{2}} \; \hat{D} (\tilde{M})    \;  \exp{\left[- \frac{\beta_-}{\beta_3} \hat{J}_+ \right]} \;  \exp{\left[ \frac{1}{2}  \left(\tilde{M} b^\ast - \tilde{M}^\dagger b  \right) \hat{J}_3\right]} \;  \tilde{U}  \;   \begin{pmatrix} \sum_{m=-j}^{j}  \tilde{\varphi}^j_{[1] m} (0)  \mid  - m  b \rangle \otimes  \mid j, m \rangle \\
  \sum_{m=-j}^{j}  \tilde{\varphi}^j_{[2] m} (0)  \mid  - m  b \rangle \otimes  \mid j, m \rangle \\
  \vdots \\
   \sum_{m=-j}^{j}  \tilde{\varphi}^j_{[K-1] m} (0)  \mid  - m  b \rangle \otimes  \mid j, m \rangle \\
    \sum_{m=-j}^{j}  \tilde{\varphi}^j_{[K] m} (0)  \mid  - m  b \rangle \otimes  \mid j, m \rangle
  \end{pmatrix}. \label{general-vector-eigenstates-b-neq-0-beta-beta3-neq-0}
 \end{equation}
By choosing $\tilde{\varphi}^j_{[s] m}, \; s=1,2,\cdots, K, $ as in equation  (\ref{choice-varphi-s}) and then taking $m_{r} = -j, \; \forall r =, 1,2, \cdots,K,$ and after that,  replacing the exponential operator depending on  $ J_+ $ by its unitary equivalent, we get the set of normalized vector coherent states
\begin{equation} 
 \mid \Psi \rangle^j_{[K]} = \frac{1}{\sqrt{K}} \; \hat{D} (\tilde{M})    
  \exp{\left[- \arctan\left( \frac{R_-}{R_3} \right) \left( e^{- i (\theta_3 - \theta_-)}  \hat{J}_+  -   e^{ i (\theta_3 - \theta_-)}  \hat{J}_- \right)          \right]}   \begin{pmatrix}  \mid  j b \rangle \otimes  \mid j, -j \rangle \\
    \mid  j  b \rangle \otimes  \mid j, -j \rangle \\
  \vdots \\
    \mid  j b \rangle \otimes  \mid j, -j \rangle \\
     \mid  j  b \rangle \otimes  \mid j, -j\rangle
  \end{pmatrix}. \label{VCS-b-neq-0-beta-beta3-neq-0}
 \end{equation}

-  On the other hand, when  $\beta_- = 0,$  $\beta_+ = R_+ e^{i \theta_+ } \neq 0  $ and $\beta_3 = R_3 e^{i \theta_3} \neq 0,$ the parameter $b= \beta_3, $  $x = - R_+^2 0$ and $\rho e^{i \nu} = R_3 R_+   e^{i (\theta_3 - \theta_+) }. $ Then, again, with help of equation (\ref{AES-j3-j--normalized-operator-factor}) and  the general structure (\ref{general-vector-eigenstates-no-unitary-b-neq-0}) we can construct the vector states verifying (\ref{vector-matrix-eigenvalue-total}), they are given by
 \begin{equation} 
 \mid \Psi \rangle^j_{[K]} = \mathcal{N}^{-\frac{1}{2}} \; \hat{D} (\tilde{M})    \;  \exp{\left[\frac{\beta_+}{\beta_3} \hat{J}_- \right]} \;  \exp{\left[ \frac{1}{2}  \left(\tilde{M} b^\ast - \tilde{M}^\dagger b  \right) \hat{J}_3\right]} \;  \tilde{U}  \;   \begin{pmatrix} \sum_{m=-j}^{j}  \tilde{\varphi}^j_{[1] m} (0)  \mid  - m  b \rangle \otimes  \mid j, m \rangle \\
  \sum_{m=-j}^{j}  \tilde{\varphi}^j_{[2] m} (0)  \mid  - m  b \rangle \otimes  \mid j, m \rangle \\
  \vdots \\
   \sum_{m=-j}^{j}  \tilde{\varphi}^j_{[K-1] m} (0)  \mid  - m  b \rangle \otimes  \mid j, m \rangle \\
    \sum_{m=-j}^{j}  \tilde{\varphi}^j_{[K] m} (0)  \mid  - m  b \rangle \otimes  \mid j, m \rangle
  \end{pmatrix}. \label{general-vector-eigenstates-b-neq-0-beta+beta3-neq-0}
 \end{equation}
By choosing again $\tilde{\varphi}^j_{[s] m}, \; s=1,2,\cdots, K, $ as in equation  (\ref{choice-varphi-s}) and then taking $m_{r} = j, \; \forall r =, 1,2, \cdots,K,$ and after that,  replacing the exponential operator depending on  $ J_- $ by its unitary equivalent, we get the set of normalized vector coherent states
\begin{equation} 
 \mid \Psi \rangle^j_{[K]} = \frac{1}{\sqrt{K}} \; \hat{D} (\tilde{M})    
  \exp{\left[- \arctan\left( \frac{R_+}{R_3} \right) \left( e^{i (\theta_3 - \theta_+)}  \hat{J}_+
   -   e^{ - i (\theta_3 - \theta_+)}  \hat{J}_- \right)          \right]}   \begin{pmatrix}  \mid  -j b \rangle \otimes  \mid j, j \rangle \\
    \mid  -j  b \rangle \otimes  \mid j, j \rangle \\
  \vdots \\
    \mid  - j b \rangle \otimes  \mid j, j \rangle \\
     \mid  -j  b \rangle \otimes  \mid j, j\rangle
  \end{pmatrix}. \label{VCS-b-neq-0-beta+beta3-neq-0}
 \end{equation}

- Finally we have the case when  $\beta_- = R_-  e^{i \theta_-} \neq 0,$  $\beta_+ = R_+ e^{i \theta_+ } \neq 0  $ and $\beta_3 = R_3 e^{i \theta_3} \neq 0,$ provided that both, $R_+ \neq R_-$ and $ R_3 \neq 2 \sqrt{R_+ \;  R_-}$ or $ \theta_3 \neq \frac{1}{2} (\theta_+ -\theta_-) + \left( k + \frac{1}{2}\right) \pi, \; k=1,2,\cdots.$ Under these conditions $$\rho e^{i \nu} =  R_3    e^{- i (\theta_+ - \theta_-) }  \sqrt{R_+^2 + R_-^2 - 2 R_+ \;  R_- \cos{(2 \varphi)}}  \exp{\left[\frac{R_+ + R_-}{R_+ - R_-} \tan{\varphi}     \right]},$$  $b= 2  \sqrt{ R_+ \; R_-} e^{  \frac{i}{2} (\theta_+ -\theta_-)  }  \sqrt{1 + \gamma^2}$ and   $x = R_-^2 - R_+^2, $  where $\gamma = \varrho e^{i \varphi}$ with $\varphi= \theta_3 -  \frac{1}{2} (\theta_+ +  \theta_-)  $ and now $\varrho = \frac{R_3}{2 \sqrt{R_+ \; R_-}}.$  Then, as in previous subsection,  from equation (\ref{phases-conditions}) we can compute the parameters $\tilde{\theta}$ and $\tilde{\phi},$ serving to construct the operator $\hat{T},$ we get the following expressions

\begin{equation}
\tan\left(\frac{\tilde{\theta}}{2} \right)= \sqrt{\frac{ b - \beta_3}{b + \beta_3}} = \sqrt{1 + \gamma^2} - \gamma, \quad 
e^{i \tilde{\phi}} =  \sqrt{\frac{R_+}{R_-}} e^{\frac{i}{2} (\theta_+ - \theta_-)}, \label{tilde-theta-phi-complex}.
\end{equation}
We note that the expression used to calculate parameter $\tilde{\theta}$ has the same structure as the one in the previous subsection, then, to get its value  we only have  to use directly  the expressions (\ref{complex-tan-tilde-theta}) and (\ref{complex-tilde-theta-dos}). On the other hand, with respect to the parameter $\tilde{\phi}$  now the  is a little different because it represent  complex phase. Indeed, this last parameter can be written  in the form $\tilde{\phi} = \tilde{\phi}_{Re} + i  \tilde{\phi}_{Im},$ where $e^{- \tilde{\phi}_{Im}} = \sqrt{\frac{R_+}{R_-}}$ and $e^{i \tilde{\phi}_{Re}} = e^{\frac{i}{2} (\theta_+ - \theta_-)}.  $

Thus, using  (\ref{master-operator-T}) we can show that the general structure of the operator $\hat{T}$ in this case is given by the product of two exponential operators, none of which is unitary or Hermitian, that is, 

\begin{eqnarray}
\hat{T} &=&  \exp\left[ \frac{\delta}{2}  \left( \sqrt{\frac{R_-}{R_+}}  e^{- \frac{i}{2} (\theta_+ - \theta_-)  }  \hat{J}_+  -  \sqrt{\frac{R_+}{R_-}}   e^{ \frac{i}{2} (\theta_+ - \theta_-)  }  \hat{J}_-    \right) \right]  \nonumber \\ &\times &\exp\left[ - \frac{i}{2}      \ln \left(   \frac{\sqrt{\left(1- \| \mathcal{Z}\|^2 \right)^2 + 4 \mathcal{Z}^2_{Re} } }{1 -2 \mathcal{Z}_{Im} + \| \mathcal{Z}\|^2    }  \right)              \left( \sqrt{\frac{R_-}{R_+}}  e^{- \frac{i}{2} (\theta_+ - \theta_-)  }  \hat{J}_+  -  \sqrt{\frac{R_+}{R_-}}  e^{ \frac{i}{2} (\theta_+ - \theta_-)  }  \hat{J}_-    \right) \right]. 
\end{eqnarray}

By inserting this operator into the general expression (\ref{general-vector-eigenstates-no-unitary-b-neq-0}) or into (\ref{master-vector-states-non-unitary}) we finally obtain the vector algebra eigenstates of the operator $\hat{\mathbb{A}}=\hat{a} + R e^{i \theta_+} \hat{J}_- + R e^{i \theta_-} \hat{J}_+  + R_3 e^{i \theta_3}, $ which parameter satisfy the condition given in this subsection:
 
 \begin{eqnarray}
 \mid \Psi \rangle^j_{[K]} &=& \mathcal{N}^{-\frac{1}{2}} \; \hat{D} (\tilde{M})  \exp\left[ \frac{\delta}{2}  \left( \sqrt{\frac{R_-}{R_+}}  e^{- \frac{i}{2} (\theta_+ - \theta_-)  }  \hat{J}_+  -  \sqrt{\frac{R_+}{R_-}}   e^{ \frac{i}{2} (\theta_+ - \theta_-)  }  \hat{J}_-    \right) \right]   \nonumber \\  &\times&     \nonumber \\ 
 &\times& \exp\left[ - \frac{i}{2}      \ln \left(   \frac{\sqrt{\left(1- \| \mathcal{Z}\|^2 \right)^2 + 4 \mathcal{Z}^2_{Re} } }{1 -2 \mathcal{Z}_{Im} + \| \mathcal{Z}\|^2    }  \right)              \left( \sqrt{\frac{R_-}{R_+}}  e^{- \frac{i}{2} (\theta_+ - \theta_-)  }  \hat{J}_+  -  \sqrt{\frac{R_+}{R_-}}  e^{ \frac{i}{2} (\theta_+ - \theta_-)  }  \hat{J}_-    \right) \right] \nonumber \\ &\times&
   \exp{\left[ \frac{1}{2}  \left(\tilde{M} b^\ast - \tilde{M}^\dagger b  \right) \hat{J}_3\right]}    \begin{pmatrix}   \mid  - m_1  b \rangle \otimes  \mid j, m_1 \rangle \\
   \mid  - m_2  b \rangle \otimes  \mid j, m_2 \rangle \\
  \vdots \\
    \mid  - m_{K-1}  b \rangle \otimes  \mid j, m_{K-1} \rangle \\
    \mid  - m_K  b \rangle \otimes  \mid j, m_K \rangle
  \end{pmatrix}. \label{AES-case-J+-J-j3}
\end{eqnarray}

From  (\ref{AES-case-J+-J-j3} ) we can find the associated vector coherent states for this case. If we choose $m_1=m_2= \cdots= m_K= \mp j,  $ and proceed as in the previous sections, we get the set of  normalized vector coherent states:
  
  \begin{eqnarray}
 \mid \Psi \rangle^j_{[K]} &=& \frac{1}{\sqrt{K}} \hat{D} (\tilde{M})   \\  &\times& 
 \exp\left[ -     \arctan \left( \| \mathcal{Z} \|      e^{ \pm \; \tilde{\phi}_{Im} }   \right)   \left( e^{-  i  \left(  \frac{\theta_+ - \theta_-}{2} \mp \arctan{ \left( \frac{ \mathcal{Z}_{Im}}{\mathcal{Z}_{Re}} \right)}   \right)}   \hat{J}_+  -   e^{ i \left( \frac{\theta_+ - \theta_-}{2} \mp \arctan{ \left(\frac{ \mathcal{Z}_{Im}}{\mathcal{Z}_{Re}} \right)}    \right)}     \hat{J}_-    \right)  \right]  \nonumber \\ &\times&   
 \begin{pmatrix}   \mid  \pm j  b \rangle \otimes  \mid j,  \mp j \rangle \\
   \mid  \pm j  b \rangle \otimes  \mid j, \mp j\rangle \\
  \vdots \\
    \mid  \pm j b \rangle \otimes  \mid j, \mp j\rangle \\
    \mid  \pm j  b \rangle \otimes  \mid j, \mp j\rangle
 \end{pmatrix}, \label{VCS-J+-J-J3-c-phase-one-two}
\end{eqnarray}
where  $ \arctan{\left( \frac{ \mathcal{Z}_{Im}}{\mathcal{Z}_{Re}} \right)}$ is replaced by  $\frac{\pi}{2}$ when $\mathcal{Z}_{Re}=0.$

 \paragraph{Some examples}
  Let us illustrate the results of this section by choosing  the same range of variation of the $\varphi$ and  $\varrho$  parameters used in the previous subsection.  Thus, when  $\varphi = \frac{\pi}{2}$ and $\varrho > 1,$  we have  $\mathcal{Z}= i ( \sqrt{\varrho^2-1} -  \varrho ),$ that implies $\delta=0,$ and by consequent $ \frac{\tilde{\theta}}{2} = - \frac{i}{2} \; \ln\left(\sqrt{\frac{R_3 + 2\sqrt{R_+ \; R_-}}{R_3 - 2\sqrt{R_+ \; R_-}}} \right)   $ and $b=i \;  e^{\frac{i}{2} (\theta_+ + \theta_-)} \sqrt{R_3^2 - 4 R_+  R_-} .$ Thus, the vector eigenstates of   $\mathbb{A} = \hat{a} + R_+ e^{i \theta_+} \hat{J}_-  + 
 R_- e^{i \theta_-}  \hat{J}_+ +  i  R_3  e^{\frac{i}{2} (\theta_+ + \theta_-)} \hat{J}_3$ with matrix eigenvalue $\tilde{M}$ are given by
 
  \begin{eqnarray}
 \mid \Psi \rangle^j_{[K]} &=& \mathcal{N}^{-\frac{1}{2}} \; \hat{D} (\tilde{M})    \exp\left[  \frac{i}{2}      \ln\left(\sqrt{\frac{R_3 + 2\sqrt{R_+ \; R_-}}{R_3 - 2\sqrt{R_+ \; R_-}}} \right)      \left( \sqrt{\frac{R_-}{R_+}}  e^{- \frac{i}{2} (\theta_+ - \theta_-)  }  \hat{J}_+  -  \sqrt{\frac{R_+}{R_-}}   e^{ \frac{i}{2} (\theta_+ - \theta_-)  }  \hat{J}_-    \right) \right]    \nonumber \\ 
 &\times&
   \exp{\left[ \frac{1}{2}  \left(\tilde{M} b^\ast - \tilde{M}^\dagger b  \right) \hat{J}_3\right]}    \begin{pmatrix}   \mid  - m_1  b \rangle \otimes  \mid j, m_1 \rangle \\
   \mid  - m_2  b \rangle \otimes  \mid j, m_2 \rangle \\
  \vdots \\
    \mid  - m_{K-1}  b \rangle \otimes  \mid j, m_{K-1} \rangle \\
    \mid  - m_K  b \rangle \otimes  \mid j, m_K \rangle
  \end{pmatrix}. \label{AES-case-J+-J-j3-special-1}
\end{eqnarray}

On the other hand, when $\varphi = \frac{\pi}{2}$ and $ 0 < \varrho < 1,$  we have $\mathcal{Z}=  \sqrt{1-\varrho^2} - i \varrho ,$  that implies $\delta = - \frac{\pi}{2},$ and by consequent  $ \frac{\tilde{\theta}}{2} = \frac{\pi}{4}  - \frac{i}{2} \ln\left( \sqrt{\frac{1+ \varrho}{1-\varrho}} \right) = \frac{\pi}{4} - \frac{i}{2} \ln\left( \sqrt{\frac{2\sqrt{R_+ \; R_-} +R_3}{2\sqrt{R_+ \; R_-} - R_3}} \right)   $ and $b=e^{\frac{i}{2} (\theta_+ + \theta_-)} \sqrt{4 R_+ \; R_-  - R_3^2} .$ Thus, the vector eigenstates of $\mathbb{A} = \hat{a} + R_+ e^{i \theta_+} \hat{J}_-  + 
 R_-  e^{i \theta_-}  \hat{J}_+ +  i  R_3  e^{\frac{i}{2} (\theta_+ + \theta_-)} \hat{J}_3 $ with matrix eigenvalue $\tilde{M}$ are given by
 
  \begin{eqnarray}
 \mid \Psi \rangle^j_{[K]} &=& \mathcal{N}^{-\frac{1}{2}} \; \hat{D} (\tilde{M})    \exp\left[ - \frac{\pi}{4}  \left( \sqrt{\frac{R_-}{R_+}}  e^{- \frac{i}{2} (\theta_+ - \theta_-)  }  \hat{J}_+  -   \sqrt{\frac{R_+}{R_-}}  e^{ \frac{i}{2} (\theta_+ - \theta_-)  }  \hat{J}_-    \right) \right]  \nonumber \\  &\times& \exp\left[  \frac{i}{2}        \ln\left( \sqrt{\frac{2\sqrt{R_+ \; R_-} +R_3}{2\sqrt{R_+ \; R_-} - R_3}} \right)           \left(  \sqrt{\frac{R_-}{R_+}}  e^{- \frac{i}{2} (\theta_+ - \theta_-)  }  \hat{J}_+  -  \sqrt{\frac{R_+}{R_-}}   e^{ \frac{i}{2} (\theta_+ - \theta_-)  }  \hat{J}_-    \right) \right]    \nonumber \\ 
 &\times&
   \exp{\left[ \frac{1}{2}  \left(\tilde{M} b^\ast - \tilde{M}^\dagger b  \right) \hat{J}_3\right]}    \begin{pmatrix}   \mid  - m_1  b \rangle \otimes  \mid j, m_1 \rangle \\
   \mid  - m_2  b \rangle \otimes  \mid j, m_2 \rangle \\
  \vdots \\
    \mid  - m_{K-1}  b \rangle \otimes  \mid j, m_{K-1} \rangle \\
    \mid  - m_K  b \rangle \otimes  \mid j, m_K \rangle
  \end{pmatrix}. \label{AES-case-J+-J-j3-special-2}
\end{eqnarray}
From these last two expression we can also extract the corresponding set of vector coherent states. Indeed, putting $m_1=m_2=\cdots=m_K= \mp \, j$   we can obtain the desired vector states: 
\begin{eqnarray}
 \mid \Psi \rangle^j_{[K]} &=& \frac{1}{\sqrt{K}} \hat{D} (\tilde{M})   \\  &\times& 
 \exp\left[ \mp  i      \arctan \left(   \frac{R_3 - \sqrt{R_3^2 - 4 R_+ \; R_-}}{2R_{\pm}}  \right)   \left(  e^{-  \frac{i}{2}  (\theta_+ - \theta_-) }  \hat{J}_+  + 
  e^{  \frac{i}{2} (\theta_+ - \theta_- )}    \hat{J}_-    \right)  \right]  \nonumber \\ &\times&   
 \begin{pmatrix}   \mid  \pm j  b \rangle \otimes  \mid j, \mp j \rangle \\
   \mid  \pm j  b \rangle \otimes  \mid j, \mp\rangle \\
  \vdots \\
    \mid  \pm j b \rangle \otimes  \mid j, \mp\rangle \\
    \mid  \pm j  b \rangle \otimes  \mid j, \mp \rangle
 \end{pmatrix}, \label{VCS-case-J+-J-j-j3-menos}
\end{eqnarray}
when $\varphi = \frac{\pi}{ 2}$ and  $R_3 > 2 \sqrt{R_+ \; R_-}$
and

\begin{eqnarray}
 \mid \Psi \rangle^j_{[K]} &=& \frac{1}{\sqrt{K}} \hat{D} (\tilde{M})   \\  &\times& 
 \exp\left[ -   \arctan\left( \frac{R_\mp}{R_\pm} \right)  \left( e^{-  i  \left(  \frac{\theta_+ - \theta_-}{2} \pm \arctan{ \left( \frac{ R_3}{\sqrt{4 R_+ \; R_-  - R_3^2} } \right)}   \right)}   \hat{J}_+  -   e^{ i \left( \frac{\theta_+ - \theta_-}{2} \pm \arctan{ \left(\frac{ R_3}{\sqrt{4 R_+ \; R_-  - R_3^2} } \right)}    \right)}     \hat{J}_-    \right)  \right]  \nonumber \\ &\times&   
 \begin{pmatrix}   \mid  \pm j  b \rangle \otimes  \mid j, \mp j \rangle \\
   \mid  \pm j  b \rangle \otimes  \mid j, \mp j\rangle \\
  \vdots \\
    \mid  \pm j b \rangle \otimes  \mid j, \mp j\rangle \\
    \mid  \pm j  b \rangle \otimes  \mid j, \mp j\rangle
 \end{pmatrix}, 
\end{eqnarray}
when $\varphi = \frac{\pi}{2}$ and $0 < R_3 < 2 R. $

%%%%%%%%%%%%%%%%%%%%%%%%%%%%%%%%%%%%%%%%%%%%%%%%%%%%%%%%%
 %%%%%%%%%%%%%%%%%%%%%%%%%%%%%%%%%%%%%%%%%%%%%%%%%%%%%%%%%

\section{Conclusions} In this article we have defined and computed the vector algebra eigenstates associated to the $h(1) \oplus su(2)$  Lie algebra. We have shown that the set of these vector states includes the subset of vector coherent states over the matrix domain of the Heisenberg-Weyl algebra as well as those of the $su(2)$ Lie algebra. Other subsets of vector coherent states have been also obtained which appear  as  consequence  of the combination  as a  direct sum  of these two algebras. We have shown that it is possible to classify  the set of all these vector eigenstates  by giving to the  elements of the algebra the role of a  generalized  annihilation operator and then using the commutator between it and its corresponding adjoint as a selection rule. Also with the help of these two operators we have constructed  Hermitian Hamiltonians to which we can  associate those vector coherent states. For example, this allowed us to find generalized  Hamiltonians which are isospectral  with the standard harmonic oscillator Hamiltonian. Moreover,  in the particular case where the  normal eigenvalue matrix  have been  constructed  of the matrix elements of a complex linear combination of the $su(2)$ Lie algebra generators and the identity operator in a given  $su(2)$ irreducible representation space, we have shown that for a special choice of the parameters, we can obtain  the so-called  quaternionic vector coherent states on the matrix domain\cite{TA-KT} and the corresponding generalized oscillator algebra  characteristic of  the coherent state quantization of quaternions \cite{BM-KT}, as well as a generalized version of all that.  The techniques used in this article and the results obtained can be easily adapted to construct families of linear and quadratic $su(2)$  and  $h(1) \oplus su(2)$ pseudo-Hermitian Hamiltonians, see \cite{PEGA} and references therein.  Also, all kind of intelligent and coherent states associated to the generalized Schrödinger-Robertson uncertainty relation \cite{HR} are included in ours results. Indeed,  by  choosing two Hermitian  operators, $ \hat{\mathcal{X}}$ and $ \hat{\mathcal{P}},$  from the set of elements of the $h(1) \oplus su(2)$ Lie algebra, it can be shown that the minimum uncertainty states in the sense of the  Schrödinger-Robertson relation  verify the matrix eigenvalue equation $ [ \hat{\mathcal{X}} + i \lambda  \hat{\mathcal{P}}] \mid \Phi \rangle = M  \mid \Phi \rangle,$ where $\lambda \in \mathbb{C}.$ Then, families of vector coherent and squeezed states can be generated from it, depending on the values of the $\lambda $ parameter, which selection rules in term of the commutator $[ \hat{\mathcal{X}},  \hat{\mathcal{P}}]$ coincide with those of the generalized creation and annihilation rules mentioned above. Finally, an important relation  involving the generators of the $su(2)$ algebra  have been  explicitly developed at the moment of disentangling the exponential operators in the case  when the parameter $b=0,$ namely, that the non-canonical  commutator $[\hat{a} +(\vec{\beta} \cdot \vec{\hat{J}}) , \hat{a}^\dagger +  (\vec{\beta} \cdot \vec{\hat{J}})^\dagger    ] = \hat{I} +  2 x \hat{J}_3 + \rho e^{i \nu} \hat{J}_+  +  \rho e^{- i \nu} \hat{J}_-,  $ where $|x| > 0,$ is equivalent to the  canonical one $[  \hat{a} +  \mathbb{B}_\mp  \hat{\mathbb{J}}_\pm ,  \hat{a}^\dagger +  \mathbb{B}_\mp^\ast   \hat{\mathbb{J}}_\mp] = \hat{I} \pm  2 \|\mathbb{B}_\mp \|^2  \hat{\mathbb{J}}_3 ,$ for  suitably defined $su(2)$ Lie algebra transformed generators $ \hat{\mathbb{J}}_+ ,$
$ \hat{\mathbb{J}}_-$ and $ \hat{\mathbb{J}}_3.$    In other words, we have shown that  to each element of the complex Lie algebra of $su(2)$, whose coefficients verify $b=0, $ we can associate its adjoint  element and the element formed from the commutator between them in such a way that the three  generators reproduce the standard commutation relations of the $su(2)$ Lie algebra.

\section*{Acknowledgment} The author dedicates this article to his little purple doll Mary and his beloved children. 

\appendix

\section{Algebra eigenstates of $h(1)\oplus su(2)$ Lie algebra}
\label{appa} 
The   $h(1) \oplus su(2)$ algebra eigenstates are defined to be the solutions of the eigenvalue equation
\begin{equation}
[\alpha_{-} \hat{a} + \alpha_{+} \hat{a}^{\dagger} + \alpha_{3} \hat{I} + \beta_{-} \hat{J}_{+} + \beta_{+} \hat{J}_{-} + \beta_{3}  \hat{J}_{3}] \mid \Phi \rangle = \beta  \mid \Phi \rangle, \label{eigenvalue-1}
\end{equation}
where $ \alpha_{\mp}, \alpha_{3},   \beta_{\mp}$ and $  \beta_{3}  $ are given complex numbers and $\beta$  a  complex number representing the associated eigenvalue.

Equation (\ref{eigenvalue-1}) have been already studied  in the literature. We will continue here the study of these states in order to prepare it suitably and to use it effectively in the construction,  simplification and  factorization of the $h(1) \oplus su(2) $ algebra eigenstates  with eigenvalues  over the matrix domain. From (\ref{eigenvalue-1}) we can see that without loss of generality we can set $\alpha_{-}=1$ and $\alpha_{3}=0.$ Then, by applying the following transformation to the states 
\begin{equation}\mid \Phi \rangle = \exp{\left[ -  \frac{1}{2} \alpha_+  \left({\hat{a}^\dagger}\right)^2 \right]}\mid \Psi \rangle, \label{sqeezed-transformation} \end{equation} (\ref{eigenvalue-1})  reduces to:

\begin{equation}
\left[ \hat{a}  + \beta_{-} \hat{J}_{+} + \beta_{+} \hat{J}_{-} + \beta_{3}  \hat{J}_{3}\right] \mid \Psi \rangle = \beta  \mid \Psi \rangle. \label{eigenvalue-2}
\end{equation} 

\vspace*{1cm}
Before continuing with the resolution of the eigenvalue problem, let us make some comments about the transformation we just performed.  The transformation (\ref{sqeezed-transformation}),  corresponds to a standard squeezed transformation provided that  $\|\alpha_+ \| <  1.$ When we apply it  to the fundamental state $\mid 0 \rangle,$ it can be replaced by the  unitary  transformation $\hat{S} (\xi) =  \exp{\left[ - \frac{1}{2} \left(  \xi  \left.{\hat{a}^\dagger}\right.^2  - \xi^\ast  \hat{a}^2    \right) \right]} \mid 0 \rangle,$ where $\xi =  \frac{ \alpha_+  }{\|\alpha_+ \|} \tanh^{-1} \left(  \|  \alpha_+ \|\right).$ On the other hand, when it is applied onto a coherent state, that is,  $ e^{- \frac{1}{2} \alpha_+ \left. \hat{a}^\dagger \right.^2}    e^{ z_1 \hat{a}^\dagger}   \mid 0 \rangle,$  where $z_1$ is an arbitrary  complex number, we can first  use   the property $ \hat{S}^\dagger (\xi) \hat{a}^\dagger \hat{S} (\xi) = \left[ \hat{a}^\dagger \cosh(\| \xi \|) -  \frac{\xi^\ast}{  \| \xi \|} \sinh(\| \xi \| ) \hat{a} \right] $ to shift the squeezed unitary operator to the left hand, and then rewrite  the resulting expression in terms of the unitary displacement operator associated to the standard harmonic oscillator to get the normalized equivalent relation $ \hat{S} (\xi) \hat{D}\left(z_1 \cosh(\| \xi \|)\right) \mid 0 \rangle.  $ Moreover, when, for example, the following situation occurs:  $ e^{- \frac{1}{2} \alpha_+ \left. \hat{a}^\dagger \right.^2}   e^{ \beta \hat{a}^\dagger}   (z a^\dagger)^n \mid 0 \rangle,$ where $n$ is a non-negative integer, by using the well know properties of the  squeezed and displacement operators we can write this state in the form

\begin{equation}
 \hat{S} (\xi) \hat{D}\left(\beta \cosh(\| \xi \|)\right)    \left[ z \left( (\hat{a}^\dagger + \beta^\ast \cosh(\| \xi \|) )     \cosh(\| \xi \|) -                       \frac{\xi^\ast}{  \| \xi \|} \sinh(\| \xi \| ) (\hat{a} +\beta \cosh(\| \xi \|) \right)  \right]^n   \mid 0 \rangle. \label{squeezed-effect}
\end{equation} 

When $\|\alpha_+ \| \geq 1,$ we can leave the transformation  ({sqeezed-transformation}) as it is.  So in this way, once  the  solutions of the eigenvalue equation (\ref{eigenvalue-2}) are know,    proceeding as we have  indicated here, we will be able to  obtain the complete set of solutions of equation (\ref{eigenvalue-1} ).

\vspace*{1cm}

Now, let us rewrite (\ref{eigenvalue-2}) in the form
\begin{equation}
 \hat{a} \mid \Psi \rangle = 
\left[\beta -  \beta_{-} \hat{J}_{+} - \beta_{+} \hat{J}_{-} - \beta_{3}  \hat{J}_{3} \right]  \mid \Psi \rangle, \label{eigenvalue-3}
\end{equation}

Projecting both sides of (\ref{eigenvalue-3}) onto the state $\mid \zeta \rangle \otimes \hat{I}^{j}  $ and using  (\ref{big-psi-zeta}) the realization  (\ref{action-annihilation-zeta}) for the $\hat{a}$ annihilator, and the relations (\ref{action-j-pm})  for the $su(2)$ algebra generators, we obtain, for fixed $j,$ the following coupled linear system of differential equations:

\begin{eqnarray}
\nonumber \frac{d}{d\zeta} \psi^{j}_{m}(\zeta)  & = &
 (\beta-m\beta_{3})  \psi^{j}_{m}(\zeta) \\ 
\nonumber 
 &-& \beta_{-} \sqrt{(j-m)(j+m+1)}  \psi^{j}_{m+1}(\zeta)  \\ 
& -&   \beta_{+}  \sqrt{(j+m)(j-m+1)}   \psi^{j}_{m-1}(\zeta), \; \; m =-j, ...., j.    
\end{eqnarray}
This system can be written in the form

\begin{equation}
\frac{d}{d\zeta}  
\begin{pmatrix}
 \psi^{j}_{-j} (\zeta) \\
 \psi^{j}_{-j+1} (\zeta) \\
\vdots\\
 \psi^{j}_{j-1} (\zeta) \\
 \psi^{j}_{j} (\zeta) 
\end{pmatrix} =  M 
\begin{pmatrix}
\psi^{j}_{-j} (\zeta)  \\
 \psi^{j}_{-j+1} (\zeta) \\
\vdots\\
 \psi^{j}_{j-1} (\zeta) \\
\psi^{j}_{j} (\zeta) 
\end{pmatrix}, \label{system-eqs}
\end{equation}
where $M$ is the $2j+1 \times 2j+1$ dimensional matrix:

\begin{equation}
M = 
 \left(
\begin{smallmatrix}
\beta + j \beta_{3} & - \sqrt{2j} \beta_{+} & 0 & 0 & \dots & 0 \\
- \sqrt{2j} \beta_{-} & \beta + (j -1) \beta_{3}  & - \sqrt{(2j-1) 2} \beta_{+}& 0 & \dots & 0 \\
0 & - \sqrt{(2j-1) 2} \beta_{-} & \beta + (j -2) \beta_{3} &   - \sqrt{(2j-2) 3} \beta_{+}
& \dots &0\\
\vdots  & \vdots & \ddots& \ddots & \ddots & \vdots \\
0& 0& - \sqrt{3 (2j-2) } \beta_{-} & \beta - (j -2) \beta_{3} & - \sqrt{2 (2j-1) } \beta_{+} &0 \\
0& 0&0&- \sqrt{2 (2j-1) } \beta_{-} & \beta - (j -1) \beta_{3}  & - \sqrt{2j} \beta_{+}\\
0& 0&0&0& - \sqrt{2j} \beta_{-}& \beta-j \beta_{3}
 \end{smallmatrix} \right) .  \label{M-matrix-general-expression}
 \end{equation}
At this time it is important to mention that the eigenvalues of  $M$ are given by 
\begin{equation}
\lambda^j_{m} = \beta + m b, \quad \mbox{where} \quad b =\sqrt{4 \beta_+ \beta_- + \beta_3^2},\quad  m=-j, \cdots,j,\label{eigenvalues-M}
\end{equation}
which implies that when $b \neq 0,$  all $\lambda^j_{m} $'s are different and when $b =0,$ all   $\lambda^j_{m} $'s  are equal to $\beta.$ Thus,  in the first case, the  matrix in (\ref{M-matrix-general-expression}) can be diagonalized by a similarity transformation, while in the second case it is not always possible to do it.

\subsection{Case $b \neq 0$}
In this subsection we will construct the $h(1) \oplus su(2)$ algebra eigenstates associated aux generators which parameters verify $b \neq 0.$  We will see that in all the different cases the structure of these states presents the form of a product of two untangled operators that act on the base states, the first one depending solely on the $h(1)$ algebra generators and the second depending exclusively on the $su(2)$ algebra generators.  The method that we will use in the different cases to solve the equation (\ref{eigenvalue-1})    will combine the techniques of the algebra of operators and that of differential equations. 

 \subsubsection{Case $\beta_+ = \beta_- =0$ and $\beta_3 \neq 0$}
 Let us start with the computation of the algebra eigenstates of the simplest generator combining operators from the  $h(1)$ sector and $su(2)$ sector, that is,
 
\begin{equation}
[ \hat{a} + \beta_3 \hat{J}_3] \mid \psi \rangle^j = \beta \mid \psi \rangle^j.  \label{AES-j3} \end{equation} 
It is clear that in this case $b=\beta_3 \neq 0.$ Then, the set of normalized  states verifying this equation is given by
\begin{equation} 
 \mid \psi \rangle^j_m = \mid \beta - m \beta_3 \rangle \otimes \mid j, m \rangle, \quad m=-j,\cdots,j .
\end{equation} 
 Thus, the general solution of equation (\ref{AES-j3}) writes
  
  \begin{equation}
  \mid \psi \rangle^j =   \sum_{m=-j}^{j}  \frac{ \tilde{\varphi}^j_m (0) }{ \sqrt{\sum_{m=-j}^{j}  \|\tilde{\varphi}^j_m (0)  \|^2    }}     \mid \beta - m \beta_3 \rangle \otimes \mid j, m \rangle, \label{AES-j3-normalized}
\end{equation} 
where  $\tilde{\varphi}^j_m (0), \; m=-1,\cdots,j,$ are  constants. 

Using the fact that $\hat{D} (\beta - m \beta_3) =\hat{D} (\beta ) \hat{D} (- m \beta_3) \exp{\left[ \frac{m}{2}  \left(\beta \beta_3^\ast - \beta^\ast \beta_3  \right) \right]}, $ the states in equation (\ref{AES-j3-normalized}) can be written in the form
\begin{equation}
  \mid \psi \rangle^j =   \hat{D} (\beta ) \exp{\left[ \frac{1}{2}  \left(\beta \beta_3^\ast - \beta^\ast \beta_3  \right) \hat{J}_3\right]}\sum_{m=-j}^{j}  \frac{ \tilde{\varphi}^j_m (0) }{ \sqrt{\sum_{m=-j}^{j}  \|\tilde{\varphi}^j_m (0)  \|^2    }}     \mid  - m \beta_3 \rangle \otimes \mid j, m \rangle. \label{AES-j3-normalized-operator-factor}
\end{equation} 
\subsubsection{Case $\beta_+  \neq 0,  \beta_- =0$ and $\beta_3 \neq 0$}
 When $\beta_+  \neq 0,  \beta_- =0$ and $\beta_3 \neq 0,$ equation (\ref{eigenvalue-1}) becomes

\begin{equation}
[ \hat{a} + \beta_+ \hat{J}_-  +  \beta_3 \hat{J}_3] \mid \psi \rangle^j = \beta \mid \psi \rangle^j.  \label{AES-j3-j-} \end{equation} 
Again, in this case $b=\beta_3 \neq 0.$ 

By performing in  (\ref{AES-j3-j+}) the following transformation on the states $\mid \psi \rangle^j = e^{\frac{\beta_+ }{\beta_3} \hat{J}_-} \mid \tilde{\psi} \rangle^j $ and acting at the same time from the left on both sides of this equation with the operator$ e^{- \frac{\beta_+ }{\beta_3} \hat{J}_-},$ and then using the fact that   $e^{ - \frac{\beta_+ }{\beta_3} \hat{J}_-}  \beta_3 \hat{J}_3  e^{  \frac{\beta_+ }{\beta_3} \hat{J}_-} \mid \tilde{\psi} \rangle= -  \beta_+ \hat{J}_- + \beta_3 \hat{J}_3$,
we get

\begin{equation}
[ \hat{a} + \beta_3 \hat{J}_3] \mid \tilde{\psi} \rangle^j = \beta \mid \tilde{\psi} \rangle^j.  \label{AES-j3-j+-transformed} \end{equation} 
Hence, the set of normalized  eigenstates verifying this last  equation is given by
 \begin{equation}
 \mid \tilde{\psi} \rangle^j_m = \mid \beta - m \beta_3 \rangle \otimes \mid j, m \rangle, \quad m=-j,\cdots,j .
\end{equation} 
Finally, using the results of the previous subsection and then returning to the original states by performing the inverse transformation,  we find that the general solution of equation (\ref{AES-j3-j+}) is given by

\begin{equation}
  \mid \psi \rangle^j =  \frac{1}{\sqrt{\mathcal{N}^j}} \hat{D} (\beta )    e^{\frac{\beta_+ }{\beta_3} \hat{J}_-}          \exp{\left[ \frac{1}{2}  \left(\beta \beta_3^\ast - \beta^\ast \beta_3  \right) \hat{J}_3\right]}\sum_{m=-j}^{j}  \tilde{\varphi}^j_m (0)     \mid  - m \beta_3 \rangle \otimes \mid j, m \rangle, \label{AES-j3-j+-normalized-operator-factor}
\end{equation}
where $\mathcal{N}^j $ is a generic normalization constant.
  \subsubsection{Case $\beta_+ =0,  \beta_-  \neq 0$ and $\beta_3 \neq 0$}
 When $\beta_+  =0,  \beta_-  \neq 0$ and $\beta_3 \neq 0,$ equation (\ref{eigenvalue-1}) becomes

\begin{equation}
[ \hat{a} + \beta_- \hat{J}_+  +  \beta_3 \hat{J}_3] \mid \psi \rangle^j = \beta \mid \psi \rangle^j.  \label{AES-j3-j+} \end{equation} 
Again, in this case $b=\beta_3 \neq 0.$
Proceeding exactly as in the previous subsection, but taking in account  that now the new transformation on the states is governed by the operator $e^{ - \frac{\beta_- }{\beta_3} \hat{J}_+},$ we find the general solution of  (\ref{AES-j3-j+}) is given by

\begin{equation}
  \mid \psi \rangle^j =  \frac{1}{\sqrt{\mathcal{N}^j}} \hat{D} (\beta )    e^{\frac{- \beta_- }{\beta_3} \hat{J}_+}          \exp{\left[ \frac{1}{2}  \left(\beta \beta_3^\ast - \beta^\ast \beta_3  \right) \hat{J}_3\right]}\sum_{m=-j}^{j}  \tilde{\varphi}^j_m (0)     \mid  - m \beta_3 \rangle \otimes \mid j, m \rangle. \label{AES-j3-j--normalized-operator-factor}
\end{equation}
 
 \subsubsection{Case $\beta_+ \neq 0,  \beta_-  \neq 0$ and $\beta_3 = 0$}
When $\beta_+  \neq 0,  \beta_-  \neq 0$ and $\beta_3 = 0,$ equation (\ref{eigenvalue-1}) becomes
\begin{equation}
[ \hat{a} + \beta_- \hat{J}_+  +  \beta_+  \hat{J}_- ] \mid \psi \rangle^j = \beta \mid \psi \rangle^j.  \label{AES-j+-j-} \end{equation} 
 Also, in this case $b= 2 \sqrt{\beta_+ \; \beta_-}  \neq 0.$
 Then, by using equations  (\ref{master-operator-T}) and (\ref{phases-conditions}), we can construct an operator $\hat{T}$ with the following characteristics:
 
 \begin{equation}
 \hat{T} = \exp{ \left[ - \frac{\pi}{4} \left(\sqrt{\frac{\beta_-}{\beta_+}} \hat{J}_+ -      \sqrt{\frac{\beta_+}{\beta_-}} \hat{J}_-   \right) \right]},
 \end{equation} 
 such that 
\begin{equation}
\hat{T}^{-1} [\beta_- \hat{J}_+  +  \beta_+  \hat{J}_- ] \hat{T} = b \hat{J}_3.
\end{equation} 
We notice that, in general, $\hat{T}$ in not unitary, it becomes unitary if and only if $\|\beta_+\|=\|\beta_- \|.$ Thus,  using the following  transformation $  \mid \psi \rangle^j = \hat{T}  \mid \psi \rangle^j $ in equation  (\ref{AES-j+-j-}), and then acting  from the left with $\hat{T}^{-1}$ on both sides of the corresponding  transformed equation we get

\begin{equation}
[ \hat{a} + b  \hat{J}_3 ] \mid \tilde{\psi} \rangle^j = \beta \mid \tilde{\psi} \rangle^j,
\end{equation}
which normalized solutions we already know from the previous sections, that is,

 \begin{equation}
 \mid \tilde{\psi} \rangle^j_m = \mid \beta - m b \rangle \otimes \mid j, m \rangle, \quad m=-j,\cdots,j .
\end{equation} 

Finally, proceeding as in the  previous subsections, we reach the general solution of (\ref{AES-j+-j-}):

\begin{equation}
  \mid \psi \rangle^j =  \frac{1}{\sqrt{\mathcal{N}^j}} \hat{D} (\beta )       \exp{ \left[ - \frac{\pi}{4} \left(\sqrt{\frac{\beta_-}{\beta_+}} \hat{J}_+ -      \sqrt{\frac{\beta_+}{\beta_-}} \hat{J}_-   \right)\right] }     \exp{\left[ \frac{1}{2}  \left(\beta b^\ast - \beta^\ast b \right) \hat{J}_3\right]}\sum_{m=-j}^{j}  \tilde{\varphi}^j_m (0)     \mid  - m b \rangle \otimes \mid j, m \rangle. \label{AES-j+-j--normalized-operator-factor}
\end{equation}
 
 \subsubsection{Case $\beta_+ \neq 0,  \beta_-  \neq 0$ and $\beta_3 \neq 0$}
 When all beta parameters of the $su(2)$ algebra sector, which in principle are complex numbers,  are different from zero, we need to select  a subset of parameters  for which $b \neq 0.$ If we define
 
 \begin{equation}
 \beta_+ = R_+ \; e^{i \theta_+}, \quad   \beta_- = R_- \;  e^{i \theta_-}, \quad \mbox{and} \quad \beta_3 = R_3 \; e^{i \theta_3}, 
 \end{equation}
 where $ R_\pm $ and $ R_3 \; \in \;  \mathbb{R}_+ ,$ and  $\theta_\pm$ and $\theta_3 \in [0,2\pi],$ the conditions for $b$ being different from zero are
 
\begin{equation}
 R_3 \neq 2 \sqrt{R_+ \; R_-} \quad \mbox{or} \quad \theta_3 \neq  \frac{1}{2} \left(  \theta_+ + \theta_-      \right) +\left(k + \frac{1}{2}\right) \pi, \quad k=0,1.
 \end{equation} 
 Then,  under these conditions, using again  (\ref{master-operator-T}) and (\ref{phases-conditions}), we can construct an exponential operator $\hat{T}$  such that
 
\begin{equation}
\hat{T}^{-1} [\beta_+ \hat{J}_-  + \beta_- \hat{J}_+  +  \beta_3 \hat{J}_3]  \hat{T}  = b \hat{J}_3. \end{equation}
Hence, proceeding exactly as in the previous section, we can reduce (\ref{eigenvalue-1}) to

\begin{equation}
[ \hat{a} + b  \hat{J}_3 ] \mid \tilde{\psi} \rangle^j = \beta \mid \tilde{\psi} \rangle^j,
\end{equation}
which normalized solutions we already know from the previous sections, that is,

 \begin{equation}
 \mid \tilde{\psi} \rangle^j_m = \mid \beta - m b \rangle \otimes \mid j, m \rangle, \quad m=-j,\cdots,j .
\end{equation} 
 and in consequence the general solution of (\ref{eigenvalue-1})  is given by
\begin{equation}
  \mid \psi \rangle^j =  \frac{1}{\sqrt{\mathcal{N}^j}} \hat{D} (\beta )    \;  \hat{T}  \;  \exp{\left[ \frac{1}{2}  \left(\beta b^\ast - \beta^\ast b  \right) \hat{J}_3\right]}\sum_{m=-j}^{j}  \tilde{\varphi}^j_m (0)     \mid  - m b \rangle \otimes \mid j, m \rangle. \label{AES-j+-j-j3-normalized-operator-factor}
\end{equation}
 Here, in general,   $\hat{T}$ is not unitary, it becomes unitary if and only if $R_+ = R_-$ and the angle  $\theta_3 = \frac{1}{2} \left(  \theta_+ + \theta_-      \right) +\left(k + \frac{1}{2}\right) \pi, \; k=0,1$ and $R_3 \neq \sqrt{R_+ R_-}.$ Remarkably, these same conditions make the matrix $M$ in  \ref{M-matrix-general-expression} normal. 
 
\subsubsection{Case M normal}

Indeed,  when $M$ in \ref{M-matrix-general-expression} is a normal matrix, i.e.,  $M^\dagger M = M M^\dagger,$ where $M^\dagger$ denotes the conjugate transpose of $M,$   the following  conditions on the parameters forming the entries of  $M$ are imposed  

 \begin{equation}
\| \beta_{+}\| = \|\beta_{-}\|,  \quad \mbox{and} \quad \beta_{3} \; \beta^{\ast}_{+}  - \beta^{\ast}_{3}\;  \beta_{-} =0, \label{normality-conditions}
\end{equation}
or if we define

\begin{equation}
\beta_{\pm} = \|\beta_{\pm}\| \; e^{\pm i \theta_{\pm}}, \; \mbox{and} \; \beta_{3}= \| \beta_{3}\| \; e^{i \theta_3}, 
\end{equation}
and insert it in (\ref{normality-conditions}) we get their equivalents: 

\begin{equation}
\| \beta_{+}\| = \|\beta_{-}\|= \|\beta_{\pm}\| \quad \mbox{and} \quad
e^{i (\theta_{+} + \theta_{-})} = e^{2i \theta_3 }.\label{normality-conditions-equivalents}
\end{equation}

 From  this last equation  we can show that  $b$  in (\ref{eigenvalues-M})  is given $b =  e^{\frac{i}{2} (\theta+ + \theta_-)} \sqrt{4 \| \beta_{\pm}  \|^2 +  \|\beta_3\|^2} $  which means that  it is equal to  zero if and only if  $\beta_+ = \beta_- = \beta_3=0,$ in which case $M= \beta I,$ i.e, a diagonal matrix. On the other hand, when $\beta_+ = \beta_- =0$ and $\beta_3 \neq 0,$ the $M$ matrix is  still diagonal and then, the unitary matrix that makes it diagonal is the identity matrix.   In a more general case, there exist a unitary matrix $U$ such that $U^\dagger M U=D,$  where  
\begin{equation}
D = \begin{pmatrix}
\lambda^j_{-j}&0&0&\cdots&0\\
0& \lambda^j_{-j+1}&0&\cdots&0\\
\vdots& \vdots&\ddots&\vdots&\vdots\\
0& \cdots&0&   \lambda^j_{j-1}&0\\
0&\cdots&0&0&\lambda^j_{j}
\end{pmatrix}
\end{equation}
  is  a diagonal matrix formed out of the eigenvalues of $M.$

Operating with $U^\dagger$ on both sides of equation (\ref{system-eqs}), we get the uncoupled linear differential equation system:

\begin{equation}
\frac{d}{d\zeta}  
\begin{pmatrix}
 \tilde{\psi}^{j}_{-j} (\zeta) \\
\tilde{ \psi}^{j}_{-j+1} (\zeta) \\
\vdots\\
\tilde{ \psi}^{j}_{j-1} (\zeta) \\
\tilde{ \psi}^{j}_{j} (\zeta) 
\end{pmatrix} =  D 
\begin{pmatrix}
\tilde{\psi}^{j}_{-j} (\zeta)  \\
 \tilde{\psi}^{j}_{-j+1} (\zeta) \\
\vdots\\
 \tilde{\psi}^{j}_{j-1} (\zeta) \\
\tilde{\psi}^{j}_{j} (\zeta) 
\end{pmatrix}, \label{system-eqs-diagonal}
\end{equation}
where the tilde vector verify

\begin{equation}
\begin{pmatrix}
\psi^{j}_{-j} (\zeta)  \\
 \psi^{j}_{-j+1} (\zeta) \\
\vdots\\
 \psi^{j}_{j-1} (\zeta) \\
\psi^{j}_{j} (\zeta) 
\end{pmatrix} = U
\begin{pmatrix}
\tilde{\psi}^{j}_{-j} (\zeta)  \\
 \tilde{\psi}^{j}_{-j+1} (\zeta) \\
\vdots\\
 \tilde{\psi}^{j}_{j-1} (\zeta) \\
\tilde{\psi}^{j}_{j} (\zeta) 
\end{pmatrix}. \label{tilde-no-tilde}
\end{equation}

The integration of equation (\ref{system-eqs-diagonal}) is direct,  we obtain
\begin{equation}
\tilde{\psi}^j_m (\zeta) = e^{\lambda^j_{-m} \zeta} \tilde{\varphi}^j_m (0), \quad  m=-j, \cdots, j,
\end{equation}
where $ \tilde{\varphi}^j_m (0), \; m=-j,\cdots,j,$ are arbitrary integration constants. With help of equation (\ref{tilde-no-tilde}), the solution of the original system (\ref{system-eqs}) can be reached, that is

\begin{equation}
\psi^j_m (\zeta) =  \;\sum_{\ell=-j}^{j} \; U_{m \ell} \; \tilde{\psi}^j_\ell (\zeta) =\sum_{m=-\ell}^{j} \;  U_{m \ell} \; e^{\lambda^j_{-\ell} \zeta} \tilde{\varphi}^j_\ell (0).
\end{equation}

Inserting this result in equation (\ref{general-state-expression}) we arrive to
\begin{equation}
 \mid \Psi (\zeta) \rangle^j =  \sum_{m=-j}^{j}  \tilde{\varphi}^j_m (0)   e^{\lambda^j_{-m} \zeta}  \otimes     \sum_{\ell= -j}^{j} \; U^t_{m \ell}      \mid j , \ell \rangle,
\end{equation} 
where $U^t$ stands for the transpose of $U.$  We can then express this last states  in terms of the energy eigenstates of the standard harmonic oscillator Hamiltonian, that is  
\begin{equation}
  \mid \Psi \rangle^j  = N^{- \frac{1}{2}}  \sum_{m=-j}^{j}  \tilde{\varphi}^j_m (0)  e^{\lambda^j_{-m} \hat{a}^\dagger}   \mid 0 \rangle \otimes     \sum_{\ell= -j}^{j} \; U^t_{m \ell} \mid j , \ell \rangle. \label{general-su2-algebra-eigenstates-solution}
 \end{equation}
where $N$ is a normalization constant which must be fixed by imposing $ {}^j\langle  \Psi \mid \Psi \rangle^j =1.$ We notice that the general solution 
(\ref{general-su2-algebra-eigenstates-solution}) is a superposition of $2j+1$ independent solutions of the $h(1) \oplus su(1)$ algebra eigenstate equation  which are orthogonal to each other. Indeed,  each solution  composing  the general state in   (\ref{general-su2-algebra-eigenstates-solution}) , in a normalized version, is given by

\begin{eqnarray}
\mid \psi \rangle^j_m & =&  e^{- \frac{1}{2} \|\lambda^j_{-m} \|^2}  e^{\lambda^j_{-m} \hat{a}^\dagger}   \mid 0 \rangle \otimes     \sum_{\ell= -j}^{j} \; U^t_{m \ell} \mid j , \ell \rangle  \nonumber \\ &=&  \mid    \lambda^j_{-m}     \rangle \otimes     \sum_{\ell= -j}^{j} \; U^t_{m \ell} \mid j , \ell \rangle \quad m=-j, \cdots,j ,  \label{solution-AES-normal-case-m} 
\end{eqnarray}
where $\mid    \lambda^j_{-m}     \rangle, m=-1, \cdots, j  $ are the canonical   coherent states of the harmonic oscillator system verifying
$ \hat{a}  \mid    \lambda^j_{-m}     \rangle =   \lambda^j_{- m} \; \mid    \lambda^j_{-m}     \rangle, m=-j,\cdots, j,$
and according with the normality condition imposed to $M,$ that implies the  unitarity  of $U,$ these states  verify

\begin{equation}
{}^j_\ell \langle \psi \mid \psi \rangle^j_m = \delta_{\ell,m}.
\end{equation}
Thus, the general state  (\ref{general-su2-algebra-eigenstates-solution}) can be written in the normalized form
 \begin{equation}
 \mid \Psi \rangle^j = N^{- \frac{1}{2}} \sum_{m=-j}^{j}   \tilde{\varphi}^j_{m} (0) \;  \mid \psi \rangle^j_m  =
 N^{- \frac{1}{2}} \sum_{m=-j}^{j}   \tilde{\varphi}^j_{m} (0)  \mid    \lambda^j_{-m}     \rangle \otimes     \sum_{\ell= -j}^{j} \; U^t_{m \ell} \mid j , \ell \rangle ,  \label{algebra-eigenstates-h1-su2}
\end{equation} 
where
 \begin{equation}
 N= \sum_{m=-j}^{j}  \|\tilde{\varphi}^j_{m} (0) \|^2.
 \end{equation}
For comparative purposes of  these states structure  with that of  the  vector algebra  eigenstates with eigenvalues on the matrix domain settled in this article, it is convenient to rewrite the general solution  (\ref{algebra-eigenstates-h1-su2}) in the matrix form, that is
 
 \begin{eqnarray}
 \mid \Psi \rangle^j &=& N^{-1/2} \nonumber \\ 
 &\times &
 \begin{pmatrix}
 \tilde{\varphi}^j_{-j} (0)  e^{\lambda^j_{j}  \hat{a}^\dagger} &  \tilde{\varphi}^j_{-j+1} (0) e^{\lambda^j_{j-1} \hat{a}^\dagger}  & \cdots & \tilde{\varphi}^j_{j-1} (0) e^{\lambda^j_{-j+1} \hat{a}^\dagger} &  \tilde{\varphi}^{j}_{j} (0) e^{\lambda^j_{-j} \hat{a}^\dagger}
 \end{pmatrix}
 U^{t} \quad
 \begin{pmatrix}
 \mid 0 ; j, -j \rangle \\
 \mid 0 ; j, -j+1 \rangle\\
\vdots \\
 \mid 0 ; j, j -1 \rangle \\
 \mid 0 ; j, j \rangle \\
 \end{pmatrix}. \nonumber \\  \label{algebra-eigenstates-h1-su2-rewriten}
 \end{eqnarray}

Equation (\ref{algebra-eigenstates-h1-su2})  gives us  the  eigenstates of the operator $[ \hat{a}  + \beta_{-} \hat{J}_{+} + \beta_{+} \hat{J}_{-} + \beta_{3}  \hat{J}_{3}]$ when    $M$ is normal. In fact, the process for computing the $h(1) \oplus su(2)$  algebra eigenstates  when $b \neq 0$ without imposing this condition  to $M$ is similar, the only difference is  that in this case  the passing matrix $U$ is not unitary, which implies that the   set of solutions, which are still linearly  independent, are in general not  orthogonal. Also, the 
non-unitary character of the passing matrix makes it more difficult to find a  simple general expression for the normalization constants.

The passing matrix  $U$ can be built using the  techniques of general linear algebra,  linear differential equations or  operator algebra. In in appendix \ref{appb}   we have followed this last method  to  explicitly construct  this matrix for all fixed values of  $j.$ In fact, there we have shown that the matrix element of $U$ are given by $U_{\ell m} =  \langle j, \ell \mid \hat{T} \mid j,m \rangle, \; \ell,m = -j, \cdots,j.$

\subsection{Case b=0}
When $b=0,$  all the eigenvalues of the $M$ matrix are equal. To obtain the $h(1) \oplus su(2)$    algebra eigenstates in this case we had better integrate systematically,  function by function,  the linear differential equation system (\ref{system-eqs}) . We distinguish here three cases, that it to say, when $\beta_- = \beta_3=0,$ and $\beta_+ \neq 0,$ when   $\beta_+ = \beta_3=0,$ and $\beta_- \neq 0,$ or when $  \beta_-  \neq 0,  \beta_- \neq 0 $ and $\beta_3 \neq 0,$  but $ \frac{\beta_3 }{2 \beta_+} = -   \frac{2 \beta_-}{\beta_3} .$

\subsubsection{Case $\beta_- = \beta_3 =0$ and $\beta_+ \neq 0$}
In the special case where $b=0,$  with $\beta_- = \beta_3 =0$ and $\beta_+ \neq 0,$ the matrix $M$ in \ref{M-matrix-general-expression}) becomes  triangular superior and the corresponding  linear differential equation system (\ref{system-eqs}) can be integrated systematically row by row, from bottom to up  or from up to bottom. As, in this case,  all diagonal elements of $M$ are equal to $\beta,$ we can try a solution of the type
$\psi^j_m (\zeta)= e^{\beta \zeta} \varphi^j_m (\zeta), \;  m=-j, \cdots, j.$  Inserting these functions in equation  (\ref{system-eqs}) and simplifying the expressions we get a simpler system of linear differential equations.  For example, starting from  bottom to up, we deduce 

\begin{equation}
\frac{d}{d \zeta} \varphi^j_j (\zeta) = 0, \quad  \mbox{then} \quad  \varphi^j_j (\zeta) =  \tilde{\varphi}^j_j (0),
\end{equation}   
where $\tilde{\varphi}^j_j (0)$ is an integration constant. Continuing the process, we find
\begin{equation}
\frac{d}{d \zeta} \varphi^j_{j-1} (\zeta) = -  \sqrt{(2j)!} \beta_+  \varphi^j_j (\zeta),
\end{equation}
then inserting in this equation the newly  obtained value of   $\varphi^j_j (\zeta)$ and integrating again we get

\begin{equation}
\varphi^j_{j-1} (\zeta) =  \varphi^j_{j-1} (0)  -  \sqrt{(2j)!} \zeta \beta_+  \varphi^j_j (0),
\end{equation}
where $\tilde{\varphi}^j_j (0)$ is a new integration constant. Continuing with the process, we can show that the $k$ th integration leads to

\begin{equation}
\varphi^j_{j-k} (\zeta) = \sum_{n=0}^{k} {(-1)}^n  \sqrt{ \frac{(2j-k+n)!(k!)}{(2j-k)! (k-n)!}}  \frac{(\zeta \beta_+)^n}{n!}  
\tilde{\varphi}^j_{j-k+n} (0),
\end{equation}
where  $\tilde{\varphi}^j_{j-k+n} (0)$ are the corresponding  integration constants. Let us recall that these function are the component of the the state representing the general solution of the linear differential equation system (\ref{system-eqs}), this latter then writes
 
\begin{equation}
\mid \Psi (\zeta) \rangle^j =  \sum_{k=0}^{2j} \psi^j_{j-k} (\zeta) \otimes \mid j, j-k\rangle =\sum_{k=0}^{2j} e^{\beta \zeta} \; \varphi^j_{j-k} (\zeta) \otimes \mid j, j-k\rangle,
\end{equation}
or more explicitly
 
\begin{equation} 
 \mid \Psi (\zeta) \rangle^j = \sum_{k=0}^{2j} \;\sum_{n=0}^{k} {(-1)}^n   \sqrt{ \frac{(2j-k+n)!(k!)}{(2j-k)! (k-n)!}} \; e^{\beta \zeta} \;  \frac{(\zeta \beta_+)^n}{n!}  
\tilde{\varphi}^j_{j-k+n} (0) \otimes \mid j, j-k\rangle.
\end{equation}
Interchanging the $ k,n$ order of summation and then defining a new index $m$ in place of $k,$  $m=j+n-k$, and again, interchanging   the $n,m$ order of summation we get the same state  but now	 factorized according to the integration constants, that is, 

\begin{equation}
\mid \Psi (\zeta) \rangle^j = \sum_{m=-j}^{j} \tilde{\varphi}^j_{m} (0) \left[  \sqrt{\frac{(j+m)!}{(j-m)!}} 
   \sum_{n=0}^{j+m} {(-1)}^n  \sqrt{\frac{ (j+n-m)!}{(j+m-n)!}} \; e^{\beta \zeta} \;  \frac{(\zeta \beta_+ )^n }{n!} 
    \otimes \mid j, m-n \rangle \right] ,
\end{equation}
from where we can extract the $2j+1$ linearly independent states  
\begin{equation}
\mid \psi (\zeta) \rangle^j_m  = \sqrt{\frac{(j+m)!}{(j-m)!}} \sum_{n=0}^{j+m} {(-1)}^n  \sqrt{\frac{(j+n-m)!}{(j+m-n)! }}  e^{\beta \zeta}  
\frac{(\zeta \beta_+ )^n}{n!} \otimes \mid j , m-n \rangle, \quad  m=-j, \cdots,j.
\end{equation} 
Finally, returning  to the Fock space representation, we get the $2j+1$ normalized linearly  independent states that verify  the eigenvalue equation      
(\ref{eigenvalue-2}) for this choice of parameters:

\begin{equation}
\mid \psi [\beta , \beta_+]  \rangle^j_m  =\frac{
\sum_{n=0}^{j+m}  {(-1)}^n \sqrt{\frac{(j+ n-m)!}{(j+m -n)! }} 
\frac{( \hat{a}^\dagger  \beta_+)^n}{n!} \mid \beta \rangle \otimes \mid  j , m-n \rangle }{
\left[ \sum_{n=0}^{j+m} \frac{(j+n-m)!}{(j+m-n)!}  \frac{\| \beta_+ \|^{2n}}{n!}  \sum_{k=0}^{n}   \binom{n}{k} \frac{\| \beta \|^{2k}}{k!})  \right]^{\frac{1}{2}}}, \quad  m=-j, \cdots,j. \label{algebra-eigenstates-beta3=0-beta-=0}
\end{equation}

These states are not orthogonal to each other, they verify the relation

 \begin{equation} 
 {}^j_{\ell} \langle \psi [\beta, \beta_+] \mid \psi [\beta , \beta_+] \rangle^j_m  =  
 \frac{ \left( - \beta^\ast \beta_+ \right)^{m - \ell}  \sum_{n=0}^{j+m}    \frac{(j+n-m)!}{(j+m-n)!}  \frac{\|\beta_+ \|^{2 (n+\ell -m)}  } {(n  + \ell -m)!}  \sum_{k=0}^{n + \ell - m }   \binom{n + \ell - m}{k}    \frac{\|\beta\|^{2k} }{ (m - \ell+ k)!} }{\left[ \sum_{n=0}^{j+ \ell} \frac{(j+n-\ell)!}{(j+\ell -n)!}  \frac{\| \beta_+ \|^{2n}}{n!}  \sum_{k=0}^{n}   \binom{n}{k} \frac{\| \beta \|^{2k}}{k!})  \right]^{\frac{1}{2}}
\left[ \sum_{n=0}^{j+m} \frac{(j+n-m)!}{(j+ m -n)!}  \frac{\| \beta_+ \|^{2n}}{n!}  \sum_{k=0}^{n}   \binom{n}{k} \frac{\| \beta \|^{2k}}{k!})  \right]^{\frac{1}{2}}}
, \end{equation}
 where $\ell \leq m,$  and when   $\ell \geq m,$ we  only take the conjugate of this expression and then we  interchange $m$ with $\ell.$    

Finally, the  general solution of  (\ref{eigenvalue-2})   in this case is given by  

\begin{equation}
\mid \Psi [\beta, \beta_+ ] \rangle^j = \sum_{m=-j}^{j} \tilde{\varphi}^j_m (0) \mid \psi [\beta , \beta_+] \rangle^j_m, \label{general-solution-beta3=0,beta-=0}
\end{equation}
with  $\mid \psi [\beta ,\beta_+] \rangle^j_m, \; m=-j, \cdots,j,$  being the normalized states  (\ref{algebra-eigenstates-beta3=0-beta-=0}). We note that by using the relation $ \hat{D}^\dagger (\beta)  \hat{a}^\dagger \hat{D}(\beta) = \hat{a}^\dagger + \beta^\ast,$ where  $\hat{D}(\beta) = \exp( \beta \hat{a}^\dagger - \beta^\ast \hat{a})$ is the displacement operator, these states   can be written in the form
 
 \begin{equation}
\mid \psi [\beta , \beta_+]  \rangle^j_m  = \hat{D} (\beta) \frac{
\sum_{n=0}^{j+m}  {(-1)}^n \sqrt{\frac{(j+ n-m)!}{(j+m -n)! }} 
\frac{( [\hat{a}^\dagger + \beta^\ast] \beta_+)^n}{n!} \mid 0 \rangle \otimes \mid  j , m-n \rangle }{
\left[ \sum_{n=0}^{j+m} \frac{(j+n-m)!}{(j+m-n)!}  \frac{\| \beta_+ \|^{2n}}{n!}  \sum_{k=0}^{n}   \binom{n}{k} \frac{\| \beta \|^{2k}}{k!})  \right]^{\frac{1}{2}}} =   \hat{D} (\beta) \mid  \tilde{\psi} [\beta , \beta_+]  \rangle^j_m,   \label{algebra-eigenstates-beta3=0-beta-=0-D} 
\end{equation}
where $ m=-j, \cdots,j, $ with the  normalized states $\mid  \tilde{\psi} [\beta , \beta_+]  \rangle^j_m $ given  explicitly  by
\begin{equation}
\mid  \tilde{\psi} [\beta , \beta_+]  \rangle^j_m =
\frac{
\sum_{n=0}^{j+m}  {(-1)}^n \sqrt{\frac{(j+ n-m)!}{(j+m -n)! }} 
\frac{  \beta_+^n}{n!}    \sum_{k=0}^{n}  \binom{n}{k}  (\beta^\ast)^k    \sqrt{(n-k)!}  \mid n-k \rangle \otimes \mid  j , m-n \rangle }{
\left[ \sum_{n=0}^{j+m} \frac{(j+n-m)!}{(j+m-n)!}  \frac{\| \beta_+ \|^{2n}}{n!}  \sum_{k=0}^{n}   \binom{n}{k} \frac{\| \beta \|^{2k}}{k!})  \right]^{\frac{1}{2}}} 
 , \quad  m=-j, \cdots,j, \label{algebra-eigenstates-beta3=0-beta-=0-D-D} 
\end{equation}
 or in a more compact form by
\begin{equation} \mid  \tilde{\psi} [\beta , \beta_+]  \rangle^j_m =
\frac{\exp\left[  - ( \hat{a}^\dagger + \beta^\ast ) \beta_+ \hat{J}_-     \right]      \; \mid 0 \rangle \otimes \mid j , m \rangle } 
{\sqrt{\tilde{\mathcal{N}}^j_m  [\beta, \beta_+]  }} , \quad  m=-j, \cdots,j, \label{algebra-eigenstates-beta3=0-beta-=0-D-J-}
\end{equation}
where
\begin{equation}
\tilde{\mathcal{N}}^j_m  [\beta, \beta_+] =  \frac{(j+m)!}{(j-m)!} \left[ \sum_{n=0}^{j+m} \frac{(j+n-m)!}{(j+m-n)!}  \frac{\| \beta_+ \|^{2n}}{n!}  \sum_{k=0}^{n}   \binom{n}{k} \frac{\| \beta \|^{2k}}{k!}  \right].
 \end{equation}
 
\subsubsection{Case $\beta_+ =\beta_3=0$ but $\beta_- \neq 0,$ }
 
In the case $\beta_+ =\beta_3=0$ but $\beta_- \neq 0,$ the process of obtaining the solution of (\ref{system-eqs}) is  analogous to the one followed above.  It is not necessary to repeat it here, it is better to adapt the  equation (\ref{algebra-eigenstates-beta3=0-beta-=0}) to obtain the solution. Indeed,  by changing in this latter  $m$ by $-m$ on the top of the summation  symbol and in the coefficients, $n$ by $-n$ in the $su(2)$ basis states and $\beta_+$ by $\beta_- ,$  we get

\begin{equation}
\mid \psi [\beta, \beta_- ] \rangle^j_m  =  \frac{ \sum_{n=0}^{j-m}  {(-1)}^n \sqrt{\frac{(j+ n+m)!}{(j-m -n)! }} 
\frac{( \hat{a}^\dagger  \beta_-)^n}{n!} \mid \beta \rangle \otimes \mid j , m+n \rangle }{
\left[ \sum_{n=0}^{j-m} \frac{(j+n+m)!}{(j-m-n)!}  \frac{\| \beta_- \|^{2n}}{n!}  \sum_{k=0}^{n}   \binom{n}{k} \frac{\| \beta \|^{2k}}{k!})  \right]^{\frac{1}{2}}}= \hat{D} (\beta) \mid  \tilde{\psi} [\beta , \beta_-]  \rangle^j_m  , \quad  m=-j, \cdots,j, \label{algebra-eigenstates-beta3=0-beta+=0}
\end{equation}
 where the normalized states $\mid  \tilde{\psi} [\beta , \beta_-]  \rangle^j_m $ are given by
\begin{equation} \mid  \tilde{\psi} [\beta , \beta_-]  \rangle^j_m =
\frac{\exp\left[  - ( \hat{a}^\dagger + \beta^\ast ) \beta_- \hat{J}_+     \right]      \; \mid 0 \rangle \otimes \mid j , m \rangle } 
{\sqrt{\tilde{\mathcal{N}}^j_m  [\beta, \beta_+]  }} , \quad  m=-j, \cdots,j, \label{algebra-eigenstates-beta3=0-beta+=0-D-J-}
\end{equation}
where
\begin{equation}
\tilde{\mathcal{N}}^j_m  [\beta, \beta_-] =  \frac{(j-m)!}{(j+m)!} \left[ \sum_{n=0}^{j-m} \frac{(j+n+m)!}{(j-m-n)!}  \frac{\| \beta_- \|^{2n}}{n!}  \sum_{k=0}^{n}   \binom{n}{k} \frac{\| \beta \|^{2k}}{k!}  \right].
 \end{equation}
Hence, in this case,  the general solution of   (\ref{eigenvalue-2})   is given by  
 \begin{equation}
\mid \Psi [\beta, \beta_- ] \rangle^j = \sum_{m=-j}^{j} \tilde{\varphi}^j_m (0) \mid \psi [\beta , \beta_-] \rangle^j_m. \label{general-solution-beta3=0,beta+=0}
\end{equation}

\subsubsection{Case  $\beta_+  \neq  0 ,   \beta_- \neq  0  $ and  $\beta_3  \neq 0$ but  $\frac{\beta_3 }{2 \beta_+} = - \frac{ 2 \beta_- }{   \beta_3}$ }
\label{appd}
When $ \beta_+ \neq 0, $ $\beta_- \neq 0 $ and $\beta_3 \neq 0,$ and  $ \frac{\beta_3 }{2 \beta_+} = - \frac{ 2 \beta_- }{   \beta_3},$     the parameter  $b=0,$ i.e.,  matrix $M$ in \ref{M-matrix-general-expression}) has again $2j+1$   repeated eigenvalues equal to $\beta.$    Then,  again,  the differential  equation system  (\ref{system-eqs}) must be integrated term by term.  As in appendix \ref{appb}, 
 we can first try a solution of the type $\psi^j_m (\zeta)= exp(\beta \zeta ) {\varphi}^j_m (\zeta), \;  m=-j, \cdots, j,$ and then manipulate the equations to  find an isolated ordinary differential equation of $2j+1$ order  for ${\varphi}^j_{-j}  (\zeta),$ that is,  $\prod_{m=-j}^{j} \left(  \frac{d}{d\zeta} + j \beta_3\right) {\varphi}^j_{-j}  (\zeta)=0. $  The solution of this latter is given by  $ {\varphi}^j_{-j}  (\zeta)= \exp( -j \beta_3 \zeta ) \sum_{q=0}^{2j}  A_q \zeta^q, $ where $A_q, \quad q=0, \cdots, 2j,$ are integration constants.  Finally, reinserting this last solution into   (\ref{system-eqs}), integrating  interactively  function   by function  and factorizing the resulting expressions in terms of the integration constants $A_q,$ we get the general solution 
 
\begin{eqnarray}
\mid \Psi (\zeta ) \rangle^j &=& \sum_{q=0}^{2j} A_{q}   \sum_{k=0}^{q}  (-1)^{k}  \binom{q}{k} \frac{(2j-k)!}{(2j)!} \nonumber \\ &\times& 
  \exp(\beta \zeta)  \; \zeta^{q-k}   \left(\frac{1}{\beta_+} \right)^{k}   \otimes  \left\{ \frac{d^k}{d\vartheta^k} \left[\sum_{r=0}^{2j}  \sqrt{\binom{2j}{r}} \vartheta^r  \mid j ,  -j+ r\rangle \right]_{\vartheta = \frac{\beta_3}{2 \beta_+} = - \frac{2 \beta_-}{\beta_3}} \right\} ,
 \end{eqnarray}
 which  after some manipulations and redefinition of the integration constants   takes the form
 
 \begin{eqnarray}
\mid \Psi (\zeta)  \rangle^j &=& \frac{1}{\sqrt{\mathcal{N}^j}} \sum_{m=-j}^{j}  \tilde{\varphi}^j_{m} (0) \sum_{n=0}^{j+m}  (-1)^n \;
\frac{(j+n-m)!}{(j+m-n)!}  \nonumber\\ &\times&   e^{\zeta \beta}\;   \frac{\left( \zeta \beta_+ \right)^n}{n!}   \otimes \sum_{\ell=0}^{j+n-m} \sqrt{\frac{(j+m - n + \ell)!}{(j+n-m-\ell)!}} 
\; \frac{\vartheta^\ell}{\ell !}  \; \mid  j, m-n + \ell \rangle,
\end{eqnarray}   
where  $\vartheta = \frac{\beta_3 }{ 2 \beta_+} = - \frac{2 \beta_- }{ \beta_3}$ and $\mathcal{N}^j $ is a generic normalization constant. Returning to the Fock space representation we finally arrive to

\begin{eqnarray}
\mid \Psi [\beta,\beta_+, \beta_-, \beta_3] \rangle^j &=& \frac{1}{\sqrt{\mathcal{N}^j}}  \sum_{m=-j}^{j}  \tilde{\varphi}^j_{m} (0) \sum_{n=0}^{j+m}   
(-1)^n \; \frac{(j+n-m)!}{(j+m-n)!}   \nonumber\\ &\times&   \frac{\left( \hat{a}^\dagger  \beta_+ \right)^n}{n!}   \mid \beta \rangle \otimes \sum_{\ell=0}^{j+n-m} \sqrt{\frac{(j+m - n + \ell)!}{(j+n-m-\ell)!}} 
\; \frac{\vartheta^\ell}{\ell !}  \; \mid  j, m-n + \ell \rangle,
\end{eqnarray}
This last equation can be written in the form

\begin{equation}
\mid \Psi [\beta,\beta_+, \beta_-, \beta_3] \rangle^j = \frac{1}{\sqrt{\mathcal{N}^j}}   \sum_{m=-j}^{j}  \tilde{\varphi}^j_{m} (0) \hat{D} (\beta) \mid \tilde{\psi} [\beta,\beta_+, \beta_-, \beta_3] \rangle^j_m \label{general-solution-AES-b=0-remaining-non-zero}
\end{equation}
 where $\mid \tilde{\psi} [\beta,\beta_+, \beta_-, \beta_3] \rangle^j_m, \; m=-j \cdots,j,$ is a set of $2j+1$ normalized linearly independent states are  given by

 \begin{eqnarray}
\mid \tilde{\psi} [\beta,\beta_+, \beta_-, \beta_3] \rangle^j_m &=& \frac{1}{\sqrt{ \tilde{\mathcal{N}}^{j}_{m}  [ \beta, \beta_+, \beta_-, \beta_3 ]  }}   \sqrt{\frac{(j+m)!}{(j-m)!}}  \sum_{n=0}^{j+m}   
(-1)^n \; \frac{(j+n-m)!}{(j+m-n)!} \label{AES-b=0-all-different-zero}   \\ &\times&   \frac{\left( (\hat{a}^\dagger + \beta^\ast)  \beta_+ \right)^n}{n!}   \mid 0 \rangle \otimes \sum_{\ell=0}^{j+n-m} \sqrt{\frac{(j+m - n + \ell)!}{(j+n-m-\ell)!}} 
\; \frac{\vartheta^\ell}{\ell !}  \; \mid  j, m-n + \ell \rangle,   \nonumber
\end{eqnarray}
where $m=-j, \cdots,j,$ or
 \begin{eqnarray}
\mid \tilde{\psi} [\beta,\beta_+, \beta_-, \beta_3] \rangle^j_m &=& \frac{1}{\sqrt{ \tilde{\mathcal{N}}^{j}_{m}  [ \beta, \beta_+, \beta_-, \beta_3 ]  }}  \sum_{n=0}^{j+m}  \quad \sum_{\ell=0}^{j+n-m}  \; \frac{(\vartheta \hat{J}_+ )^\ell}{\ell !}   \nonumber \\
  &\times&   (-1)^n \;     \frac{\left( (\hat{a}^\dagger + \beta^\ast)  \beta_+ \hat{J_-} \right)^n}{n!} 
   \mid 0  \rangle \otimes 
\mid  j, m  \rangle, \quad m=-j, \cdots,j,  \label{AES-b=0-all-different-zero-operator-form} 
\end{eqnarray}
where

\begin{eqnarray}
  \tilde{\mathcal{N}}^{j}_{m}  [ \beta, \beta_+, \beta_-, \beta_3 ]   &=& \frac{(j+m)!}{(j-m)!}  \sum_{n=0}^{j+m} \sum_{\tilde{n}}^{j+m} (-1)^{(\tilde{n} - n)} \frac{(j + n -m)! (j+\tilde{n} -m)!}{(j+ m -n)!(j+m -\tilde{n})!}  \frac{(\beta^\ast_{+})^{\tilde{n}}  (\beta_{+})^{n}}{\tilde{n}! n!}
 \nonumber \\ &\times & \sum_{\ell =0}^{j+n-m} \frac{(j+ m - n + \ell)!}{(j+ n-m - \ell)!} \frac{\|\vartheta\|^{2 \ell} {\vartheta^\ast}^{(\tilde{n} - n)}}{  (\ell + \tilde{n} - n)! \ell! } \sum_{k=0}^{min(\tilde{n},n)} \binom{\tilde{n}}{k} \binom{n}{k} k! \beta^{(\tilde{n} -k)} {\beta^{\ast}}^{(n-k)},
 \end{eqnarray}
where $min \; (\tilde{n},n)$ means the smaller between $\tilde{n}$ and $n.$

\section{su(2) algebra eigenstates, a useful operator}
\label{appb}
\numberwithin{equation}{section}
\renewcommand{\theequation}{\thesection\arabic{equation}}
Let us recall that, in the case  $\beta_- \neq 0, \beta_+ \neq 0$  and for arbitrary $\beta_3,$  such that  
\begin{equation}  b=\sqrt{4 \beta_+ \beta_- + \beta_3^2} \neq 0,
\end{equation}  in the process of computing  the $su(2)$ algebra eigenstates, we have built an operator $T$ with the following characteristics  

\begin{equation}
\hat{T} = \exp\left( - \frac{\tilde{\theta}}{2} [ e^{-i \tilde{\phi}} \hat{J}_+ -  e^{i \tilde{\phi}} \hat{J}_-] \right), \label{master-operator-T}\end{equation}
where
\begin{equation}
\frac{\tilde{\theta}}{2} = \arctan \left( \sqrt{\frac{b - \beta_3}{ b + \beta_3}} \right),  \quad  \mbox{and} \quad  e^{i \tilde{\phi}}= \sqrt{\frac{\beta_+}{\beta_-}},  \label{phases-conditions} \end{equation}
so that applied on a pure state  $ \mid j , m \rangle$ of the irreducible representation $j,$ gives as result  an  eigenstate of the the general operator $ [\beta_{-} \hat{J}_+ + \beta_{+} \hat{J}_- + \beta_3 \hat{J}_3]$ associated  to the eigenvalue $mb,$ that is
 \begin{equation}
[\beta_{-} \hat{J}_+ + \beta_{+} \hat{J}_- + \beta_3 \hat{J}_3] \left(\hat{T} \mid j , m \rangle\right) = mb \left(\hat{T} \mid j , m \rangle \right).
\end{equation}
 An explicit and disentangled form of it is given by

\begin{equation}
\hat{T} = \exp\left( - \frac{2 \beta_-}{ b + \beta_3} \hat{J}_+  \right) \;
 \exp\left( \log  \left(\frac{2 b}{ b + \beta_3}\right) \hat{J}_3  \right)\;
\exp\left(  \frac{2 \beta_+}{ b + \beta_3} \hat{J}_-  \right).
\end{equation}
Then, the action of $\hat{T}$ on a pure state $ \mid j , \ell \rangle$ is given explicitly by
\begin{align}
\hat{T} \mid j , \ell \rangle = {\left(\frac{2 b}{b + \beta_3}\right)}^\ell
\exp\left( - \frac{2 \beta_-}{ b + \beta_3} \hat{J}_+  \right) \;
\exp\left(  \frac{\beta_+}{ b } \hat{J}_-  \right)\;  \mid j , \ell \rangle \nonumber \\
= {\left(\frac{2 b}{b + \beta_3}\right)}^\ell 
 \sqrt{\frac{(j + \ell )!}{(j-\ell)!}}\nonumber \\ \times \left[ \sum_{u=\ell +1}^{j}
  {(\frac{- 2 \beta_{-}}{b + \beta_3})}^{(u-\ell)}  \sqrt{\frac{(j + u )!}{(j-u)!}} 
  \sum_{n=0}^{(j + \ell)}  (-1)^n \left(\frac{(1-\beta_3 / b)}{2}\right)^n \frac{(j-\ell+n)!}{n!(n+u-\ell)!(j+\ell-n)!} \right. \nonumber\\
 + \left.  \sum_{u=-j}^{\ell}
  {(\frac{ \beta_{+}}{b})}^{(\ell-u)}  \sqrt{\frac{(j + u )!}{(j-u)!}}  \sum_{n=0}^{(j + u)}
     (-1)^n \left(\frac{(1-\beta_3 / b)}{2}\right)^n \frac{(j- u +n)!}{n!(n+ \ell -u)!(j+ u -n)!} \right] \mid j, u \rangle, \nonumber \\ 
 \end{align}
which after some algebraic manipulations can be written in the form
 
 \begin{eqnarray}
 \hat{T} \mid j , \ell \rangle = 
   {\left(\frac{2 b}{b + \beta_3}\right)}^\ell  \nonumber \\ \times 
 \left[ \sqrt{\frac{(j+\ell)!(j-\ell)!}{(2j)!}}  \sum_{u=\ell+1}^{j} 
    \sqrt{\frac{(2j)!}{(j+u)!(j-u)!}} \left(\frac{- 2 \beta_{-}}{b + \beta_3}\right)^{u- \ell}  P_{j+\ell}^{-\ell+ u,-\ell-u} \left(\frac{\beta_3}{b}\right) \right.
 \nonumber\\  + 
 \left. \sqrt{\frac{(2j)!}{(j+\ell)!(j-\ell)!}} \sum_{u=-j}^{\ell}
    \sqrt{\frac{(j+u)!(j-u)!}{(2j)!}}  {\left(\frac{\beta_+}{b}\right)}^{\ell -u}  P_{j+u}^{-u+ \ell,-u-\ell} \left(\frac{\beta_3}{b}\right) \right] \mid j, u \rangle, \nonumber
\end{eqnarray}
 where
 \begin{equation}
  P_{j+u}^{-u+ \ell,-u-\ell} \left(\frac{\beta_3}{b}\right) = \frac{(j+\ell)!}{(j-u)!} \sum_{n=0}^{(j + u)} (-1)^n \left(\frac{(1-\beta_3 / b)}{2}\right)^n  \frac{(j- u +n)!}{n! (n+ \ell -u)!(j+ u -n)!} .
 \end{equation}
are the Jacobi polynomials.  

We are interested in the matrix elements of $\hat{T},$ that is 

\begin{equation}
T^j_{m \ell}=\langle j , m \mid \hat{T} \mid j,\ell \rangle,
\end{equation}
which are equal to
\begin{eqnarray}
T^j_{m \ell}[\beta_+,\beta_- ,\beta_3] =  \begin{cases} 
 \left(\frac{- 2 \beta_-}{b+\beta_3}\right)^m  \left(\frac{-b}{\beta_-} \right)^\ell
 \sqrt{\frac{(j+\ell)!(j-\ell)!}{(j+m)!(j-m)!}}  P_{j+\ell}^{-\ell+ m,-\ell-m} \left(\frac{\beta_3}{b}\right), \; \;
  \ell +1  \leq m \leq j \nonumber \\ \\
   \left(\frac{ 2 \beta_+}{b+\beta_3}\right)^\ell  \left(\frac{b}{\beta_+} \right)^m
  \sqrt{\frac{(j+m)!(j-m)!}{(j+\ell)!(j-\ell)!}}  P_{j+m}^{-m+\ell,-m-\ell} \left(\frac{\beta_3}{b}\right),
  \; -j \leq m \leq \ell . \end{cases}  \nonumber \\ \label{matrix-elements-T}
\end{eqnarray}
 
\subsection{Unitary case} 
 
 In the special case when   $\|\beta_+\| = \| \beta_-\|$ and $\beta_3 \beta^{\ast}_{-} - \beta_{3}^{\ast} \beta_+ =0,$ these matrix elements correspond to the  matrix elements  of a unitary matrix. Indeed, taking into account these conditions in equation (\ref{phases-conditions}), we realize that the parameters  $\tilde{\theta}$ and $\tilde{\phi}$ become real and then $\hat{T}$ becomes unitary.

Now we will show that the  unitary matrix $T,$ which matrix elements are given in (\ref{matrix-elements-T}), makes  the normal matrix $M$ given in (\ref{M-matrix-general-expression}) diagonal. Indeed, if we express the $su(2)$ algebra eigenstate  $\mid \psi \rangle^j$  in terms of the $2j+1$ basis states spanning the $j$ irreducible representation space in the form $ \mid \psi \rangle^j = \sum_{m=-j}^{j} C^j_m \mid j , m \rangle$ and insert it
in the equation   

\begin{equation}
[\beta_{-} \hat{J}_+ + \beta_{+} \hat{J}_- + \beta_3 \hat{J}_3]  \mid  \psi \rangle^j = \Gamma  \mid  \psi \rangle^j,  \label{su(2)-eigenvalue-equation}
\end{equation}
we can see that the algebraic linear equation system for the coefficients $C^j_m, -j \leq m \leq j, $ can be written in the form

\begin{equation}
(\beta I - M)
\begin{pmatrix}
 C^{j}_{-j}  \\
C^{j}_{-j+1}  \\
\vdots \\
 C^{j}_{j-1} \\
C^{j}_{j} \rangle
\end{pmatrix} =\Gamma \begin{pmatrix}
 C^{j}_{-j}  \\
C^{j}_{-j+1}  \\
\vdots \\
 C^{j}_{j-1} \\
C^{j}_{j}  \label{eq-coefficients}
\end{pmatrix},\end{equation}
where $M$ is given by equation (\ref{M-matrix-general-expression}). It is clear from this last equation that the unitary matrix which diagonalize $\mathcal{M}= \beta I - M,$ also diagonalize $M.$

 As $\hat{T} \mid j; u\rangle,  $ for all fixed $u$ such that $ u =  -j, \cdots, j,$ is a normalized $su(2)$ algebra eigenstate  verifying  
(\ref{su(2)-eigenvalue-equation}), with eigenvalue $\Gamma^j_u = u b,$

\begin{equation}
 \mid \psi\rangle^j_u = \sum_{\ell=-j}^{j} C^j_{\ell} \mid j,\ell \rangle = {\hat T} | j , u \rangle, \label{eq-two-ways} 
\end{equation}
 Projecting both sides of equation (\ref{eq-two-ways}) on a generic pure state $\mid j, \ell \rangle$ we obtain a connection between the $C^j_\ell$ coefficients and the matrix elements of $\hat{T}$

\begin{equation}
C^j_{\ell} = T_{\ell u}, \quad \ell =  -j, \cdots, j. \label{u-coeffients}
\end{equation}
On the other hand, from  (\ref{eq-coefficients}) we can see that the explicit form of the coefficient equation system   is given by
\begin{equation}
\sum_{\ell = -j}^{j} \mathcal{M}_{m \ell} C^j_{\ell} = \Gamma^{j} C^j_{m}, \quad m =  -j, \cdots, j  ,  
\end{equation} 
which  is verified by (\ref{u-coeffients})  for a given $u,$ i.e.,  
\begin{equation}
\sum_{\ell=-j}^{j} \mathcal{M}_{m \ell} T_{\ell u} = \Gamma^{j}_u  T_{m u},  \quad m =  -j, \cdots, j, \quad \Gamma^{j}_u = u b. \label{equation-coefficients-eigenvalues}
\end{equation} 
 Finally operating from the left  with the unitary matrix $T^\dagger $ on both sides of equation (\ref{equation-coefficients-eigenvalues}) we obtain 
 
\begin{equation}
\sum_{m=-j}^{j} \sum_{\ell=-j}^{j} T^\dagger_{\varrho m}  \tilde{M}_{m \ell}  T_{\ell u} = \Gamma^{j} \sum_{m=-j}^{j} {T^\dagger}_{\varrho m} T_{m u}= \Gamma^{j}_{u} \delta_{\varrho u}, 
\end{equation} 
for all chosen $u,$ then we conclude that $T$ diagonalize $\tilde{M} ,$   and consequently also diagonalize $M.$ 

In the case when $M$ is not normal but diagonalizable,   the same argument is valid, the only thing to do in the argumentation of the above process is to change $T^\dagger$ by $T^{-1}.$

%\printbibliography

\end{document}